\documentclass[preprint,aip,pof]{revtex4-1}
\usepackage[utf8]{inputenc}
\usepackage{amssymb}
\usepackage{graphicx}
\usepackage{subfigure}
\usepackage{bm}
\usepackage{amsmath}
\usepackage{threeparttable}
\usepackage{CJK}
\usepackage{hyperref}
\usepackage{color}
\usepackage{booktabs}
\usepackage{bookmark}
\usepackage{array}
\usepackage{multirow}
\usepackage{natbib}
\usepackage{setspace}

\begin{document}
\begin{CJK*}{UTF8}{gbsn}
\title{A direct relaxation process for particle methods in gas kinetic theory}

\author{Sirui Yang (杨思睿)}
\email[]{ysr1997@mail.nwpu.edu.cn}
\affiliation{School of Aeronautics, Northwestern Polytechnical University, Xi'an, Shaanxi 710072, China}
\author{Sha Liu (刘沙)}
\email[Corresponding author:]{shaliu@nwpu.edu.cn}
\affiliation{National Key Laboratory of Science and Technology on Aerodynamic Design and Research, Northwestern Polytechnical University, Xi'an, Shaanxi 710072, China}
\author{Chengwen Zhong (钟诚文)}
\email[]{zhongcw@nwpu.edu.cn}
\affiliation{National Key Laboratory of Science and Technology on Aerodynamic Design and Research, Northwestern Polytechnical University, Xi'an, Shaanxi 710072, China}
\author{Junzhe Cao(曹竣哲)}
\affiliation{School of Aeronautics, Northwestern Polytechnical University, Xi'an, Shaanxi 710072, China}
\author{Congshan Zhuo(卓丛山)}
\affiliation{National Key Laboratory of Science and Technology on Aerodynamic Design and Research, Northwestern Polytechnical University, Xi'an, Shaanxi 710072, China}

\begin{abstract}
The multi-scale flow mechanism is crucial for force and heat loaded on near-space and reentry vehicles, the control of spacecraft by thrusters, the propelling and cooling of MEMS, etc. Since the continuum flow and rarefied flow often exist simultaneously, the modeling and prediction for such a multi-scale flow field is very complicated. One important and efficient way of predicting the multi-scale flow is constructing numerical methods by adapting the multi-scale properties of the temporal integral solutions (or equivalent characteristic line solutions) for the model equations in the gas-kinetic theory. The model equations can be classified into FP-type and BGK-type, and the numerical methods can have deterministic form or stochastic particle form. Since these numerical methods are strictly based on model equations, they are also restricted by the model equations. The difficulty and complexity in constructing model equation that has complete asymptotic preserving property for gas mixture with non-equilibrium internal energy will prevent the further extension of these methods. Therefore, this paper addresses the question whether a multi-scale numerical method can be established not on the model equations, but directly adapts the relaxation rates of macroscopic variables such as stress and heat flux, because these relaxation rates are the aggregate effect of microscopic particle collisions, and are the essential constrains when constructing model equations. Since the algorithm of multi-scale particle method based on BGK-type equation is clear and simple, its collision step is modified in this work, where the macroscopic variables firstly evolve according to their relaxation rates, and then the molecules that participate in inter-molecular collisions get their after-collision velocities from the after-evolution macroscopic variables. Therefore, this direct relaxation process does not depend on the model equations anymore. Finally, the validity and accuracy of the particle method with direct relaxation process are examined with homogenous relaxation case, Sod shock tube in both continuum and rarefied regime, high non-equilibrium normal shock structure, low speed cavity flow, and hypersonic cylinder flow in transitional regime.
\end{abstract}

\maketitle
\end{CJK*}

\section{\label{sec:introduction}Introduction}
The multi-scale mechanism from the continuum flow to the rarefied flow (or from the macroscopic flow to the microscopic flow) is crucial for the force and heat loaded on near-space and reentry vehicles, the control of spacecraft by thrusters~\cite{rv1, rv2}, the propelling and cooling of Micro-Electro-Mechanical System (MEMS)~\cite{mems1, mems2}, etc. Since the continuum flow and rarefied flow often exist simultaneously in a singe multi-scale flow field, the modeling and prediction become complicated: The continuum flow is governed by the Navier-Stokes (N-S) equation, and the rarefied flow is governed by the Boltzmann equation, while there is no multi-scale governing equation for the transitional flow between continuum and rarefied ones. In this transitional regime, the N-S methods are inaccurate, and rarefied flow methods such as DSMC are computationally unaffordable.

Therefore, besides the hybrid continuum and rarefied methods~\cite{sun2004hybrid, kolobov2007unified} that are directly based on the flow field decomposition technologies, several multi-scale methods are proposed based on the model equations in the gas-kinetic theory during the past decade. These methods are based on the Bhatnagar-Gross-Krook-type (BGK-type) equations~\cite{Bhatnagar1954A, Holway1966New, 1968Generalization, Rykov2007Numerical} or Fokker-Planck-type (FP-type) equations~\cite{Cercignani1990The, gorji2011fokker-planck, mathiaud2016a}. The most important and apparent feature of these methods is utilizing the temporal integral solution (or equivalent difference solution along the characteristic-line). These multi-scale methods can be categorized into deterministic methods and stochastic methods.

The deterministic methods for all flow regimes from rarefied to continuum ones, such as Unified Gas-Kinetic Scheme (UGKS)~\cite{xu2010unified, Chen2012A, liu2014unified, Li2018A}, Discrete UGKS (DUGKS)~\cite{wang2015a, guo2013discrete, Chen2019Conserved}, Gas-Kinetic Unified Algorithm (GKUA)~\cite{li2009gas, peng2016implicit, Wu2020On} and Improved Discrete Velocity Method (IDVM)~\cite{yang2018improved, yang2019an}, use the BGK-type equations. The temporal integral or characteristic-line solution of BGK-type model equations is utilized for constructing the numerical flux at the cell interface, achieving their multi-scale property. Benefitting from this multi-scale property and concrete Asymptotic Preserving (AP) property or even Unified Preserving (UP) property~\cite{guo2020On}, the scope of these methods is extended from flow predictions to the predictions of radiative transport~\cite{sun2017Amultidimensional}, plasma gas~\cite{Liu2017AUnified}, photon transport~\cite{Guo2016Discrete} and neutron transport~\cite{Shuang2019A} in multi-scale cases. Recently, the UGKS is translated into a particle version Unified Gas-Kinetic Wave-Particle (UGKWP) method~\cite{LiuUnified2020, zhu2019unified, ChenY2020A, Shi2020An}. By quantifying the model competition mechanism between the particle model and the N-S model, which is implied in multi-scale flows, the Simplified Unified Wave Particle (SUWP) method~\cite{2020Simplified} is proposed which can also be viewed as a simple version of UGKWP or an hybrid continuum and rarefied method in the algorithm level.

The stochastic methods based on the FP equations~\cite{jenny2010a, Gorji2012A, sadr2017a} utilize the temporal integral solutions of the corresponding Langevin equations to describe the transportation of particles. The transport and collision process are coupled in the temporal integrals, whose intrinsic mechanism is close to the physical reality, and leads to its multi-scale property. By coupling with the DSMC solver~\cite{Gorji2015Fokker} and reducing the numerical viscosity~\cite{fei2017a}, the validity and accuracy of the particle-FP method in all flow regimes are ensured. On the other hand, the stochastic methods based on the BGK equation~\cite{macrossan2001nu, gallis2011investigation, pfeiffer2018particle-based, tumuklu2016particle, Pfeiffer2018Extending} split the transportation of particle into two decoupled processes, a free transport process in the flow field and a collision process in local cells. This operator splitting treatment is the same as that of the DSMC, except that its collision process is obtained from the temporal integral solution of the homogenous model equation. The particle-BGK methods can also be coupled with the DSMC solver~\cite{Fei2021hybrid} and reduce their numerical viscosity by considering the coupled transport and collision process in a DUGKS-way~\cite{Fei2020A}.

At the present stage, these multi-scale methods are all based on the model equations. However, establishing accurate model equation for heterogeneous (monatomic, diatomic and polyatomic) gas mixture with non-equilibrium internal energies (such as rotational and vibrational ones) is a difficult task~\cite{Bisi2020A, Baranger2020A}. The only accurate BGK-type gas-mixture model equation that can fulfill the conservative property, H theorem (second law of thermodynamics), positive property (positive distribution, density and temperature), and concrete AP property (the correct constitutive relation with the correct transport coefficient) is that proposed in 2019 for simple monatomic gas mixtures~\cite{Todorova2019Derivation}. On the other hand, the FP-type one is also proposed for monatomic gas mixtures~\cite{Hepp2020A} in 2020. As to the thermal non-equilibrium gas mixture, if too many coefficients need to be fitted, the model equations will be mathematically very complicated (this complexity can be seen from the work in Ref.~\cite{1981Kinetic}).

In this paper, we address the question whether a multi-scale numerical method can be established not based on model equations, but based on the relaxation rates of macroscopic variables (such as stress, heat flux and non-equilibrium internal energies), because these relaxation rates are more essential and they are actually the constrains that should be fulfilled when constructing model equations. Therefore, the corresponding Direct Relaxation (DR) process is designed, from which the after-collision velocities of particles can be determined without knowing the model equation. Moreover, since the algorithm of particle-BGK method is clear and concise, it is chosen in this work, and its collision process is replaced by the DR process. More specially, the modified particle method has the same free transport process as both the DSMC and the particle-BGK method, while its collision process is based on the DR process proposed in this paper. The remainder of this paper is arranged as follows: Section~\ref{sec:sp} is a quick overview of the gas-kinetic theory and the particle-BGK method; Section~\ref{sec:DR} is the DR process proposed in this paper and the details of the modified numerical method; Section~\ref{sec:Test} is the numerical validation and verification of the present method with a wide range of Mach (Ma) and Knudsen (Kn) numbers, the discussion and conclusion are in Section~\ref{sec:conclusion}.

\section{\label{sec:sp}Gas-kinetic theory and the particle-BGK method}
\subsection{The gas-kinetic theory}
Molecular motions are described in terms of the distribution function $f(\mathbf{x},\bm{\xi},t)$ in the gas-kinetic theory, which is the number density of molecules with the velocity $\bm{\xi}$ that arrive at the location $\mathbf{x}$ at time $t$. The evolution of $f$ is governed by the following Boltzmann equation~\cite{Kremer2010An}:
\begin{equation}\label{eq:Boltzmann}
\frac{{\partial f}}{{\partial t}} + {\bm{\xi }} \cdot \frac{{\partial f}}{{\partial {\mathbf{x}}}} + {\mathbf{a}} \cdot \frac{{\partial f}}{{\partial {\bm{\xi }}}} = C\left( {f,f} \right),
\end{equation}
where $\mathbf{a}$ is the acceleration of molecule. The left-hand side of Eq.~\ref{eq:Boltzmann} is the free-transport part, while the right-hand side is the five-fold nonlinear integral collision part. In multi-scale methods, the BGK-type model equation is often used. Taking the simple BGK model equation for example, it is in the following form:
\begin{equation}\label{eq:BGK}
\frac{{\partial f}}{{\partial t}} + {\bm{\xi }} \cdot \frac{{\partial f}}{{\partial {\mathbf{x}}}} + {\mathbf{a}} \cdot \frac{{\partial f}}{{\partial {\bm{\xi }}}} = \frac{\left( {g - f} \right)}{\tau},
\end{equation}
where the Boltzmann collision term in Eq.~\ref{eq:Boltzmann} is replaced by a simple relaxation term on the right-hand side of Eq.~\ref{eq:BGK}. $\tau$ is the relaxation time defined as $\mu/P$, where $\mu$ and $P$ are the temperature-dependent dynamic viscosity and the pressure, respectively. The equilibrium distribution $g$ is the Maxwellian distribution with the maximum local entropy, which is in the form below:
\begin{equation}\label{eq:Maxwell}
g_{M} = {\left( {\frac{m}{{2\pi kT}}} \right)^{3/2}}\exp \left( { - \frac{\left|\mathbf{c}\right|^{2}}{{2kT}}} \right),
\end{equation}
where $n$, $\mathbf{c}$, $\mathbf{u}$, $T$, $k$ and $m$ are the number density, the peculiar velocity defined as $\bm{\xi}-\mathbf{u}$, the macroscopic velocity, the thermodynamic temperature, the Boltzmann constant, and the mass of molecule, respectively. For other models like Shakhov~\cite{1968Generalization}, ES-BGK~\cite{Holway1966New} and Rykov model~\cite{Rykov2007Numerical}, the stress and heat flux will be evolved in the expression of $g$ to get the right Prandtl (Pr) number.

In the gas-kinetic theory, the macroscopic mass density $\rho$, momentum density $\rho \mathbf{U}$, energy density $\rho\left|{\mathbf{U}}\right|^{2}/2 + \rho E$ (here $E$ is inertial energy per unit mass), stress tensor $\bm{S}$ and heat flux $\mathbf{Q}$ can be obtained from the distribution function $f$ with the following equations:
\begin{equation}\label{eq:constrain}
\begin{aligned}
&\rho  = \left\langle {mf} \right\rangle,\\
&\rho {\mathbf{U}} = \left\langle {m{\bm{\xi }}f} \right\rangle,\\
&\frac{1}{2}\rho\left|{\mathbf{U}}\right|^{2} + \rho e  = \frac{1}{2}\left\langle {m{\bm{\xi }} \cdot {\bm{\xi }}f} \right\rangle,\\
&{\mathbf{S}} =  - \left\langle {m{\mathbf{cc}}f} \right\rangle  + P{\mathbf{I}},\\
&{\mathbf{Q}} = \frac{1}{2}\left\langle {m{\bm{c }}\left( {{\bm{c }} \cdot {\bm{c }}} \right)f} \right\rangle,
\end{aligned}
\end{equation}
where $\mathbf{I}$ is an identity matrix, and the operator $\left\langle \cdot \right\rangle$ denotes an integral over of the whole velocity space as the following:
\begin{equation}\label{eq:operator}
\left\langle\cdot\right\rangle  = \int_{{R^3}} {\left(  \cdot  \right)} d{\bm{\xi }}.
\end{equation}

\subsection{The particle-BGK method}
The particle-BGK methods adopt the operator splitting treatment that splits the particle motion into free transport process in the flow field and collision process in local cells. The collision process is described by utilizing the integral solution of the homogenous BGK equation:
\begin{equation}\label{Eq:HomoBGK}
\frac{{\partial f}}{{\partial t}} = \frac{{g - f}}{\tau},
\end{equation}
which is actually the BGK equation Eq.~\ref{eq:BGK} without spacial gradient, and its temporal integral solution can be written in a discrete form as:
\begin{equation}\label{Eq:HomoIntegral}
{f^{i + 1}} = {e^{ - \frac{{\Delta t}}{\tau }}}{f^i} + \left( {1 - {e^{ - \frac{{\Delta t}}{\tau }}}} \right){g^i},
\end{equation}
where the superscript $i$ and $i+1$ are the indices of time step, and $\Delta t$ is the iteration time step of a numerical method (it can also be viewed as the time scale of the flow problem).

Using this integral solution in Eq.~\ref{Eq:HomoIntegral}, ${e^{ - \frac{{\Delta t}}{\tau }}}$ portion of particles are categorized as free particles and can keep their velocities, while $1-{e^{ - \frac{{\Delta t}}{\tau }}}$ portion of particles are categorized as the colliding particles and their velocities are renewed by sampling from $g^{i}$. After collision process, all particles transport freely in the flow field like the DSMC. The sampling technology can be chosen from direct sampling~\cite{Bird2003Molecular}, Acceptance-Rejection sampling~\cite{Bird2003Molecular}, Metropolis-Hastings sampling~\cite{pfeiffer2018particle-based} and importance sampling~\cite{Liu1996Metropolized} according to the mathematical expression of equilibrium state $g$ (such as in the form of Hermitian expansion or Gaussian distribution).

\section{\label{sec:DR}Particle methods with Direct Relaxation (DR) process}
In this section, the general framework of DR process is proposed first, in which the thermal non-equilibrium gas mixture is considered. Then, it is used for the simple single component monatomic gas to give right relaxation rates of stress and heat flux in the collision process. Finally, the route and important details of the modified particle methods with DR process are provided.

\subsection{General DR process}
To illustrate the DR process in a general way, we firstly assume that the precise BGK-type model equations for arbitrary thermal non-equilibrium gas mixtures exist and can be written in the following form (DR process dose not need the exact equation. Writing a model equation here is only for sake of clarity and convenience):
\begin{equation}
\begin{aligned}
\left\{ {\begin{array}{*{20}{c}}
{\dfrac{{\partial {f_1}}}{{\partial t}} = \dfrac{{{g_1} - {f_1}}}{{{\tau _1}}}}\\
{\begin{array}{*{20}{c}}
{...}\\
{\dfrac{{\partial {f_n}}}{{\partial t}} = \dfrac{{{g_n} - {f_n}}}{{{\tau _n}}}}\\
{...}
\end{array}}\\
{\dfrac{{\partial {f_N}}}{{\partial t}} = \dfrac{{{g_N} - {f_N}}}{{{\tau _N}}}}
\end{array}} \right.
\end{aligned},
\end{equation}
where $N$ components gas mixture is considered, and the subscripts are the indices of species. Here, the effects of self-collision, cross-collision and thermal non-equilibrium are all included in the equilibrium state $g_{n}$. Therefore, in the construction of model equations, the relaxation time $\tau$ which is related to the collision frequency can be chosen conveniently, while the equilibrium state $g$ is hard to design. In the DR treatment, only the relaxation time is needed. In the following procedure, we suppose the relaxation time $\tau$ is given. Using the simple backward Euler method to construct an implicit expression (the implicit expression is chosen for adapting large time scale), the discrete model equations become:
\begin{equation}
\begin{aligned}
\left\{ {\begin{array}{*{20}{c}}
{f_1^{i + 1} = \dfrac{{{\tau _1}}}{{{\tau _1} + \Delta t}}f_1^i + \dfrac{{\Delta t}}{{{\tau _1} + \Delta t}}g_1^{i + 1}}\\
{\begin{array}{*{20}{c}}
{...}\\
{f_n^{i + 1} = \dfrac{{{\tau _n}}}{{{\tau _n} + \Delta t}}f_n^i + \dfrac{{\Delta t}}{{{\tau _n} + \Delta t}}g_n^{i + 1}}\\
{...}
\end{array}}\\
{f_N^{i + 1} = \dfrac{{{\tau _N}}}{{{\tau _N} + \Delta t}}f_n^i + \dfrac{{\Delta t}}{{{\tau _N} + \Delta t}}g_n^{i + 1}}
\end{array}} \right.
\end{aligned}.
\end{equation}
This equation system indicates that, for arbitrary specie $n$, $\frac{{{\tau _n}}}{{{\tau _n} + \Delta t}}$  portion of molecules are free molecules, while $\frac{{\Delta t}}{{{\tau _n} + \Delta t}}$ portion of molecules are colliding ones that should be assigned a new after-collision velocity, according to the relaxation process. In DR treatment, the relaxation rate of macroscopic is directly used. For thermal non-equilibrium gas mixtures, the non-conservative macroscopic variables for each component are:
\begin{equation}
{{\bf{\Phi }}_n}{\rm{ = }}\left( {\rho_{n}{{\bf{U}}_n},{\rho_{n}E_{tran,n}},{\rho_{n}E_{rot,n}},{\rho_{n}E_{vib,n}},{{\bf{S}}_n},{{\bf{Q}}_{tran,n}},{{\bf{Q}}_{rot,n}},{{\bf{Q}}_{vib,n}}} \right),
\end{equation}
where the subscripts ``tran", ``rot", ``vib" stand for translational, rotational and vibrational degrees of freedom, respectively. Since this BGK-type model is accurate, its relaxation rate of ${{\bf{\Phi }}_n}$ is the same with that of the Boltzmann equation or its extended equation such as WCU. In other words, the relaxation rate of ${{\bf{\Phi }}_n}$ from the Boltzmann equation can be directly used in the DR process.

By writing all ${{\bf{\Phi}}_n}$ together as:
\begin{equation}
{\bf{\Phi}}{\rm{ = }}{\left( {{{\bf{\Phi }}_1},{{\bf{\Phi }}_2},...{{\bf{\Phi }}_n}}\right)},
\end{equation}
the strongly coupled relaxation rate can be abbreviated to:
\begin{equation}
\frac{{\partial {\bf{\Phi }}}}{{\partial t}}{\rm{ = }}{\bf{F}}\left( {\bf{\Phi }} \right).
\end{equation}
Using a backward Euler method for consistency, the non-conservative macroscopic variables after relaxation can be obtained by solving the following nonlinear (simple polynomial) system:
\begin{equation}
{{\bf{\Phi }}^{i + 1}} - \Delta {\rm{t}}{\bf{F}}\left( {{{\bf{\Phi }}^{i + 1}}} \right) - {{\bf{\Phi }}^i}{\rm{ = 0}}.
\end{equation}
After ${{\bf{\Phi }}^{i + 1}}$ is obtained, the non-conservative macroscopic variables for colliding molecules denoted by ${\bf{\Phi }}_n^*$ can be directly derived as:
\begin{equation}
{\bf{\Phi }}_n^*{\rm{ = }}{\bf{\Phi }}_n^{i + 1} - \frac{{{\tau _n}}}{{{\tau _n} + \Delta t}}{\bf{\Phi }}_n^i.
\end{equation}
Once ${\bf{\Phi }}_n^*$ is obtained, the after-collision velocity of component $n$ can be directly sampled from ${\bf{\Phi }}_n^*$. Given the up to third order moments, several mathematical form can be chosen for the distribution, such as generalized Gaussian and Hermitian distributions. In the next section, the DR method will be used in single component monatomic gas flows.

\subsection{DR process for single component gas}
For the single component monatomic case, the general BGK-type model equation can be written as Eq.~\ref{Eq:HomoBGK}, and the time-implicit discrete form becomes:
\begin{equation}
{f^{i + 1}} = \frac{\tau }{{\tau  + \Delta t}}{f^i} + \frac{{\Delta t}}{{\tau  + \Delta t}}{g^{i + 1}}.
\end{equation}
The non-conservative macroscopic variables to be recorded are stress and heat flux, therefore,
\begin{equation}
{\bf{\Phi }}{\rm{ = }}\left( {{\bf{S}},{\bf{Q}}} \right),
\end{equation}
and the relaxation rates of stress and heat from the Boltzmann equation are
\begin{equation}
\begin{aligned}
&{\dfrac{{\partial {\mathbf{S}}}}{{\partial t}} =  - \dfrac{1}{\tau }{\mathbf{S}}},\\
&{\dfrac{{\partial {\mathbf{Q}}}}{{\partial t}} =  - \dfrac{{\Pr }}{\tau }{\mathbf{Q}}}.
\end{aligned}
\end{equation}
For this simple case, the relaxation rates are not coupled. Using the same time-implicit scheme, the stress and heat flux after relaxation can be obtained as:
\begin{equation}\label{eq:non-con_renew}
\begin{aligned}
&{{{\bf{S}}^{i + 1}} = \dfrac{\tau }{{\tau {\rm{ + }}\Delta t}}{{\bf{S}}^i}},\\
&{{{\bf{Q}}^{i + 1}} = \dfrac{\tau }{{\tau {\rm{ + Pr}}\Delta t}}{{\bf{Q}}^i}}.
\end{aligned}
\end{equation}
Therefore, the non-conservative macroscopic variables for after-collision molecules can be directly obtained as:
\begin{equation}\label{eq:non-con_star}
\begin{aligned}
&{{\bf{S}}^{*} = {{\bf{S}}^{i + 1}} - \dfrac{\tau }{{\tau {\rm{ + }}\Delta t}}{{\bf{S}}^i} = 0},\\
&{{\bf{Q}}^{*} = {{\bf{Q}}^{i + 1}} - \dfrac{\tau }{{\tau {\rm{ + }}\Delta t}}{{\bf{Q}}^i} = \dfrac{{\Delta t\tau \left( {1 - \Pr } \right)}}{{\left( {\tau  + \Delta t} \right)\left( {\tau  + \Pr \Delta t} \right)}}{{\bf{Q}}^i}}.
\end{aligned}
\end{equation}
Also, the conservative macroscopic variables for after-collision molecules (denoted by ${{\bf{W}}^{\rm{*}}}$) are
\begin{equation}\label{eq:con_star}
{{\bf{W}}^{\rm{*}}}{\rm{ = }}\frac{{\Delta t}}{{\tau {\rm{ + }}\Delta t}}{{\bf{W}}^i}.
\end{equation}
In this paper, the third order Hermitian distribution is utilized to express the after-collision distribution as follows:
\begin{equation}\label{eq:gstar}
{g^*}_{H}\left( {\bf{\xi }} \right) = {g^*}_{M}\left( {\bf{\xi }} \right)\left\{ {1{\rm{ + }}\frac{1}{{2R{T^{*}}{P^{*}}}}\left[ { - \bf{S}^*:\bf{c}^*\bf{c}^* + \frac{2}{5}Q^{*} \cdot \bf{c}^{\rm{*}}\left( {\frac{{{{\left| {{\bf{c}^*}} \right|}^2}}}{{R{T^{*}}}} - 5} \right)} \right]} \right\},
\end{equation}
where $\bf{c}^{*}$ is $\bm{\xi}-\bf{U}^{*}$.
Up to now, the whole DR process has been finished. It can be summarized as:
\begin{itemize}
  \item Categorize the particles in cells into free transport ones and colliding ones by conducting the following test for each particle: if $rand < \frac{{\Delta t}}{\tau+\Delta t}$, the particle is a colliding particle, or else it is a free transport particle. Here $rand$ is a random real number between $0$ and $1$.
  \item Calculate both the conservative and non-conservative macroscopic variables for colliding particles in cells according to Eq.~\ref{eq:non-con_renew}, Eq.~\ref{eq:non-con_star} and Eq.~\ref{eq:con_star}.
  \item Assign new velocities to colliding particles according to Eq.~\ref{eq:gstar}.
\end{itemize}
After the DR process, the relaxation (or collision) process is complete, then the transport process can be conducted, which is the same with both the DSMC and the original particle-BGK methods.

\subsection{The algorithm of particle method with DR process}

\subsubsection{The framework of the present method}
\begin{figure}
\centering
\includegraphics[width=0.75\textwidth]{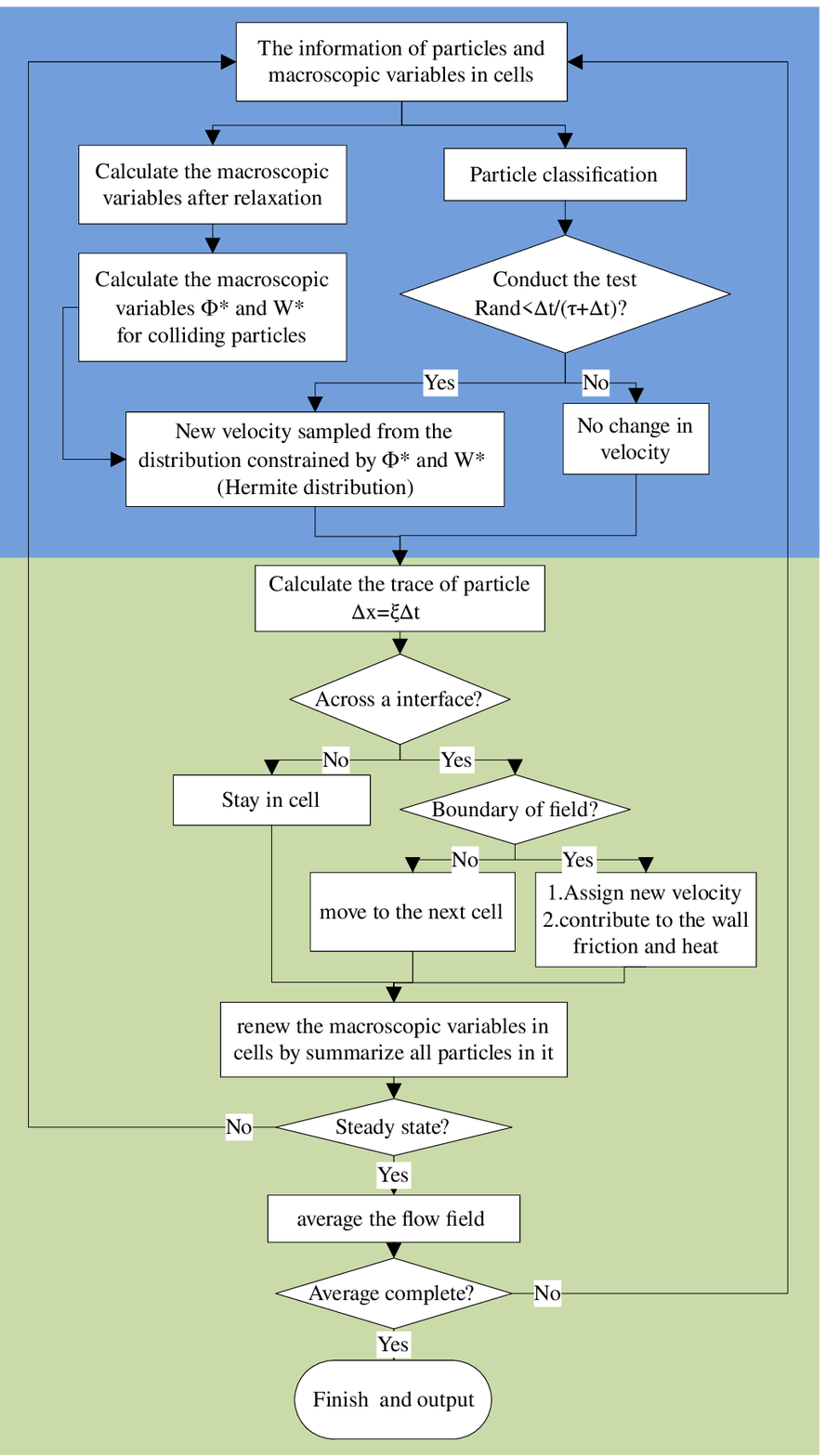}
\caption{\label{Fig:route} The route of the present method. The blue regime is the DR method for collision, the green regime is the standard DSMC treatment for free transport}
\end{figure}

The present method can be seen as a modification of the original particle-BGK method, where the sampling of after-collision velocity is from the DR process, instead of using the equilibrium distribution of model equations. Therefore, the route of present method is actually a DR process for collision and a standard free transport process. In consideration of integrity, the whole route is illustrated in details in Fig.~\ref{Fig:route}.

\subsubsection{Sampling from Hermitian distribution}
The importance sampling is chosen in this paper to sample the after-collision velocity. The thought of importance sampling is based on the following equation:
\begin{equation}
\int_{ - \infty }^{ + \infty } {A\left( x \right)f\left( x \right)} dx = \int_{ - \infty }^{ + \infty } {A\left( x \right)\frac{{f\left( x \right)}}{{g\left( x \right)}}g\left( x \right)} dx,
\end{equation}
where the moment of $f$ can be transformed to the moment of $g$ when multiplying $A$ with a weight $f/g$. In the particle method of this paper, $g$ is the Maxwellian distribution and $f$ is the Hermitian distribution, then $f/g$ is actually the Hermitian polynomial. The sampling procedure is that: Firstly the after-collision velocity is sampled from the Maxwellian distribution, then a new mass is assigned whose value is the standard particle mass multiplied by the value of Hermitian polynomial. The sampling from Maxwellian distribution adapts the following direct sampling:
\begin{equation}\label{eq:sample_maxwell}
\begin{aligned}
{\xi _x} = {U_x} + \sqrt {2RT} \cos \left( {2\pi {\varepsilon _1}} \right)\sqrt { - \ln \left( {{\varepsilon _2}} \right)}, \\
{\xi _y} = {U_y} + \sqrt {2RT} \cos \left( {2\pi {\varepsilon _3}} \right)\sqrt { - \ln \left( {{\varepsilon _4}} \right)}, \\
{\xi _z} = {U_z} + \sqrt {2RT} \sin \left( {2\pi {\varepsilon _3}} \right)\sqrt { - \ln \left( {{\varepsilon _4}} \right)},
\end{aligned}
\end{equation}
where $\varepsilon$ with different subscripts are different random real numbers between 0 and 1.

\subsubsection{Boundary conditions}
Let the normal direction of the boundary surface (pointing into the inner flow field) denoted by subscript ``x" without loss of generality.

For hypersonic inlet boundary condition, the velocity in this normal direction is obtained through the Acceptance-Rejection sampling of $\xi_{x}g_{x}$, where $g_{x}=\int_{{R^2}} {gd{\xi _y}d{\xi _z}}$ is the Maxwellian distribution in $x$ direction (a marginal distribution). The maximum value of $\xi_{x}g_{x}$ is denoted by $A_{max}$. $\xi_{x}$ can be randomly chosen from $\left(0,U_{x}+5\sqrt{RT}\right)$. If $\xi_{x}g_{x}>{\varepsilon_{1}}A_{max}$, this velocity is accepted. The velocities in the tangential directions y and z are sampled according to those in Eq.~\ref{eq:sample_maxwell}. The free transport time of inlet particle is $\varepsilon_{2}\Delta t$. The mass importing through the boundary surface is $\int_0^{ + \infty } {m{\xi _x}{g_x}d{\xi _x}}$. Since the particle number is integer, the left mass (less than one particle) is accumulated to the next iteration time.

For hypersonic outlet boundary, no backflow particle needs to be sampled, and the particle transports out of the flow filed is directly deleted.

For the Maxwell wall boundary condition with full accommodation used in this paper, the normal velocity is sampled from
\begin{equation}
{\xi _x} = U_{x} + \sqrt {2RT}\sqrt { - \ln \left( {{\varepsilon}} \right)}.
\end{equation}
The tangential velocities y and z are also sampled according to those in Eq.~\ref{eq:sample_maxwell}, where the macroscopic velocity is the velocity of the moving wall, such as the cavity case in this paper.

\section{\label{sec:Test}Test cases}
\subsection{Homogenous relaxation for Maxwell molecule}
This 0-Dimensional (0-D) homogenous case is used to examine the validity and accuracy of the DR process. The initial distribution consists of two Maxwellian distributions to mimic the high non-equilibrium distribution in the normal shock wave~\cite{liu2019Conservative}. The macroscopic variables of the two Maxwellian distributions are as follows:
\begin{equation}
\begin{aligned}
&\rho_{A} = 0.9, \rm{U}_{A}=\left(10.328, 0, 0\right), T_{A}=1.0,\\
&\rho_{B} = 0.1, \rm{U}_{B}=\left(2.703,0,0\right), T_{B}=20.8721,
\end{aligned}
\end{equation}
where the high speed low temperature distribution is denoted by the subscript "A", and the low speed high temperature one is denoted by subscript "B". Then, the macroscopic variables of the initial distribution is:
\begin{equation}
\rho = 1, \rm{U}=\left(9.56546, 0, 0\right), T=4.7314.
\end{equation}
The anisotropic temperature ${\bf{T}}=\left(P{\bf{I}}-{\bf{S}}\right)/RT$ and heat flux are
\begin{equation}
{\bf{T}} = \left[ {\begin{array}{*{20}{c}}
{8.28154}&0&0\\
0&{3.0125}&0\\
0&0&{3.0125}
\end{array}} \right],{\bf{Q}} = \left[ {\begin{array}{*{20}{c}}
{ - 50.5902}\\
0\\
0
\end{array}} \right].
\end{equation}
The analytical solutions of anisotropic temperature and heat flux for Maxwell molecules are as follows:
\begin{equation}
\begin{aligned}
&{\bf{T}}(t) = {e^{ - t/\tau }}\left\{ {{\bf{T}}(0) - T(0){\bf{I}}} \right\} + T(0){\bf{I}},\\
&{\bf{Q}}(t) = {e^{ - \Pr t/\tau }}{\bf{Q}}(0).
\end{aligned}
\end{equation}
By the way, this equation is a precise analytical solution for Maxwell molecules, and high-quality approximation for other molecular potential models.

$10^{5}$ molecules are used for simulation. With the same weight for colliding molecules ($\frac{\Delta t}{\Delta t + \tau}$), the DR process and Shakhov collision model are compared. The iteration time is chosen as $0.1\tau$, $0.5\tau$ and $2.0\tau$, respectively. By comparing with the analytical solution, it is found that when iteration time step is small, such as $\Delta t=0.1\tau$, both DR and Shakhov match well with the analytical solution. With time step increasing, DR behaves better than Shakhov. Since DR can be viewed as an implicit Euler scheme, it has deviations from the analytical solution when the time step is large. However, the numerical method is stable (benefitting from the implicit treatment), and this deviation can be ignored in multi-scale flow simulations in the following tests in this paper. This is also supported by the success of multi-scale deterministic methods such as UGKS, GKUA, DUGKS, and IDVM, where the implicit collision term is used.

\subsection{Sod shock tube}
The Sod shock tube cases with Kn$=10^{-5}$ and $0.1$ are conducted in this section. The Sod shock tube case is an unsteady flow, therefore, 100 times of ensemble average is carried out to reduce the influence of statistic fluctuation. The length of the whole flow field is set to unity, which is also the reference length. The flow field is discretized into 100 uniform cells for Kn$=0.1$ case, and 500 uniform cells for Kn$=10^{-5}$ case. The initial condition is:
\begin{equation}
\begin{aligned}
(\rho ,U,p) = \left\{ {\begin{array}{*{20}{c}}
{(1,\;(0,\;0),\;1),}\\
{(0.125,\;(0,\;0),\;0.1),}
\end{array}} \right.\,\;\begin{array}{*{20}{c}}
{\;0 < x < 0.5}\\
{\;0.5 < x < 1}
\end{array}
\end{aligned}.
\end{equation}

\begin{figure}
\centering
\subfigure[\label{Fig:sod_N1_rho}]{
\includegraphics[width=0.45\textwidth]{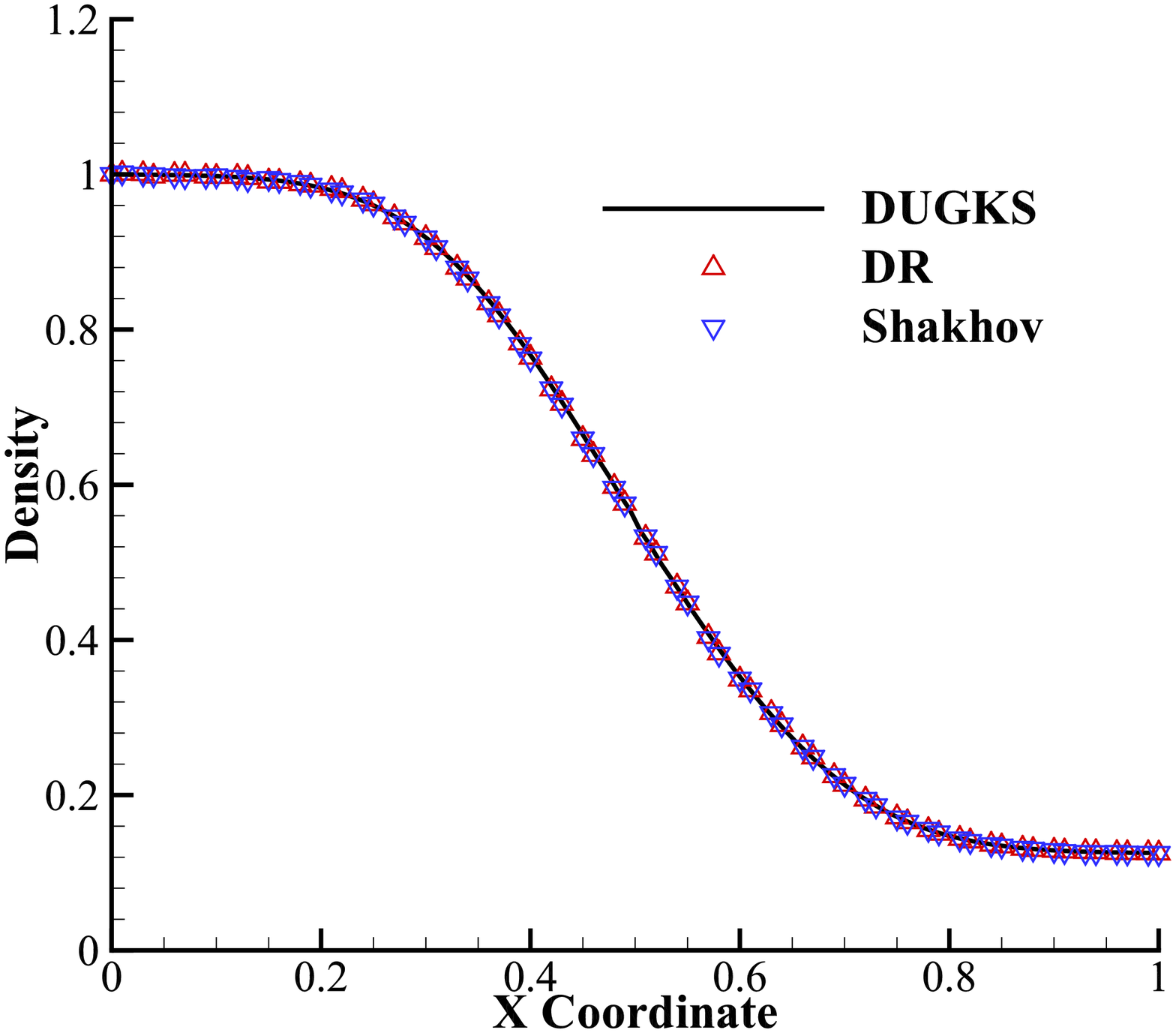}
}\hspace{0.05\textwidth}%
\subfigure[\label{Fig:sod_N1_V}]{
\includegraphics[width=0.45\textwidth]{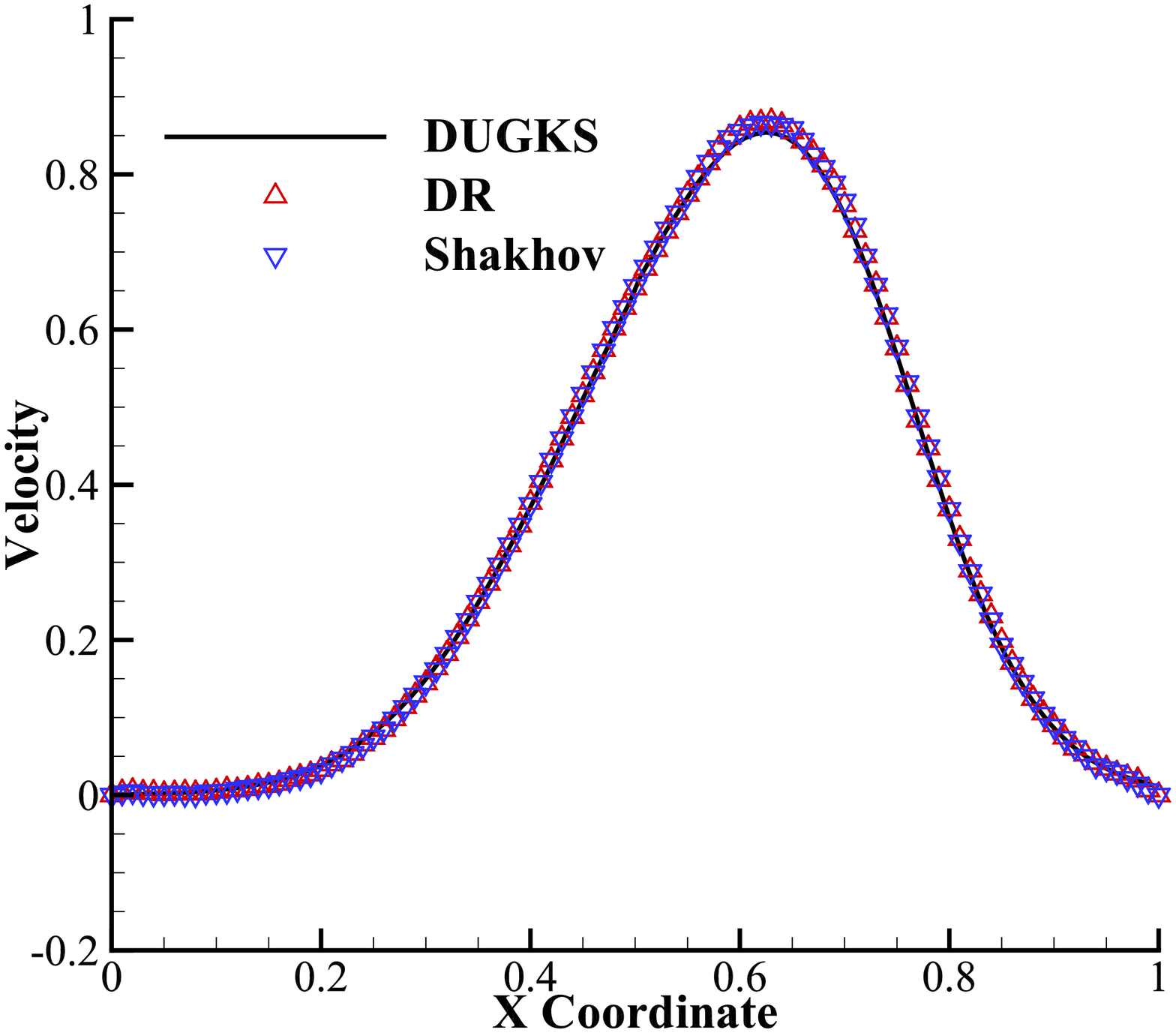}
}\\
\subfigure[\label{Fig:sod_N1_T}]{
\includegraphics[width=0.45\textwidth]{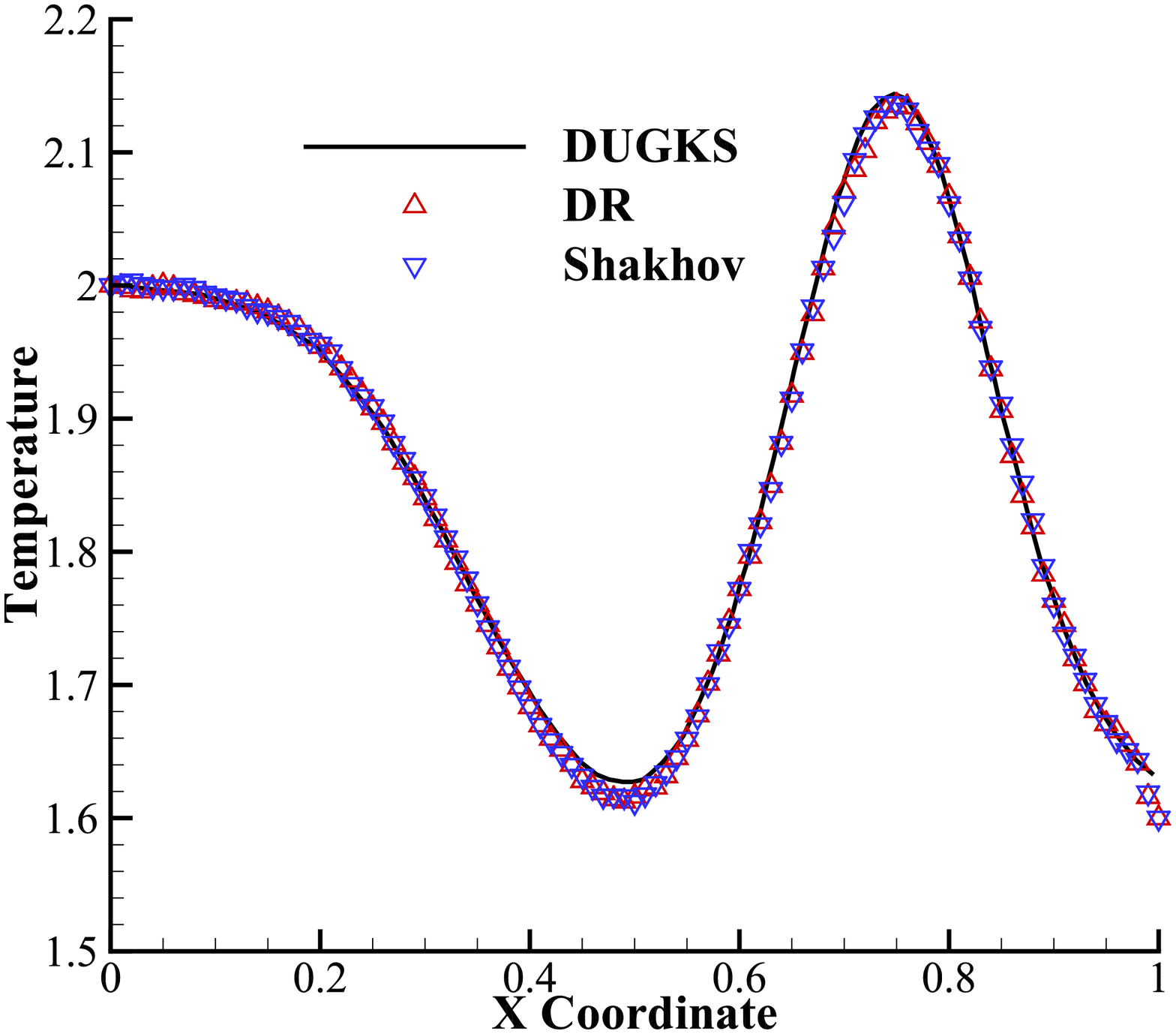}
}\hspace{0.05\textwidth}%
\caption{\label{Fig:sod_N1} The profiles of Sod shock tube with Kn$=0.1$, (a) density profile, (b) velocity profile, (c) temperature profile}
\end{figure}

\begin{figure}
\centering
\subfigure[\label{Fig:sod_N5_rho}]{
\includegraphics[width=0.45\textwidth]{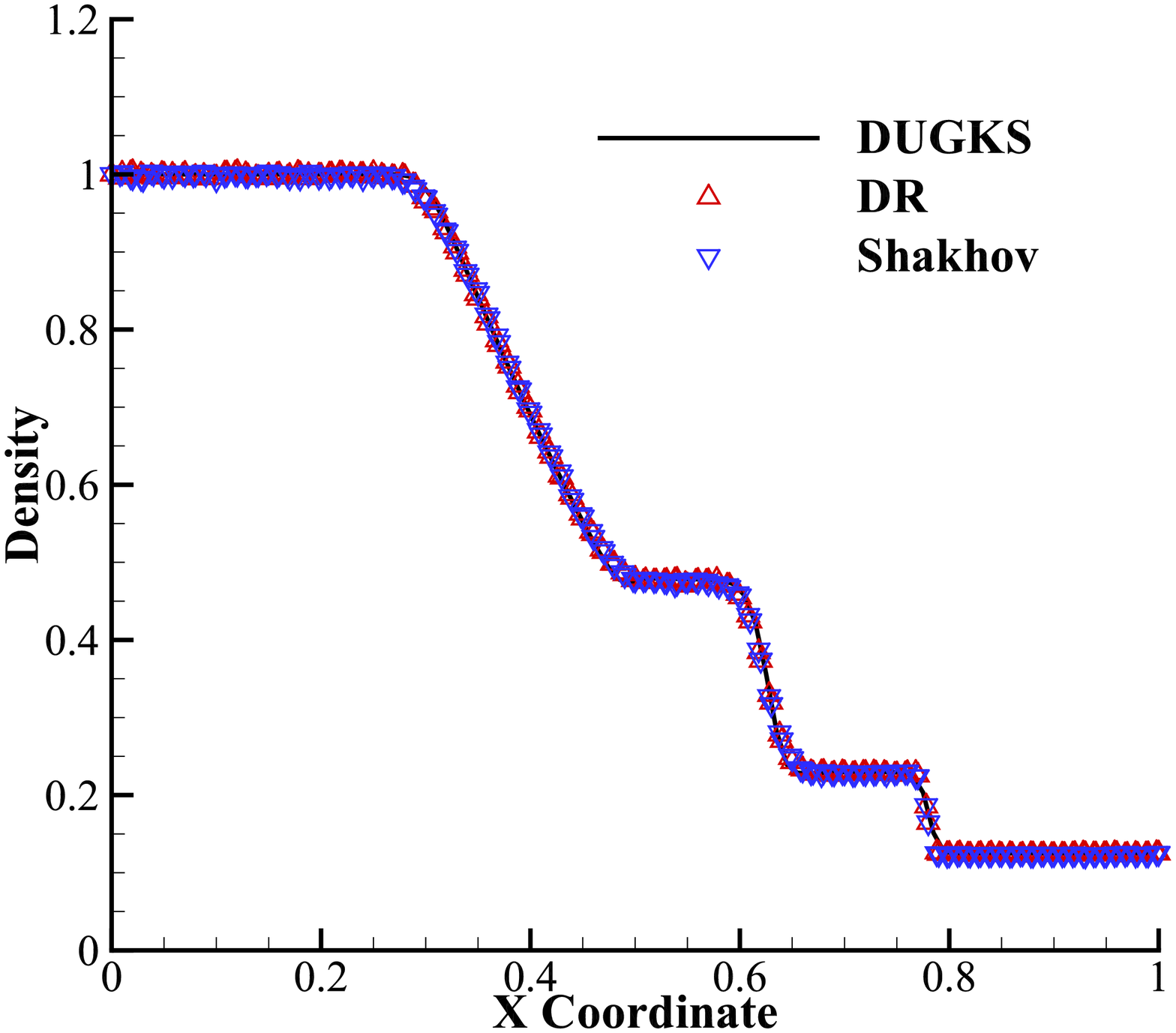}
}\hspace{0.05\textwidth}%
\subfigure[\label{Fig:sod_N5_V}]{
\includegraphics[width=0.45\textwidth]{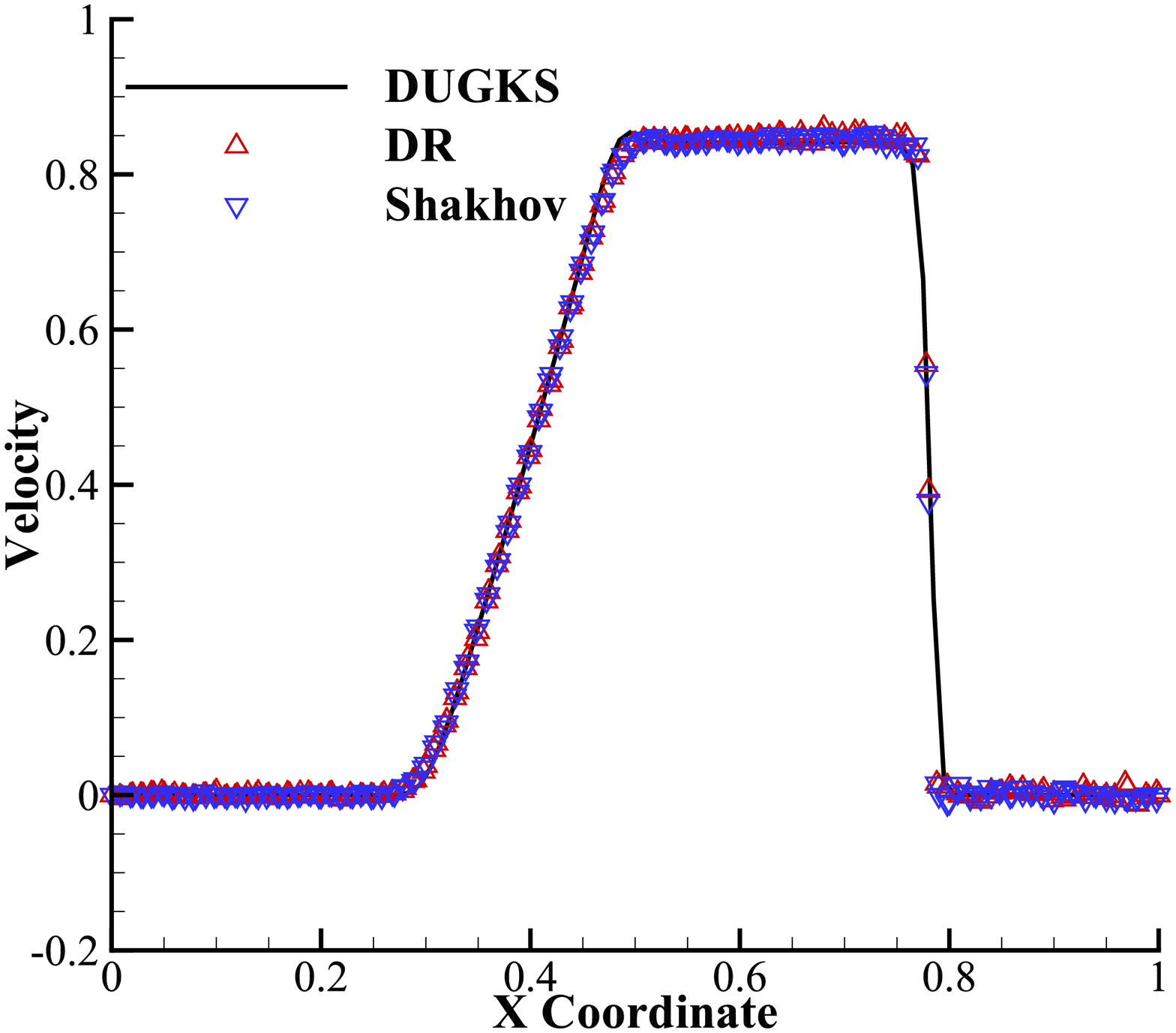}
}\\
\subfigure[\label{Fig:sod_N5_T}]{
\includegraphics[width=0.45\textwidth]{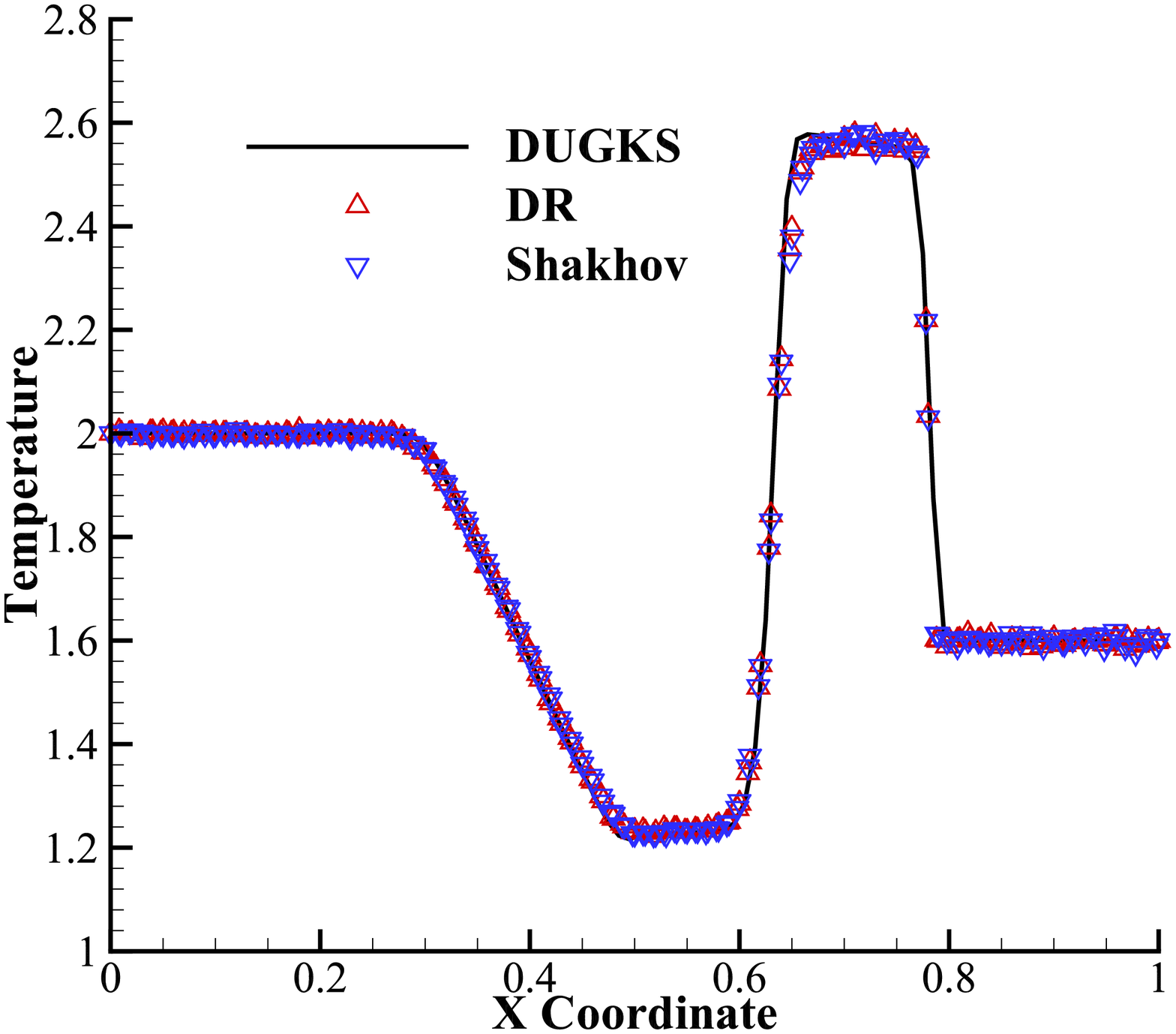}
}\hspace{0.05\textwidth}%
\caption{\label{Fig:sod_N5} The profiles of Sod shock tube with Kn$=10^{-5}$, (a) density profile, (b) velocity profile, (c) temperature profile}
\end{figure}

The VHS (Variable Hard Sphere) model with heat index $\omega=0.81$ for Argon gas is adopted, and molecule weight is $10^{-4}$ (when calculating the weight of molecules, the cell is considered as $0.01 \times 1 \times 1$ cubic). The iteration time step is set to $\Delta t=2.0 \times 10^{-3}$. In this paper, the viscosity and reference viscosity for VHS model can be calculated by the following equations in Ref.~\cite{Bird2003Molecular}:
\begin{equation}
\begin{aligned}
&\mu  = {\mu _{ref}}{\left( {\frac{T}{{{T_0}}}} \right)^\omega },\\
&{\mu _{ref}} = \frac{{15\sqrt \pi  }}{{2(5 - 2\omega )(7 - 2\omega )}}{\rm{Kn}},
\end{aligned}
\end{equation}
and the relation between the mean free path and the viscosity can be written as
\begin{equation}\label{eq:mfp}
\lambda  = \frac{{2\mu (7 - 2\omega )(5 - 2\omega )}}{{15\rho {{(2\pi RT)}^{1/2}}}}.
\end{equation}
The stochastic particle methods with DR and Shakhov model are imposed for numerical prediction, respectively. The DUGKS data obtained by the code in Ref~\cite{Chen2019Conserved} is used as the benchmark solution. It can be found in Fig.~\ref{Fig:sod_N1} and Fig.~\ref{Fig:sod_N5} that DR and Shakhov also overlap with each other and match well with the benchmark solutions in both Kn numbers for rarefied flows and continuum flows, respectively.

\subsection{Normal shock structure}
The normal shock is counted as a discontinuity from the macroscopic point of view. If zooming into the normal shock, the thickness of which is about 20 times of molecular mean free path (m.f.p.), and it has a smooth structure for the profiles of macroscopic variables. Since the distribution function is in high non-equilibrium in the case of high Mach number, this case is often used as a benchmark and challenging case for numerical methods aiming at predicting rarefied and multi-scale flows. The inlet and outlet boundary conditions are set as follows according to the Rankine-Hugoniot relation for normal shock:
\begin{equation}
\begin{aligned}
&\frac{{{\rho _2}}}{{{\rho _1}}} = \frac{{(\gamma  + 1){\rm{M}}{{\rm{a}}^2}}}{{(\gamma  - 1){\rm{M}}{{\rm{a}}^2} + 2}},\\
&\frac{{{T_2}}}{{{T_1}}} = \frac{{\left( {1 + \frac{{\gamma  - 1}}{2}{\rm{M}}{{\rm{a}}^2}} \right)\left( {\frac{{2\gamma }}{{\gamma  - 1}}{\rm{M}}{{\rm{a}}^2} - 1} \right)}}{{{{{\mathop{\rm Ma}\nolimits} }^2}\left( {\frac{{2\gamma }}{{\gamma  - 1}} + \frac{{\gamma  - 1}}{2}} \right)}},
\end{aligned}
\end{equation}
where the subscripts "1" and "2" denote the variables before and after the shock wave, $\gamma=5/3$ is the specific heat ratio, and Ma is the inlet Mach number.

In this case, to compare with the benchmark data in Ref.~\cite{guo2013discrete}, the Argon gas is chosen as the working gas, and the VHS model is adopted, and $\gamma$ is set to $5/3$. To be consistent with the data in Ref.~\cite{guo2013discrete}, the mean free path before the shock denoted by $\lambda_{1}$ is calculated from the HS model (Hard Sphere model, the same as the VHS model with $\omega$=0.5). Then, the range of computational domain is set to $\left(-50\lambda_{1},50\lambda_{1}\right)$. The cell length $\Delta x$ is equal to $\lambda_{1}$, and the CFL number for iteration time is set to $0.5$. The weight of molecules is chosen as $1 \times 10^{-3}$ (about 700 particles in the cell element before the shock wave). In this section, shock waves with Ma number 3.0 and 8.0 are considered.

\begin{figure}
\centering
\subfigure[\label{Fig:shock3_rho}]{
\includegraphics[width=0.45\textwidth]{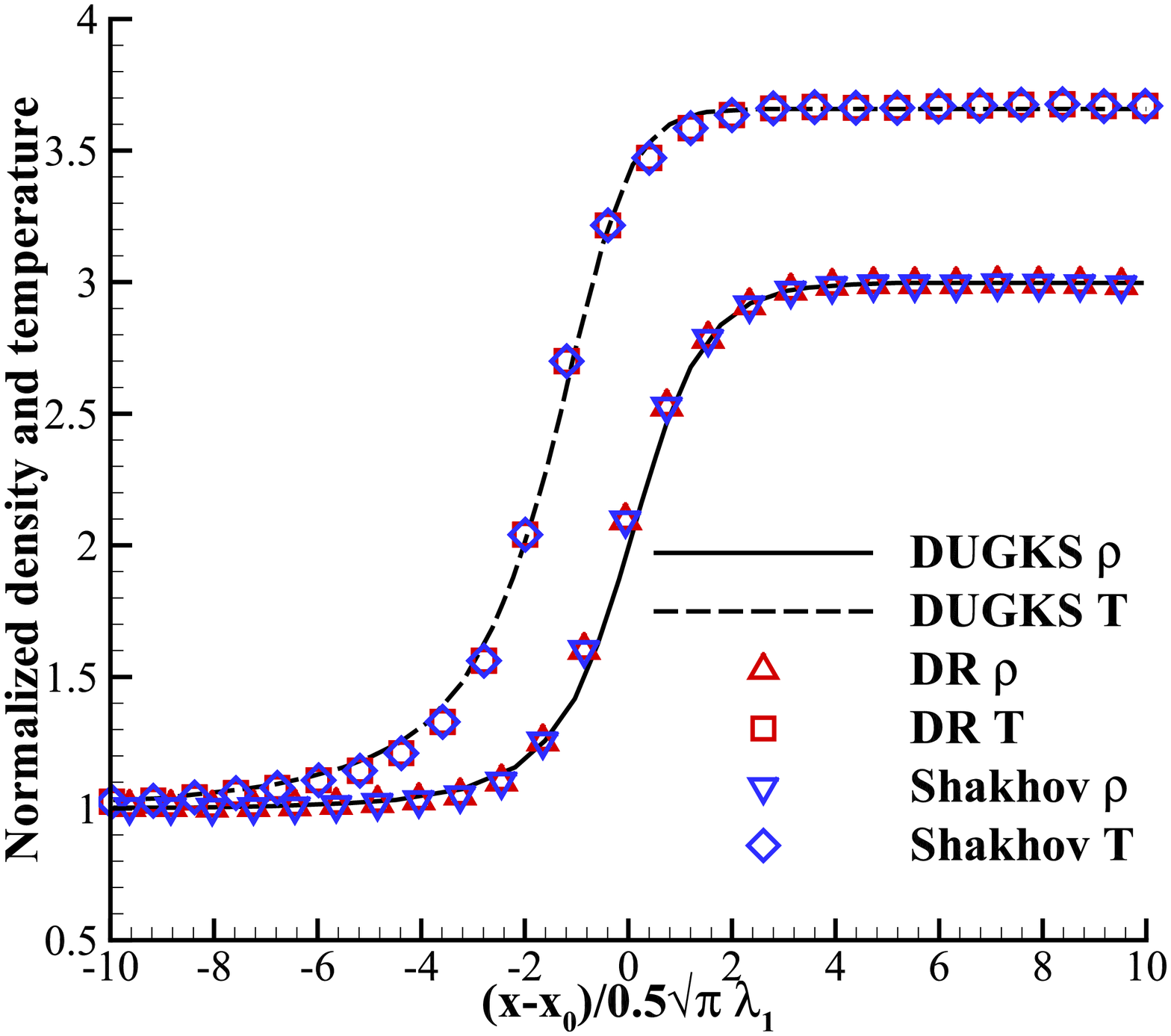}
}\hspace{0.05\textwidth}%
\subfigure[\label{Fig:shock3_stress}]{
\includegraphics[width=0.45\textwidth]{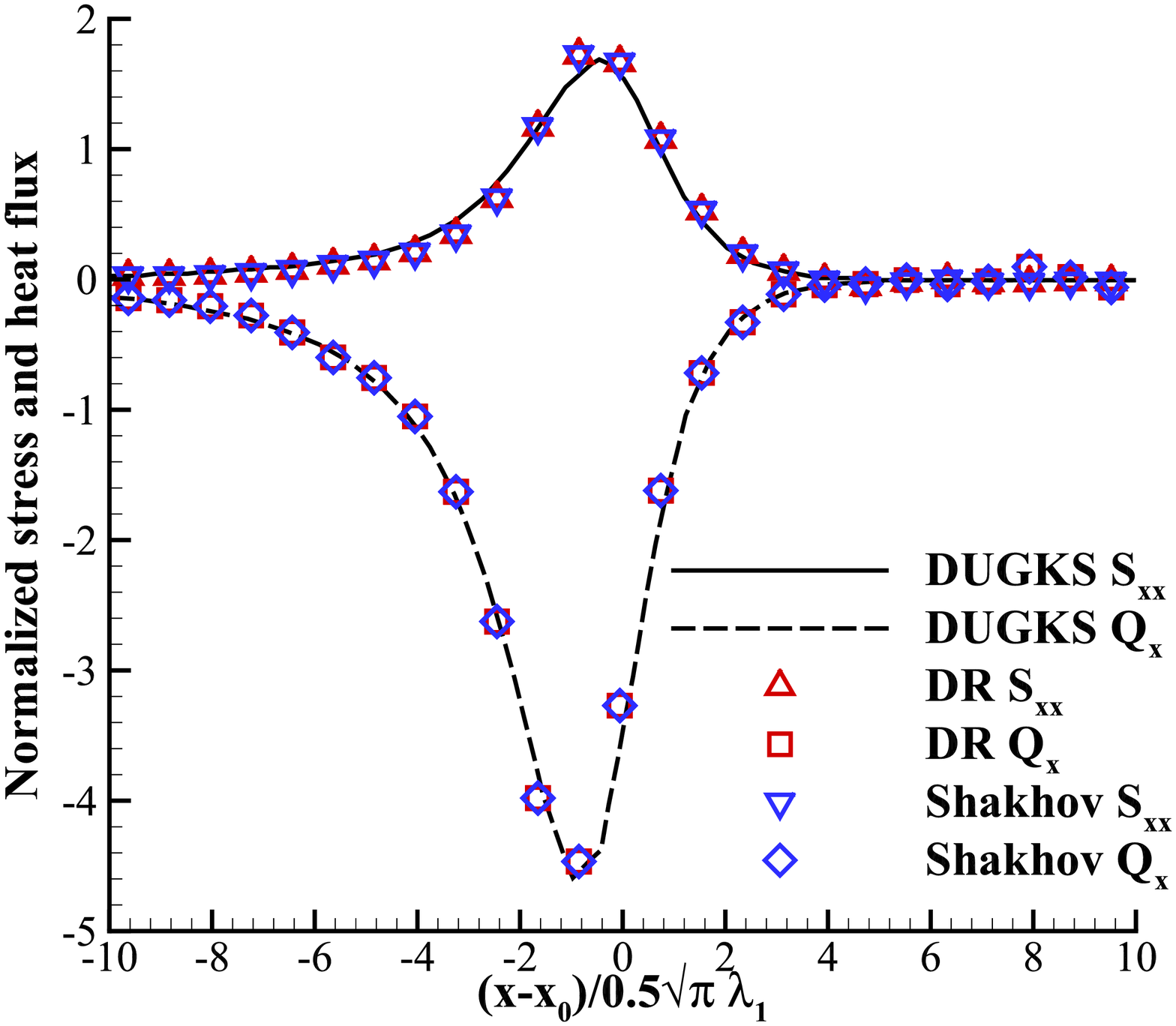}
}
\caption{\label{Fig:shock3} The profiles of Argon shock structure with Ma$=3.0$, (a) density and temperature profiles, (b) stress and heat flux profiles}
\end{figure}

For the Ma=3 case, the heat index $\omega$ is set to 0.5 to compare with the DUGKS data in Ref.~\cite{guo2013discrete}. The profiles of density, temperature, stress and heat flux predicted by stochastic particle method with DR and Shakhov are illustrated in Fig.~\ref{Fig:shock3}, where these variables are nondimensionalized using the following equations
\begin{equation}
\hat \rho  = \frac{\rho }{{{\rho _1}}},\;\hat T = \frac{T}{{{T_1}}},\;\hat S_{xx}^{} = \frac{{{S_{xx}}}}{{{\rho _1}R{T_1}}}, \hat Q_x^{} = \frac{{{Q_x}}}{{{\rho _1}R{T_1}{{(2R{T_1})}^{3/2}}}}.
\end{equation}
In Fig.~\ref{Fig:shock3}, the results predicted by DR almost overlap with those predicted by Shakhov model, and match well with the DUGKS-Shakhov method.

\begin{figure}
\centering
\subfigure[\label{Fig:shock8_rho}]{
\includegraphics[width=0.45\textwidth]{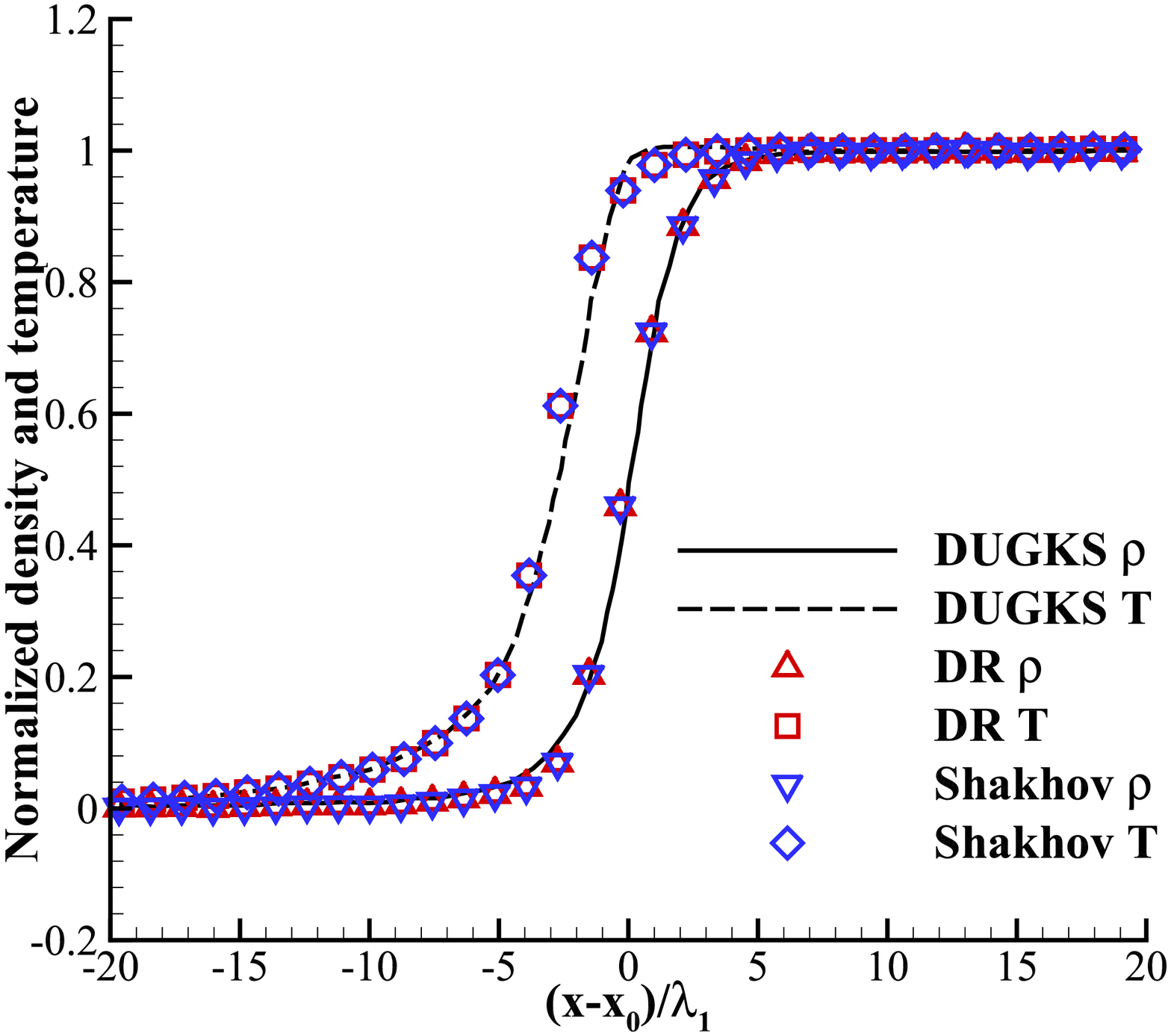}
}\hspace{0.05\textwidth}%
\subfigure[\label{Fig:shock8_stress}]{
\includegraphics[width=0.45\textwidth]{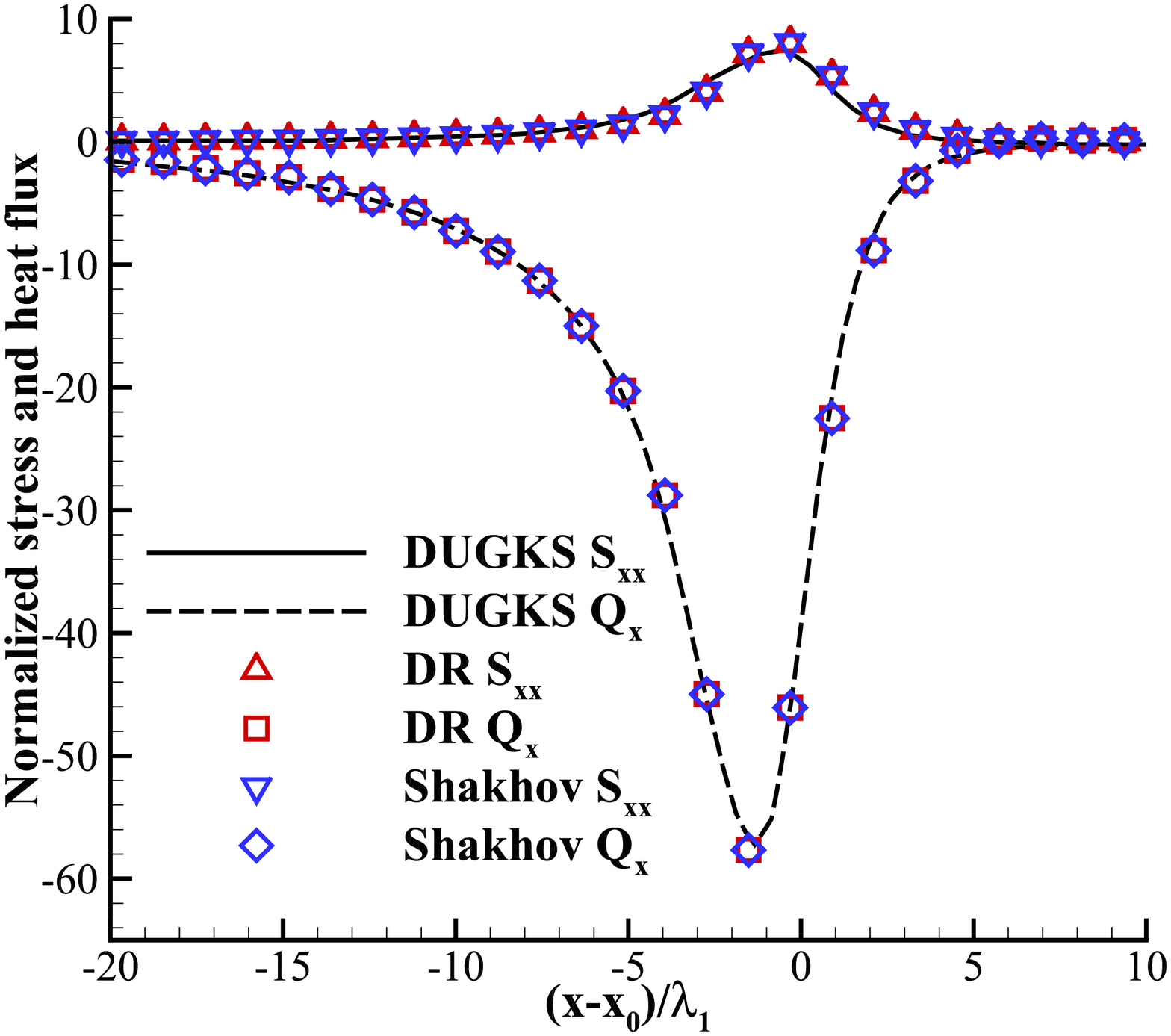}
}
\caption{\label{Fig:shock8} The profiles of Argon shock structure with Ma$=8.0$, (a) density and temperature profiles, (b) stress and heat flux profiles}
\end{figure}

For the Ma=8.0 case, the heat index $\omega$ is set to 0.68 to compare with the settings in Ref.~\cite{guo2013discrete}, the profiles of density, temperature, stress and heat flux are normalized or nondimensionalized as follows:
\begin{equation}
\hat \rho  = \frac{{\rho  - {\rho _1}}}{{{\rho _2} - {\rho _1}}},\;\hat T = \frac{{T - {T_1}}}{{{T_2} - {T_1}}},\;\hat S_{xx}^{} = \frac{{{S_{xx}}}}{{{\rho _1}R{T_1}}},\;\hat Q_x^{} = \frac{{{Q_x}}}{{{\rho _1}R{T_1}{{(2R{T_1})}^{3/2}}}}.
\end{equation}
Similar with the Ma=3 case, these variables predicted by the DR almost overlap with those predicted by Shakhov model, and match well with the DUGKS data in Fig.~\ref{Fig:shock8}. Since the $\omega$ is set to 0.68, there is no overshoot in temperature profiles such as the case with $\omega=0.78$ in high Mach number~\cite{liu2014Investigation}.

\subsection{Cavity flow}
The cavity flow case has a rectangle flow field enclosed by three static walls and a top moving wall towards right. It is a benchmark for low-speed viscous flows, and it is difficult for stochastic particle methods, since the macroscopic velocity is less than the peculiar velocity.

The initial density and temperature are chosen as the reference density and temperature, respectively. The length of edge is chosen as the reference length. The wall temperature is equal to the initial one. The velocity of top wall is in a Mach number of 0.20975. The VHS molecular model with $\omega=0.81$ is used as the working gas. Two Kn numbers 10 and 0.075 are considered. In this case, a $64 \times 64$ uniform mesh is used. The iteration time is set to 0.015.

The weight of particles is set to $0.5 \times 10^{-5}$ (about 48 particles in a cell). After achieving the steady state, $2 \times 10^{7}$ steps are used for statistical averaging in Kn=10 case, and $6 \times 10^{6}$ steps are used in Kn=0.075 case. In this case, to achieve the correct result, the conservative sampling is adopted. For this low speed and closed flow field, the conservative sampling that forcing the conservation in cells is needed.

\begin{figure}
\centering
\subfigure[\label{Fig:2D_Cavity_Kn10_contour_U}]{
\includegraphics[width=0.45\textwidth]{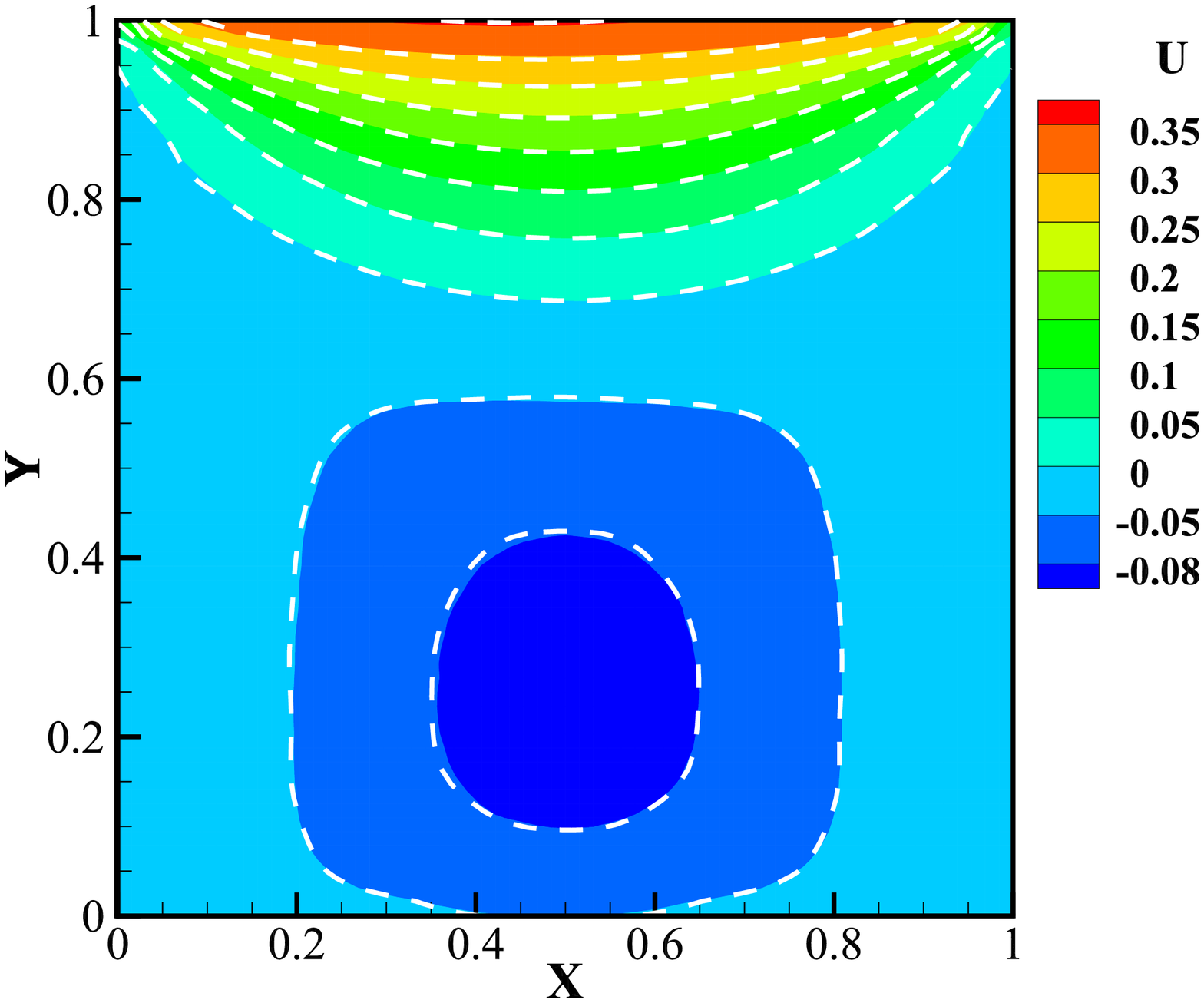}
}\hspace{0.05\textwidth}%
\subfigure[\label{Fig:2D_Cavity_Kn10_contour_V}]{
\includegraphics[width=0.45\textwidth]{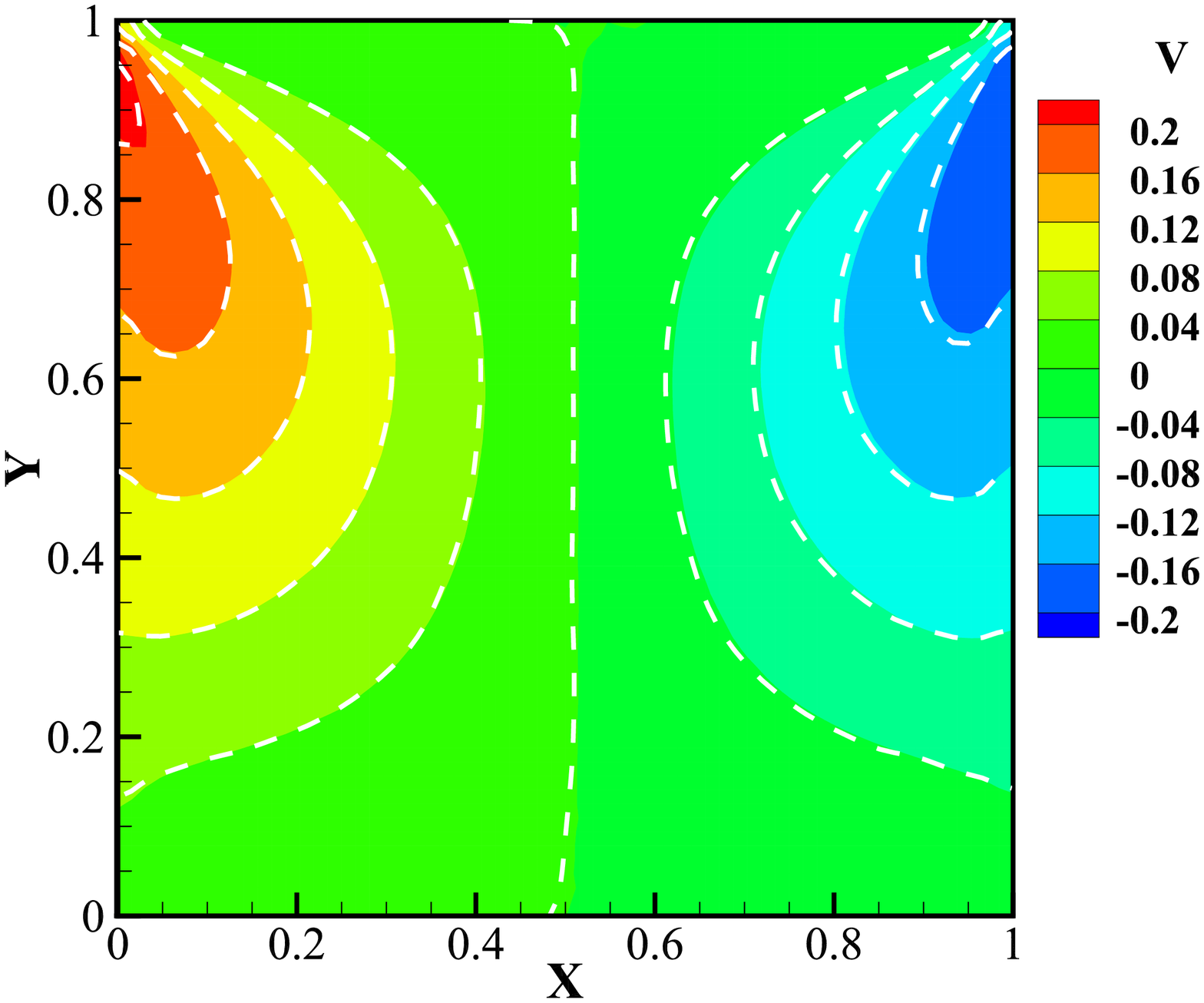}
}\\
\subfigure[\label{Fig:2D_Cavity_Kn10_StreamLine}]{
\includegraphics[width=0.45\textwidth]{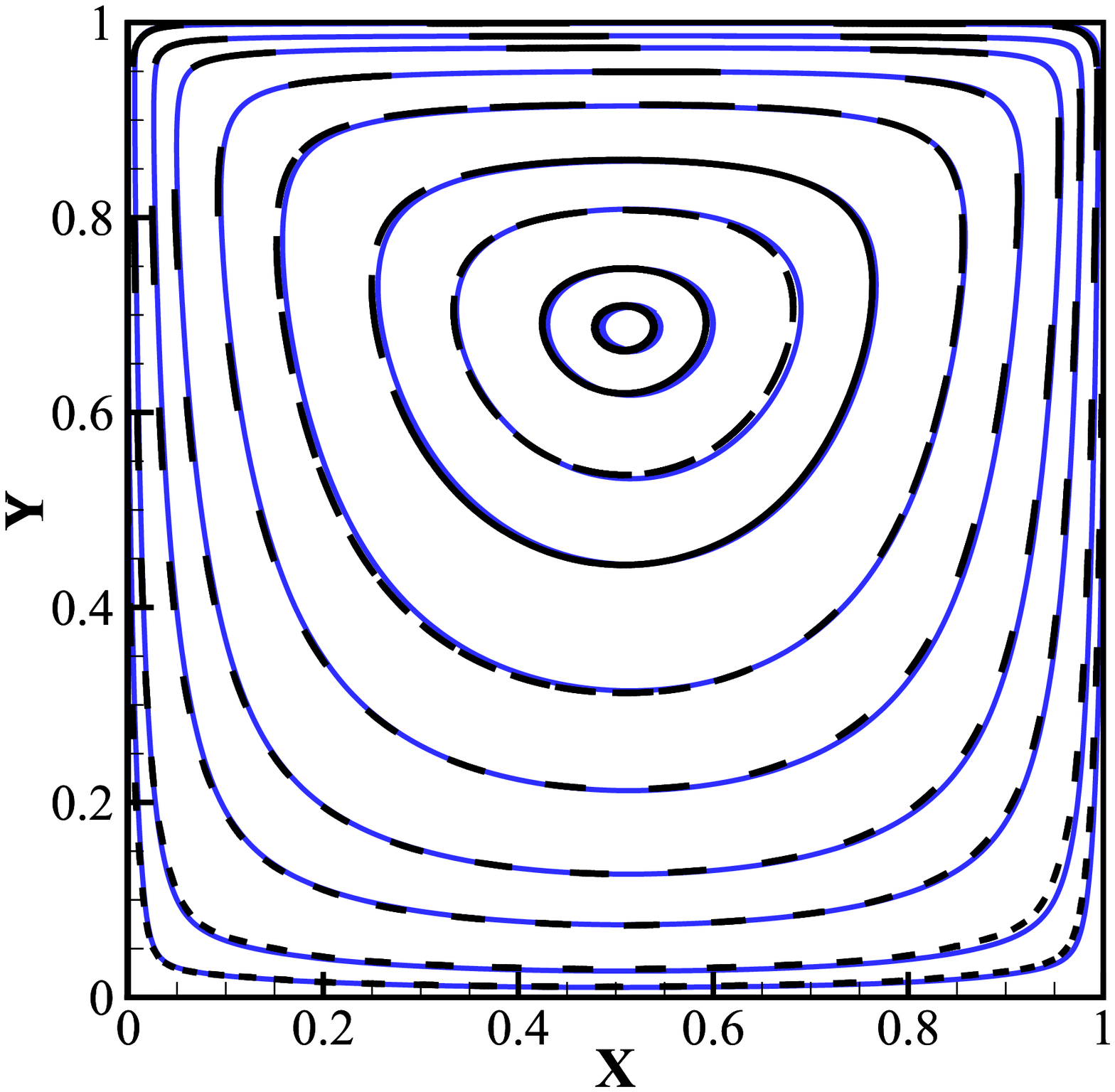}
}\hspace{0.05\textwidth}%
\subfigure[\label{Fig:2D_Cavity_Kn10_YU_XV}]{
\includegraphics[width=0.45\textwidth]{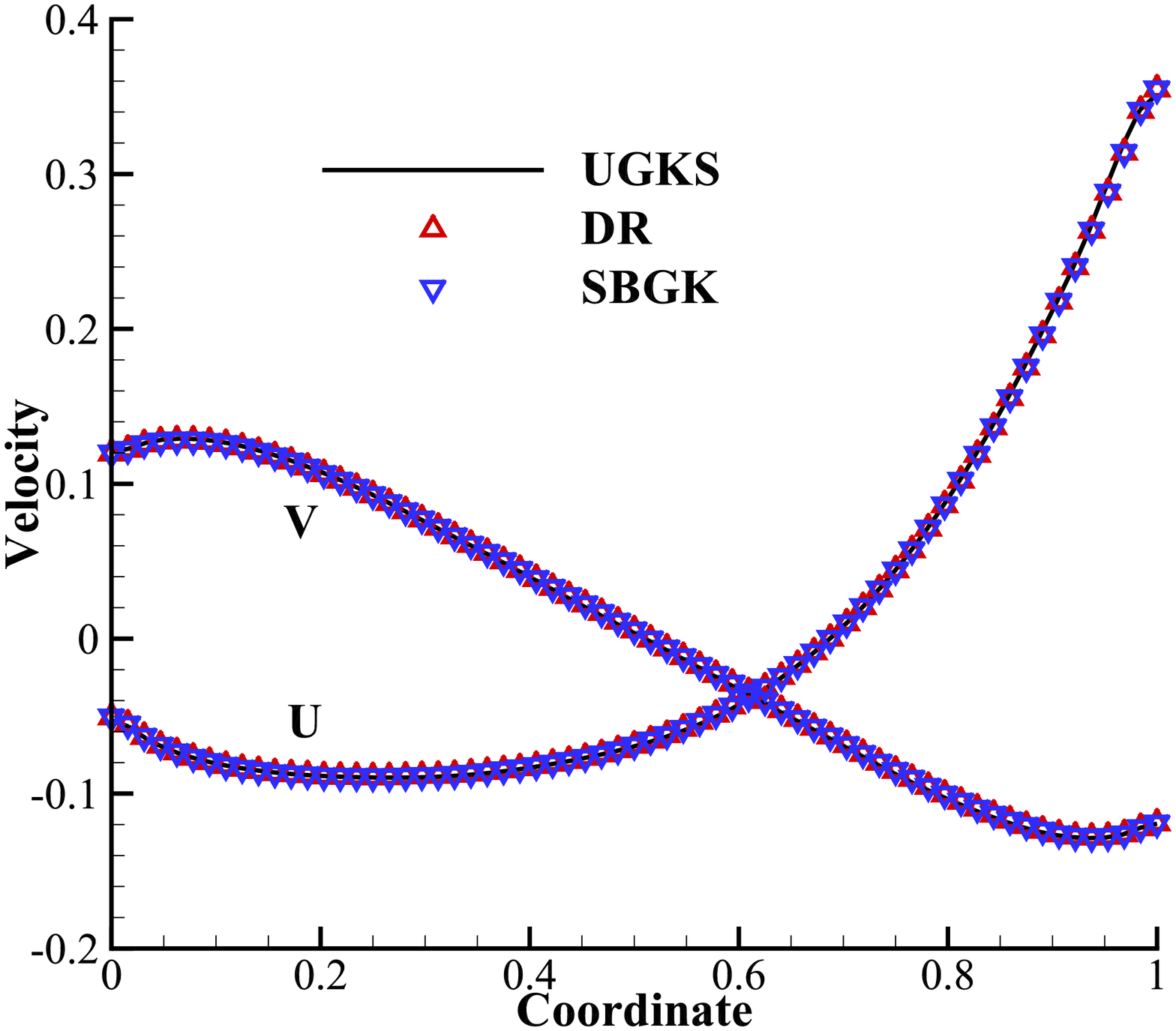}
}
\caption{\label{Fig:2D_Cavity_Kn10} The cavity flow with Kn$=10$, (a) U-velocity contour, (b) V-velocity contour, (c) streamline, (d) U-velocity along the central vertical line and V-velocity along the central horizontal line. In subfigure (a) and (b), the solid contour line with colored band are DR results, the dash-line contour is from UGKS results. In subfigure (c), the solid streamline is from DR, and dashed streamline is from UGKS.}
\end{figure}

\begin{figure}
\centering
\subfigure[\label{Fig:2D_Cavity_Kn0075_contour_U}]{
\includegraphics[width=0.45\textwidth]{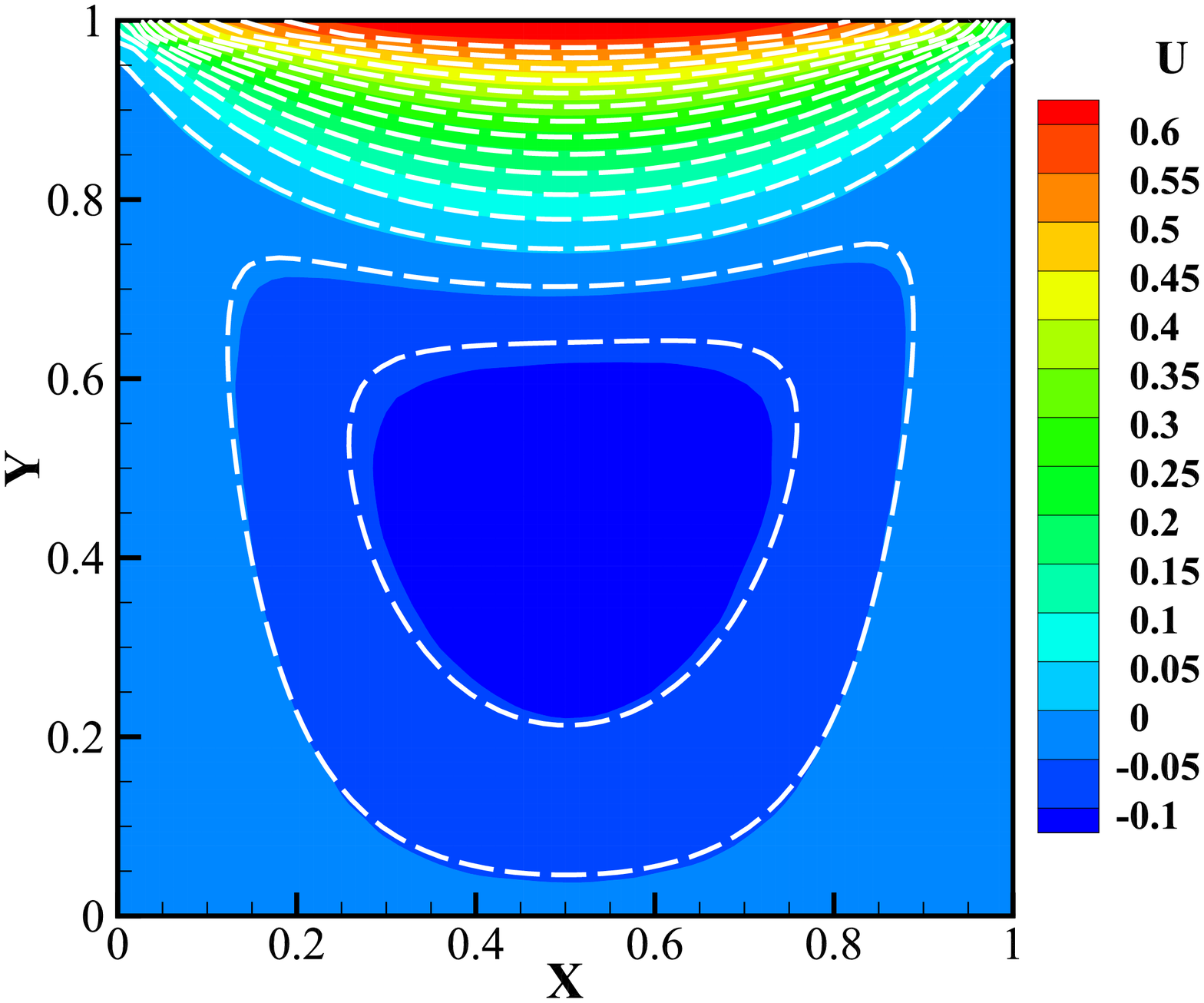}
}\hspace{0.05\textwidth}%
\subfigure[\label{Fig:2D_Cavity_Kn0075_contour_V}]{
\includegraphics[width=0.45\textwidth]{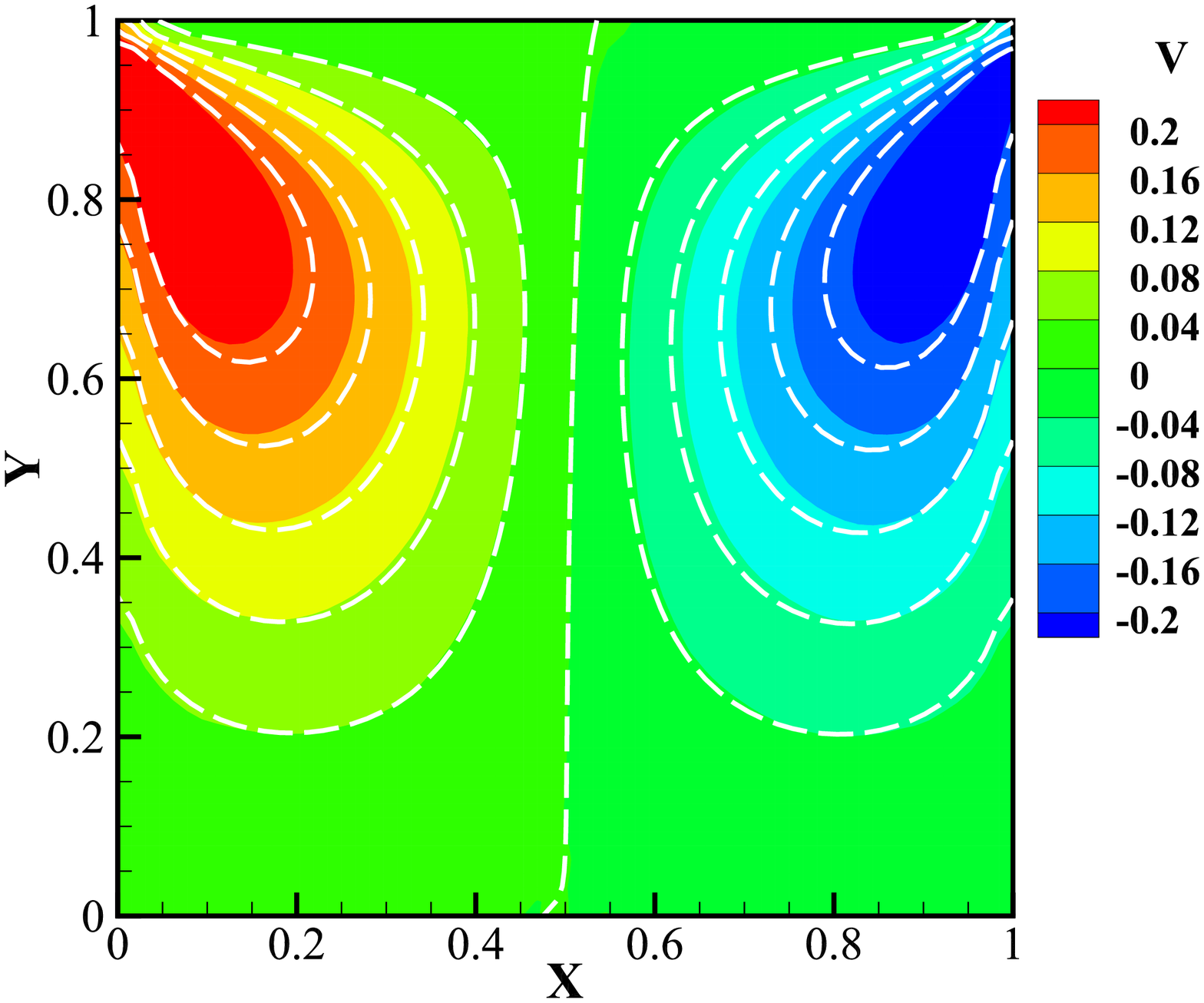}
}\\
\subfigure[\label{Fig:2D_Cavity_Kn0075_StreamLine}]{
\includegraphics[width=0.45\textwidth]{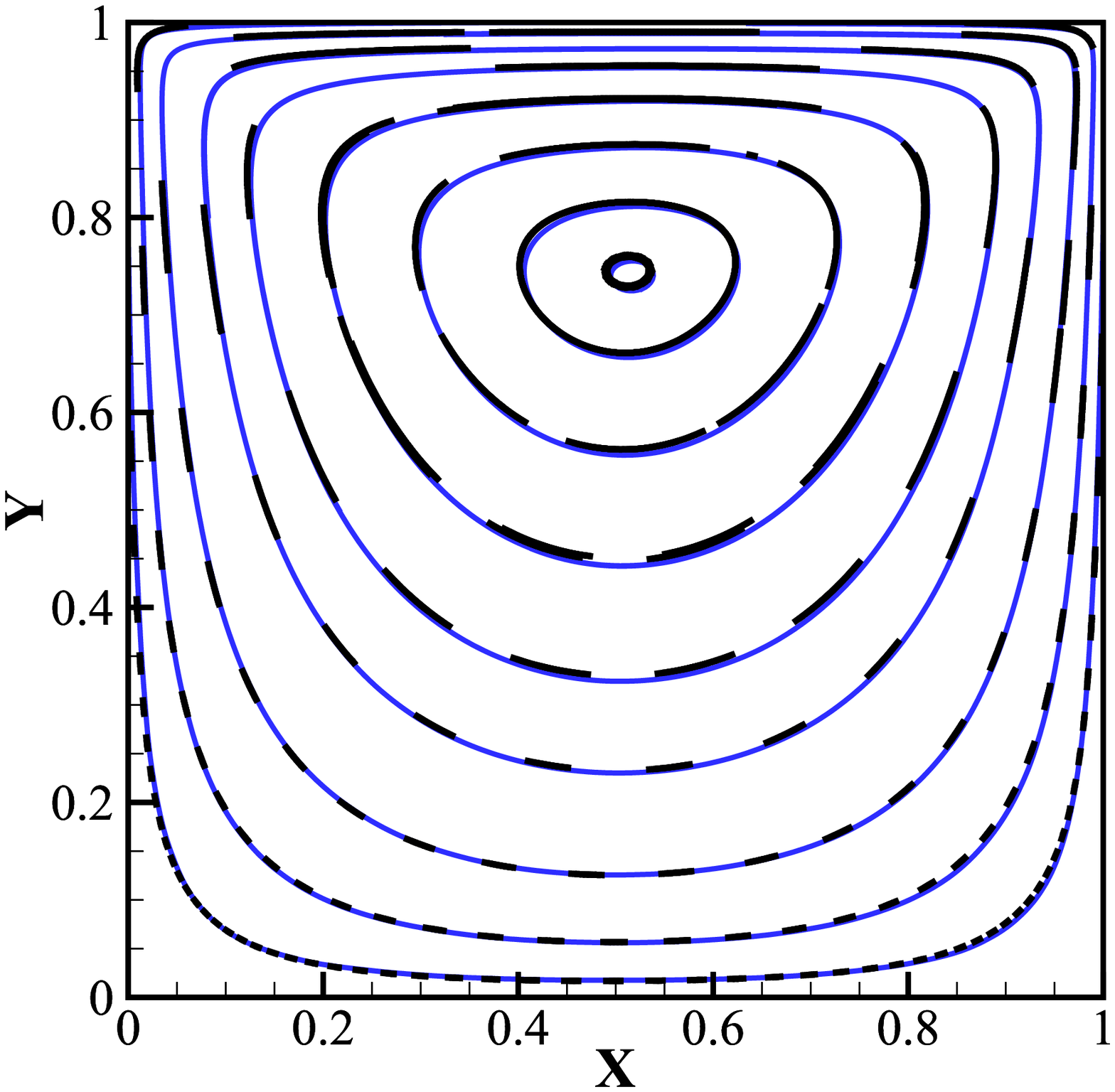}
}\hspace{0.05\textwidth}%
\subfigure[\label{Fig:2D_Cavity_Kn0075_YU_XV}]{
\includegraphics[width=0.45\textwidth]{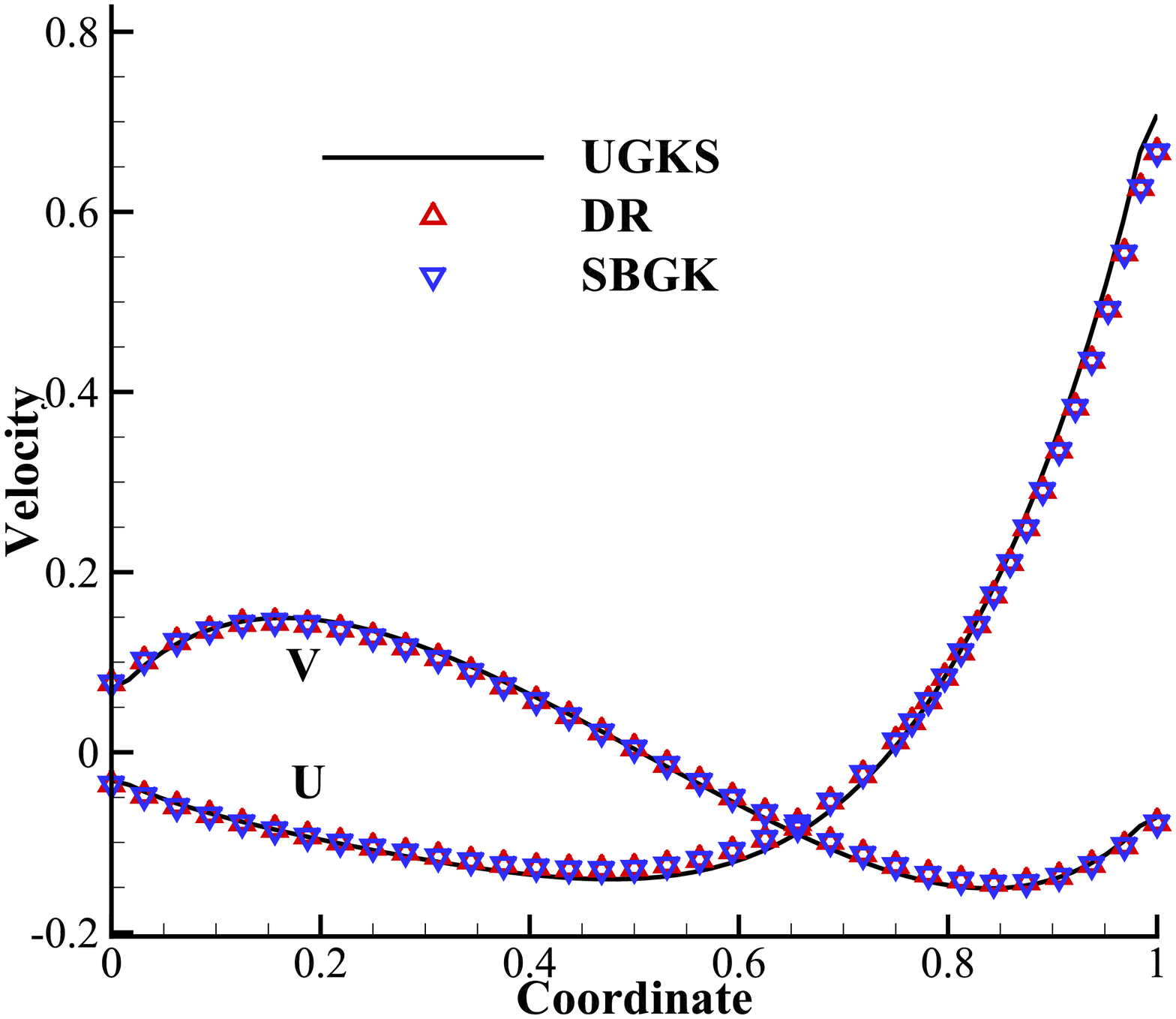}
}
\caption{\label{Fig:2D_Cavity_Kn0075} The cavity flow with Kn$=0.075$, (a) U-velocity contour, (b) V-velocity contour, (c) streamline, (d) U-velocity along the central vertical line and V-velocity along the central horizontal line. In subfigure (a) and (b), the solid contour line with colored band are DR results, the dash-line contour is from UGKS results. In subfigure (c), the solid streamline is from DR, and dashed streamline is from UGKS.}
\end{figure}

Fig.~\ref{Fig:2D_Cavity_Kn10} illustrates the contours of velocities, streamline and the velocity distribution along central horizontal and central vertical lines for Kn$=10$ case. The results given by stochastic particle method with DR process almost overlap with those predicted by the deterministic UGKS method~\cite{Yuan2020Amulti}. In the Kn$=0.075$ case near continuum regime, the contours, streamline and distributions of macroscopic velocity also match well with the benchmark solution in Fig.~\ref{Fig:2D_Cavity_Kn0075}. The smooth contours and distributions given by particle method benefit from sufficient ensemble average times.


\subsection{hypersonic flow past cylinder}
The hypersonic cylinder flow in this section is used to examine the validity and accuracy of the present method for high speed viscous flows. Test cases with two Ma numbers 5.0 and 2.0 are calculated, where Kn$=10$ and $1$ are considered in Ma $5.0$ case;  Kn$=0.1$ is considered in Ma $=20.0$ case.

In this case, the diameter of cylinder $d$ is chosen as the reference length, and the whole flow field is enclosed by a concentric circle with a diameter of $15d$. The VHS model for Argon gas with $\omega=0.81$ is the working gas. The density and temperature of the inlet boundary are chosen as the reference values. The wall temperature is the same with the inlet one, and Maxwell boundary with full accomodation coefficient is adopted.

The weight of molecular is set to $10^{-3}$. The computational domain is decomposed into $75 \times 62$ mesh cells, where 75 cells are in the radial direction, 62 cells are in the circular direction. The smallest cell length is the height of cells adjacent the wall, and its value is 0.05.

The stress and heat flux are normalized as follows:
\begin{equation}
{\hat {\bf{S}}} = \frac{\bf{S}}{{{\rho _\infty }U_\infty ^2}},{\hat{\bf{Q}}} = \frac{\bf{Q}}{{{\rho _\infty }U_\infty ^3}}.
\end{equation}

\begin{figure}
\centering
\subfigure[\label{Fig:Ma5Kn10_line_rho}]{
\includegraphics[width=0.45\textwidth]{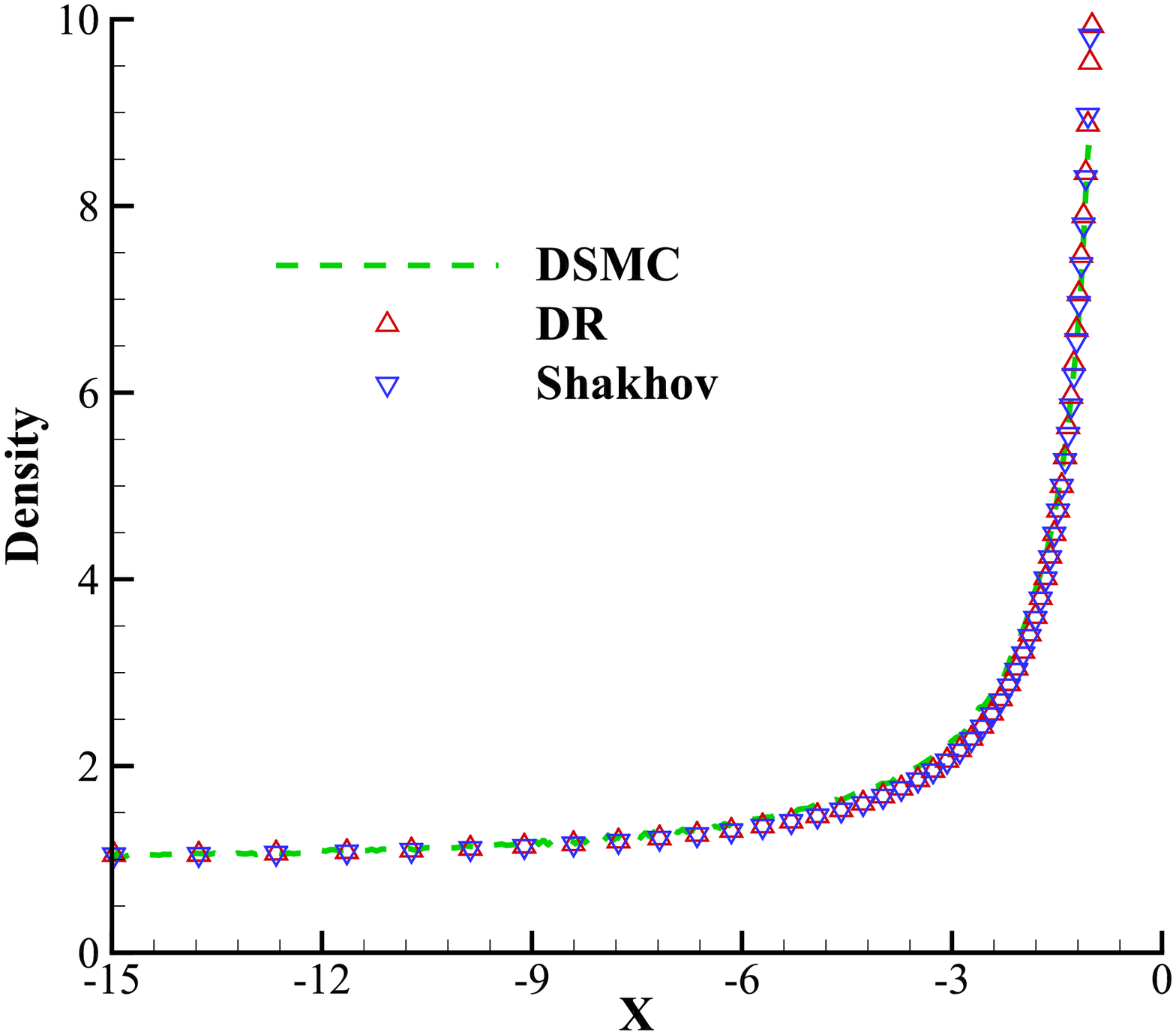}
}\hspace{0.05\textwidth}%
\subfigure[\label{Fig:Ma5Kn10_line_V}]{
\includegraphics[width=0.45\textwidth]{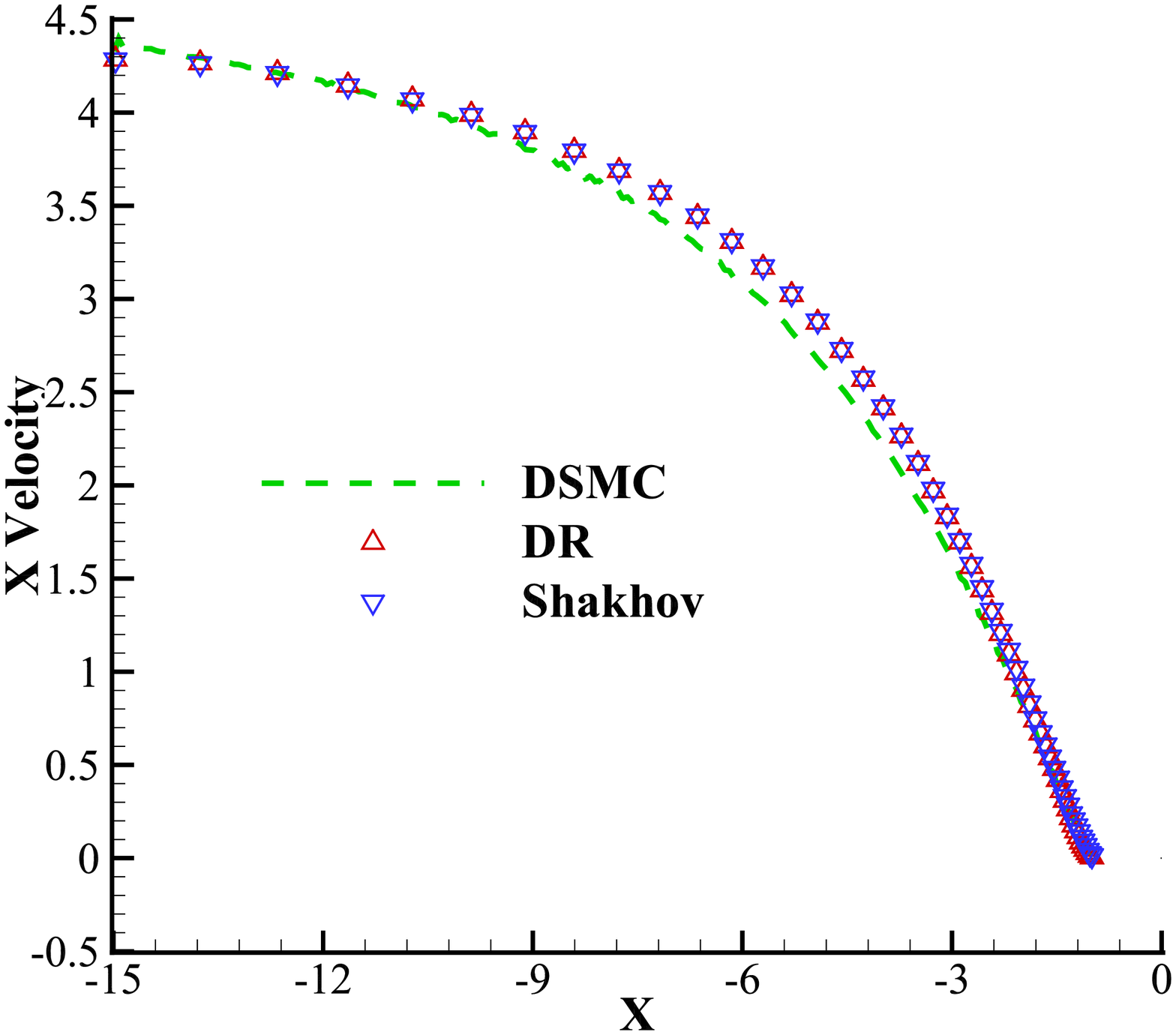}
}\\
\subfigure[\label{Fig:Ma5Kn10_line_T}]{
\includegraphics[width=0.45\textwidth]{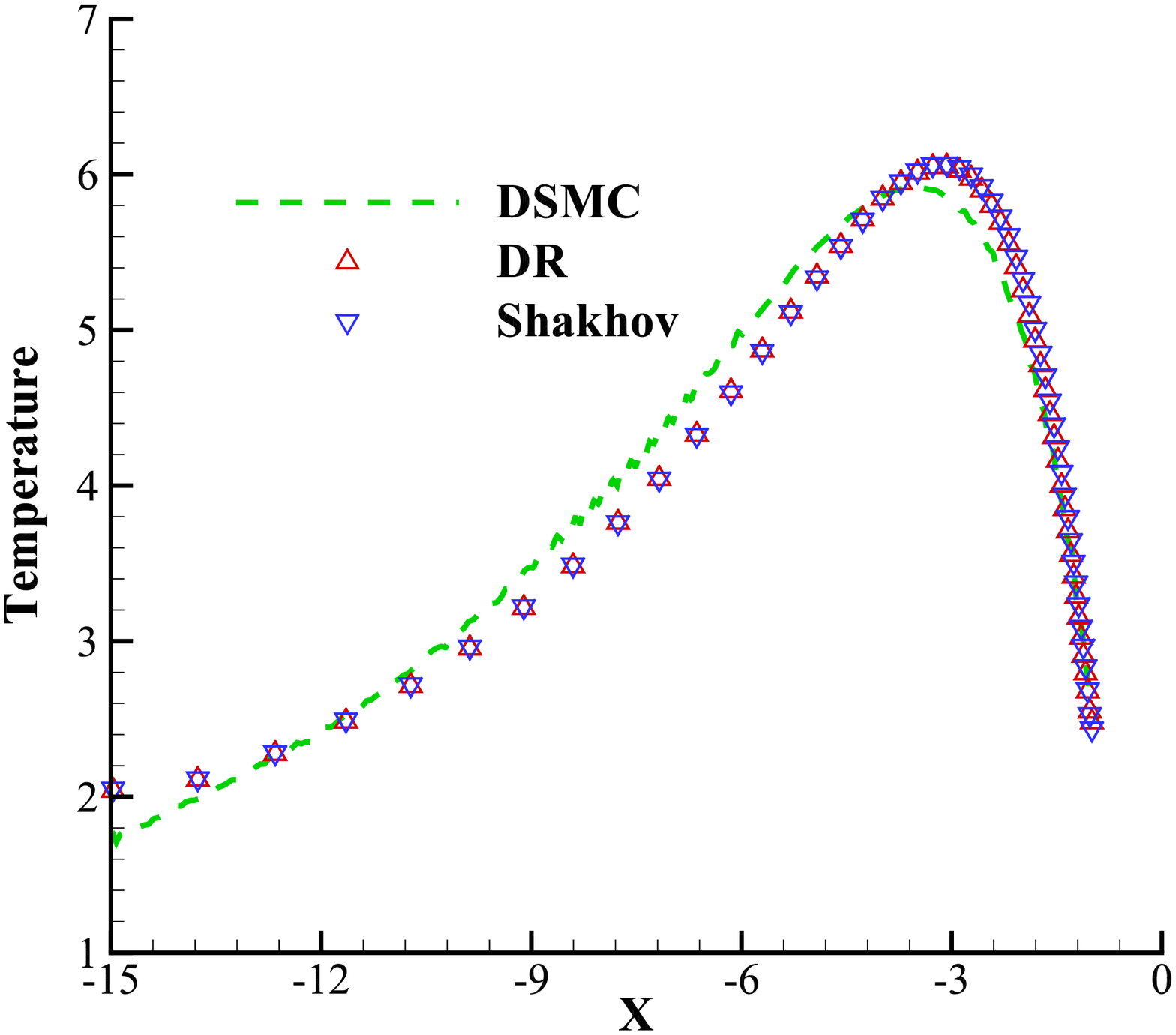}
}\hspace{0.05\textwidth}%
\caption{\label{Fig:Ma5Kn10_line} The macroscopic variables along the stagnation line of the hypersonic cylinder flow with Ma$=5$ and Kn=$10$, (a) density, (b) U-velocity, (c) temperature}
\end{figure}

\begin{figure}
\centering
\subfigure[\label{Fig:Ma5Kn1_line_rho}]{
\includegraphics[width=0.45\textwidth]{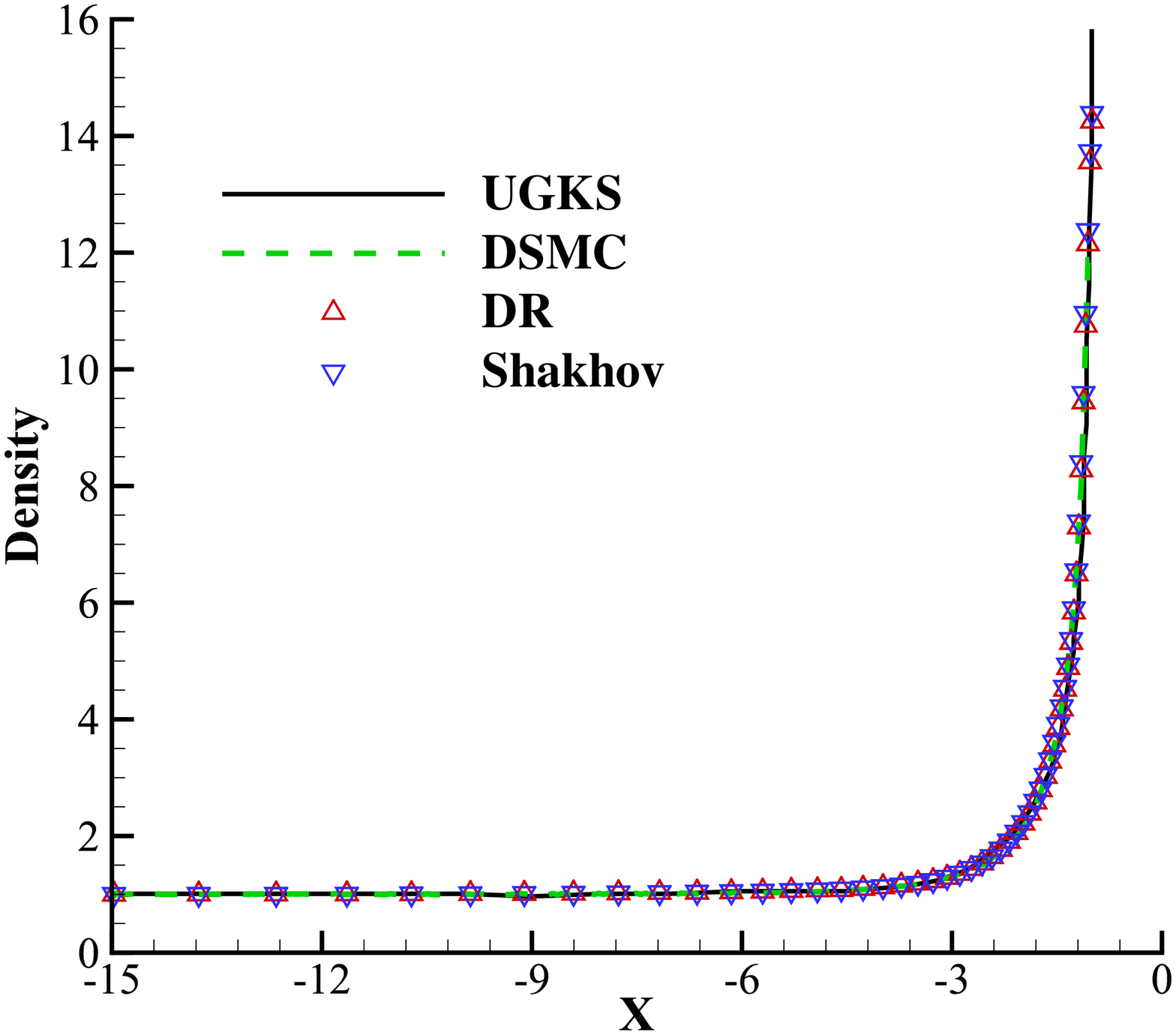}
}\hspace{0.05\textwidth}%
\subfigure[\label{Fig:Ma5Kn1_line_V}]{
\includegraphics[width=0.45\textwidth]{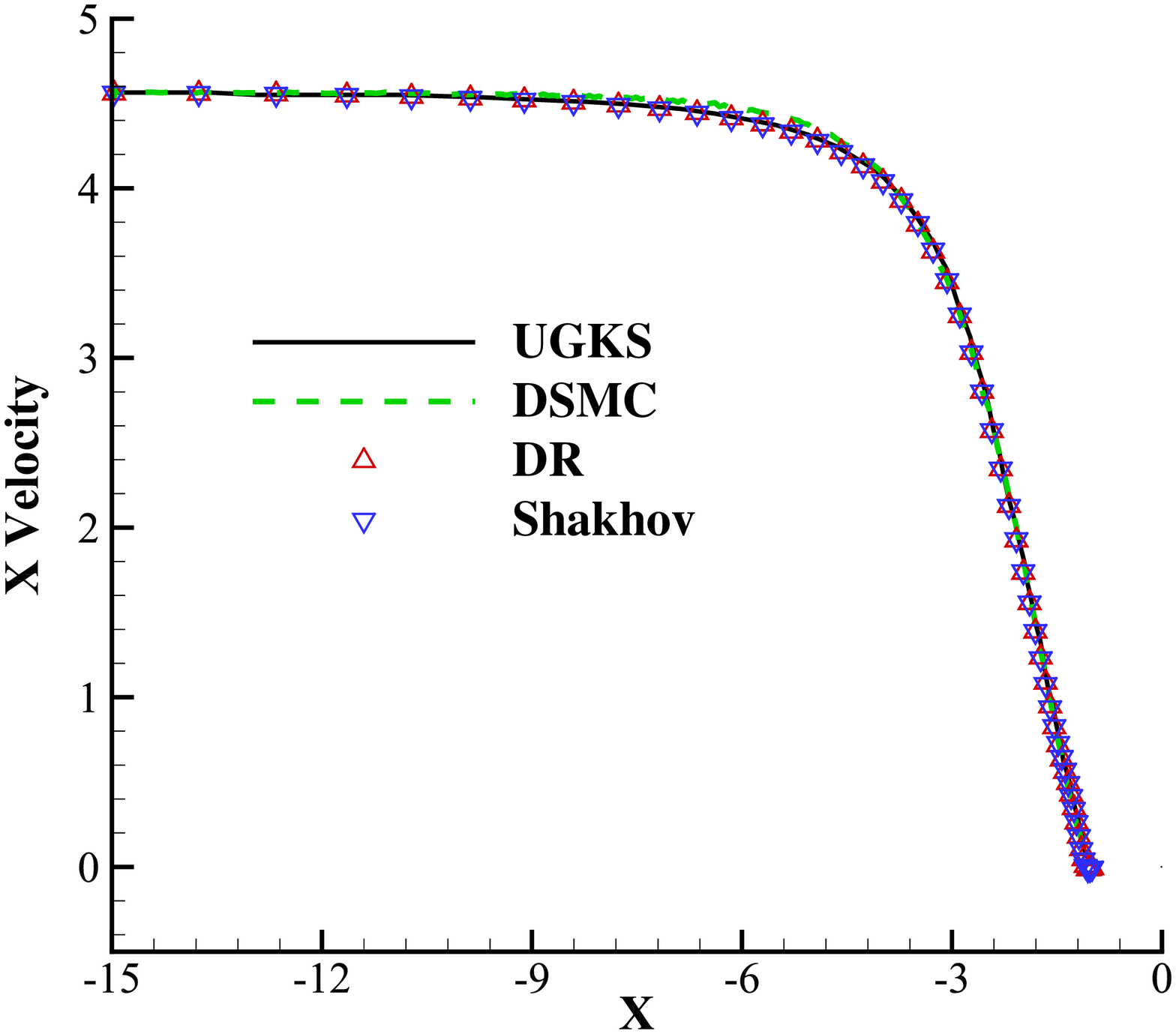}
}\\
\subfigure[\label{Fig:Ma5Kn1_line_T}]{
\includegraphics[width=0.45\textwidth]{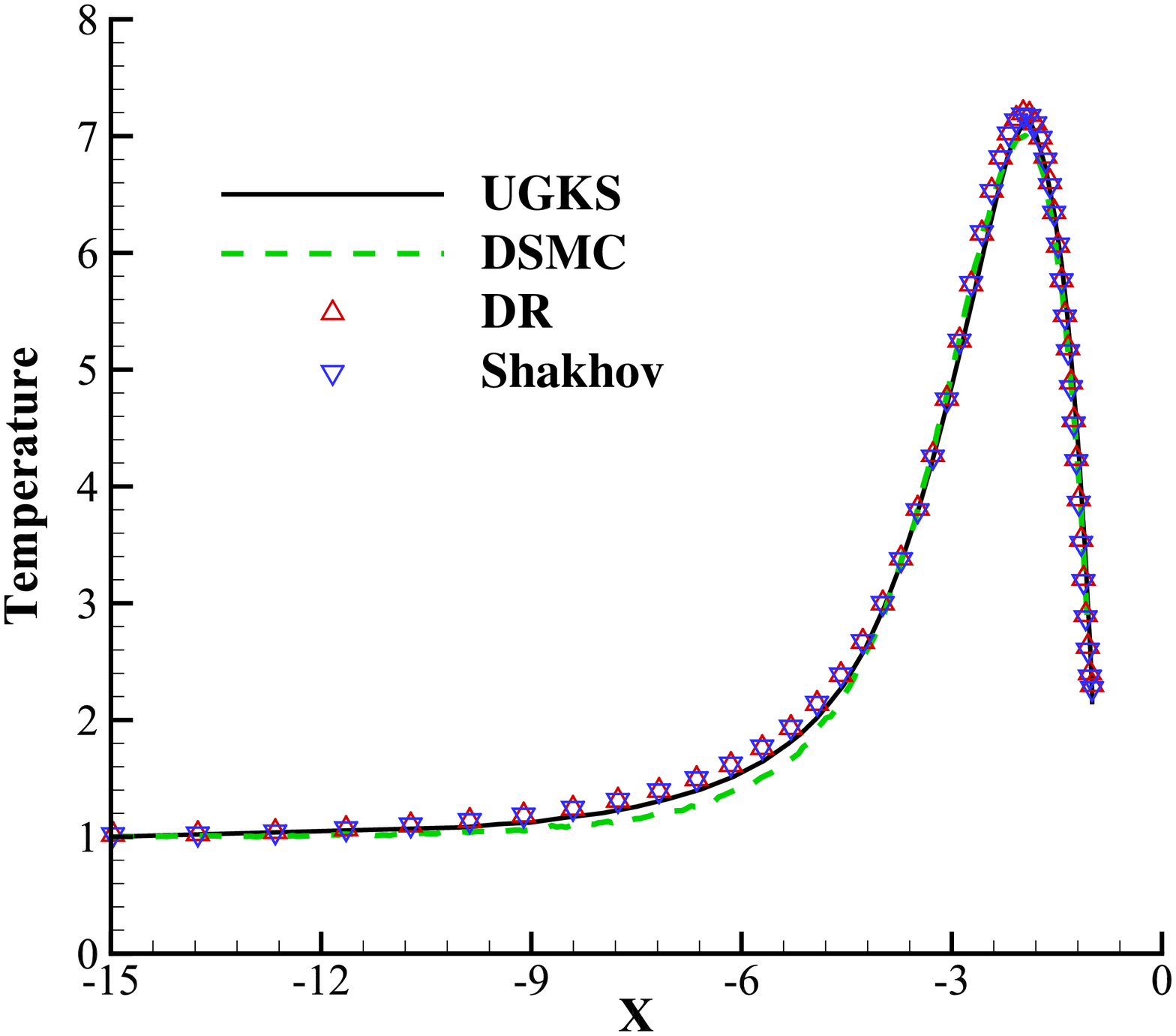}
}\hspace{0.05\textwidth}%
\caption{\label{Fig:Ma5Kn1_line} The macroscopic variables along the stagnation line of the hypersonic cylinder flow with Ma$=5$ and Kn=$1$, (a) density, (b) U-velocity, (c) temperature}
\end{figure}

\begin{figure}
\centering
\subfigure[\label{Fig:Ma5Kn10_rho}]{
\includegraphics[width=0.45\textwidth]{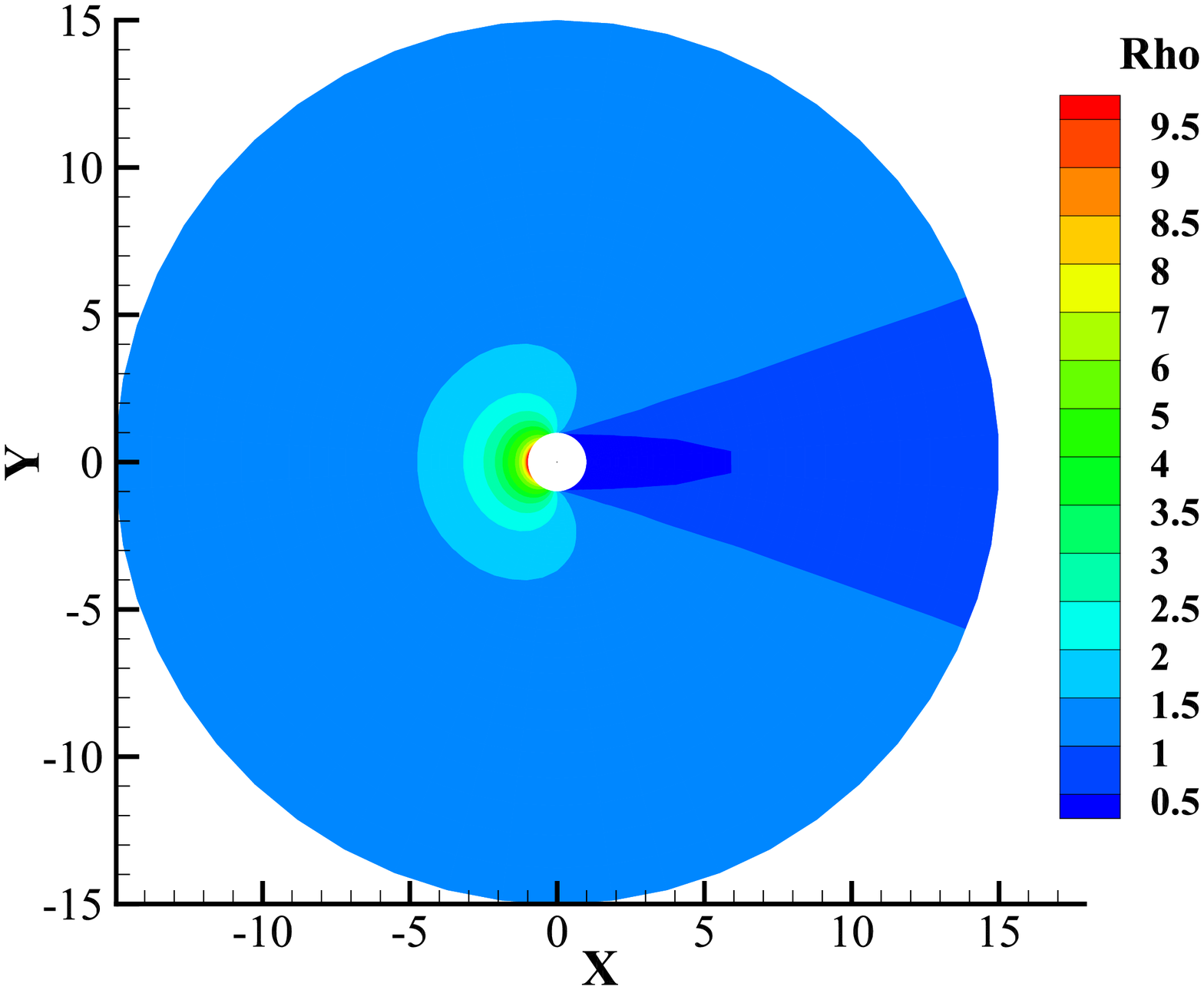}
}\hspace{0.05\textwidth}%
\subfigure[\label{Fig:Ma5Kn10_V}]{
\includegraphics[width=0.45\textwidth]{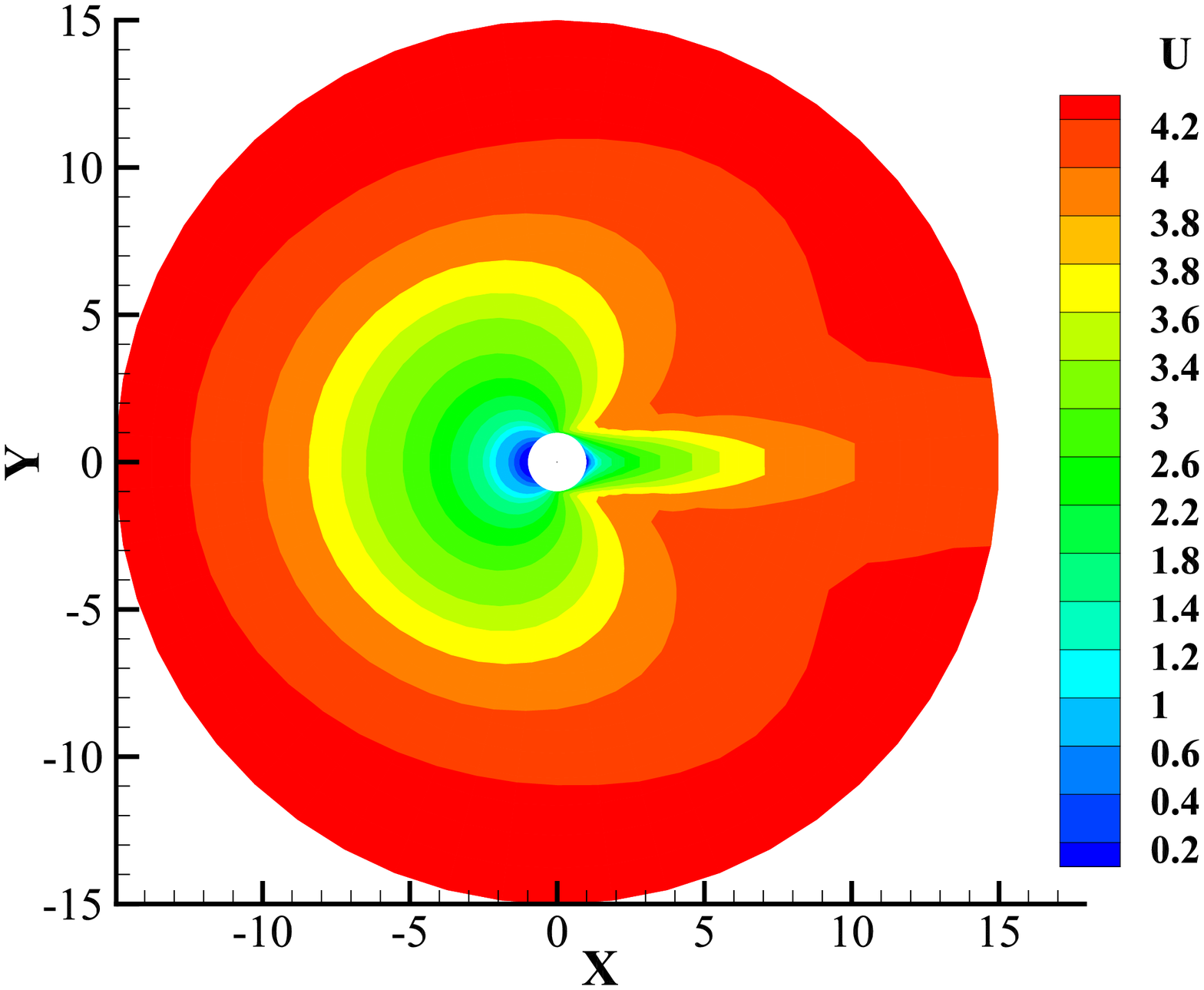}
}\\
\subfigure[\label{Fig:Ma5Kn10_V2}]{
\includegraphics[width=0.45\textwidth]{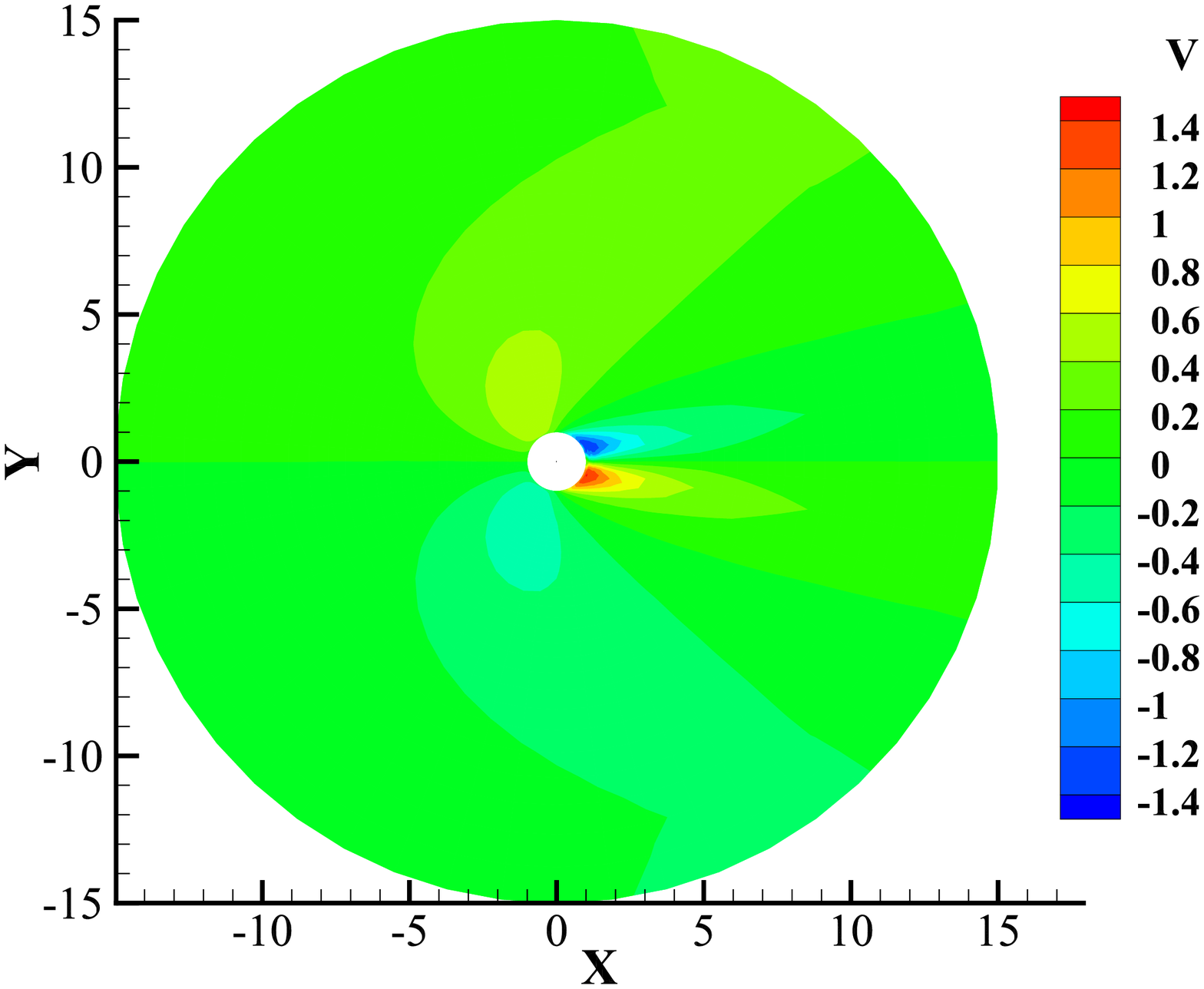}
}\hspace{0.05\textwidth}%
\subfigure[\label{Fig:Ma5Kn10_T}]{
\includegraphics[width=0.45\textwidth]{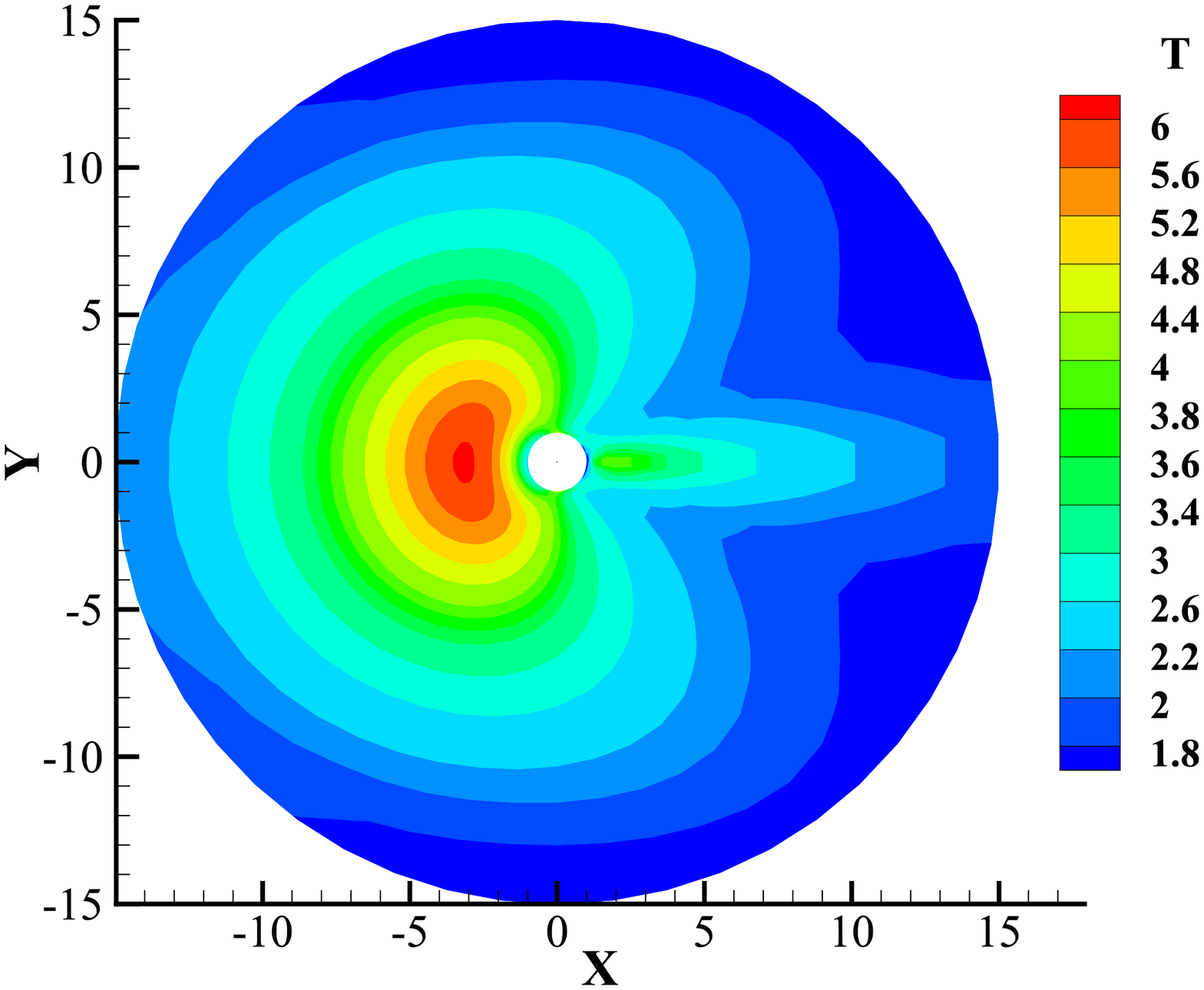}
}
\caption{\label{Fig:Ma5Kn10} The flow field of the hypersonic cylinder flow with Ma$=5$ and Kn=$10$, (a) density contour, (b) U-velocity contour, (c)V-velocity contour, (d) temperature contour}
\end{figure}

\begin{figure}
\centering
\subfigure[\label{Fig:Ma5Kn1_rho}]{
\includegraphics[width=0.45\textwidth]{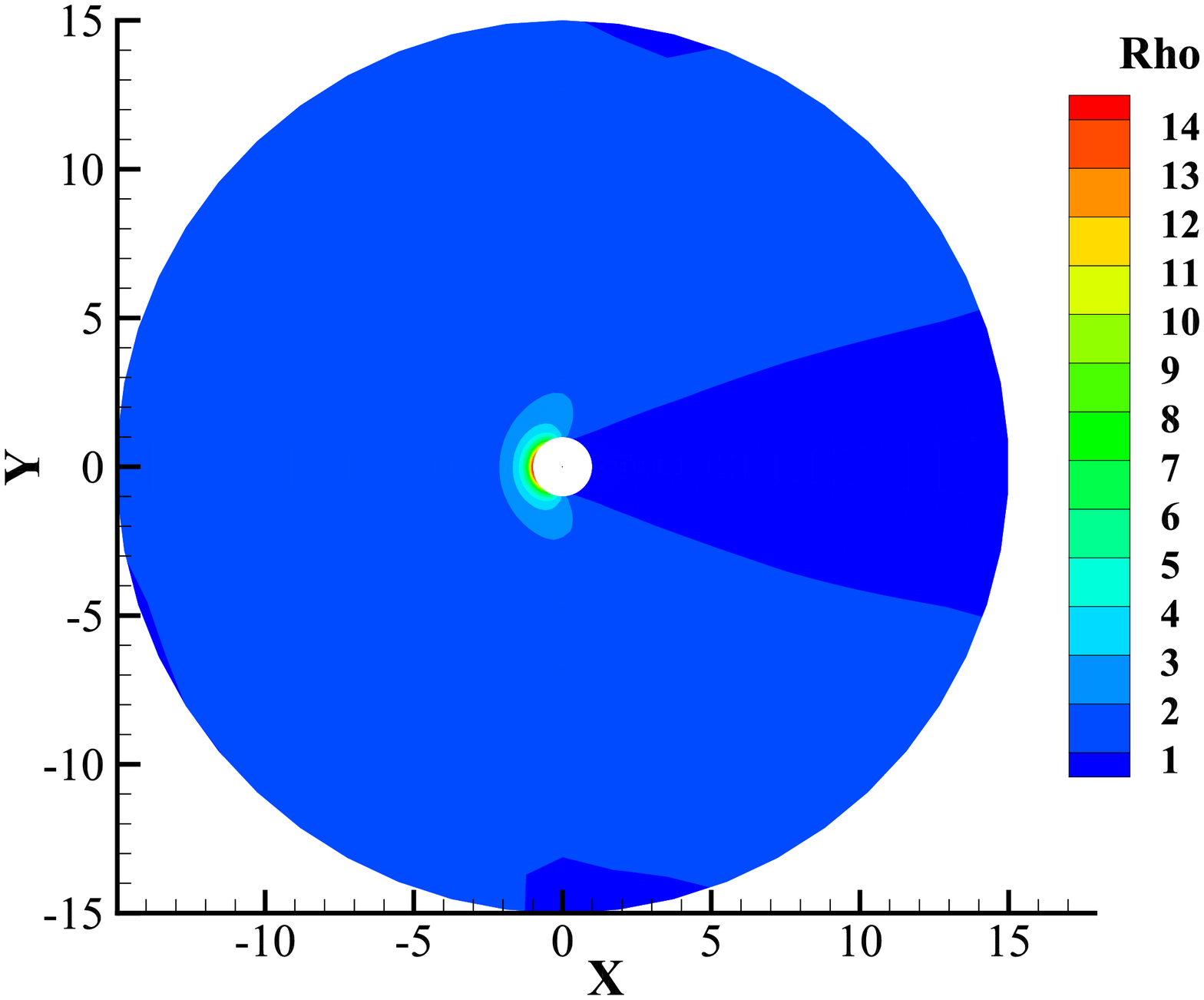}
}\hspace{0.05\textwidth}%
\subfigure[\label{Fig:Ma5Kn1_V}]{
\includegraphics[width=0.45\textwidth]{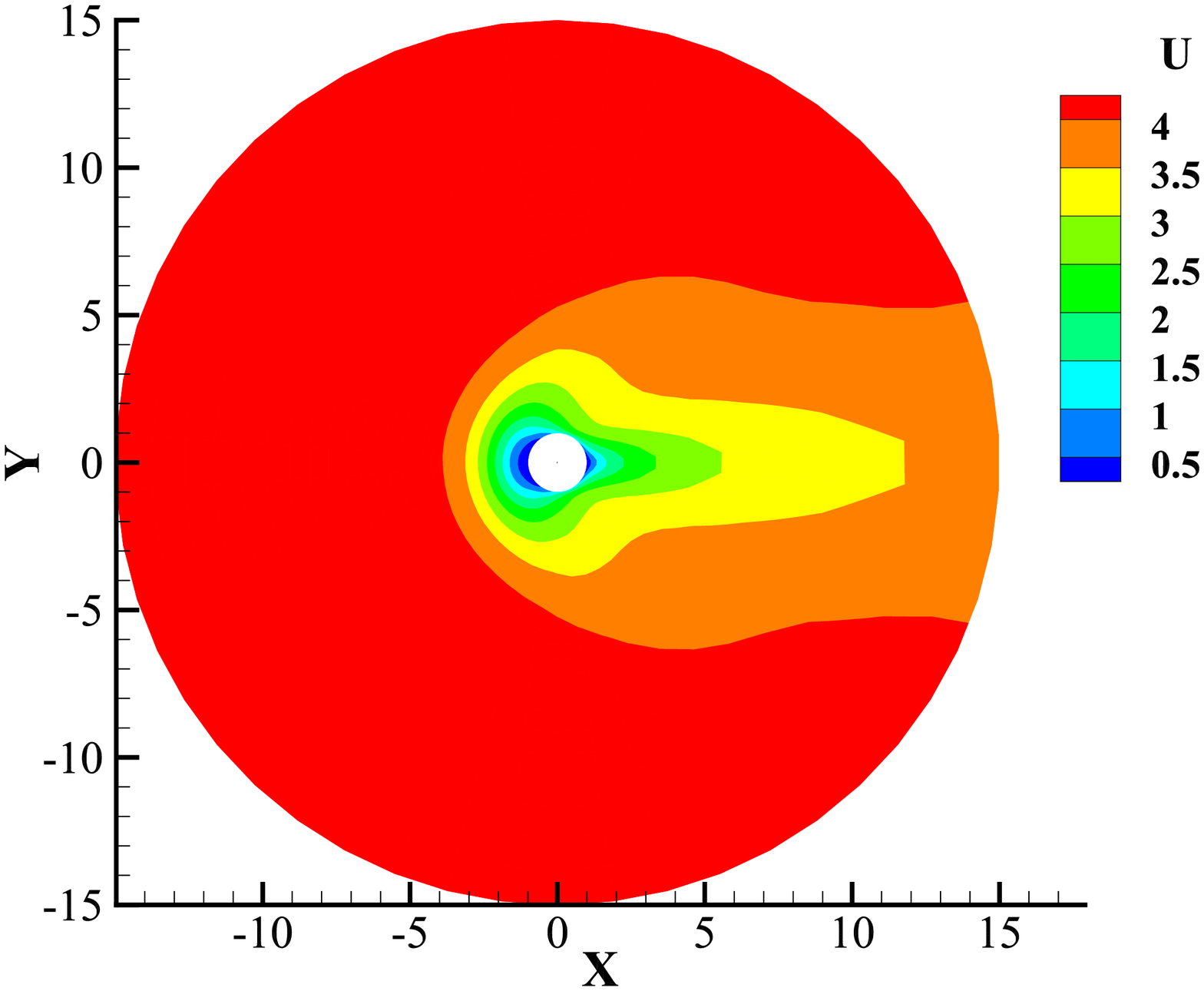}
}\\
\subfigure[\label{Fig:Ma5Kn1_V2}]{
\includegraphics[width=0.45\textwidth]{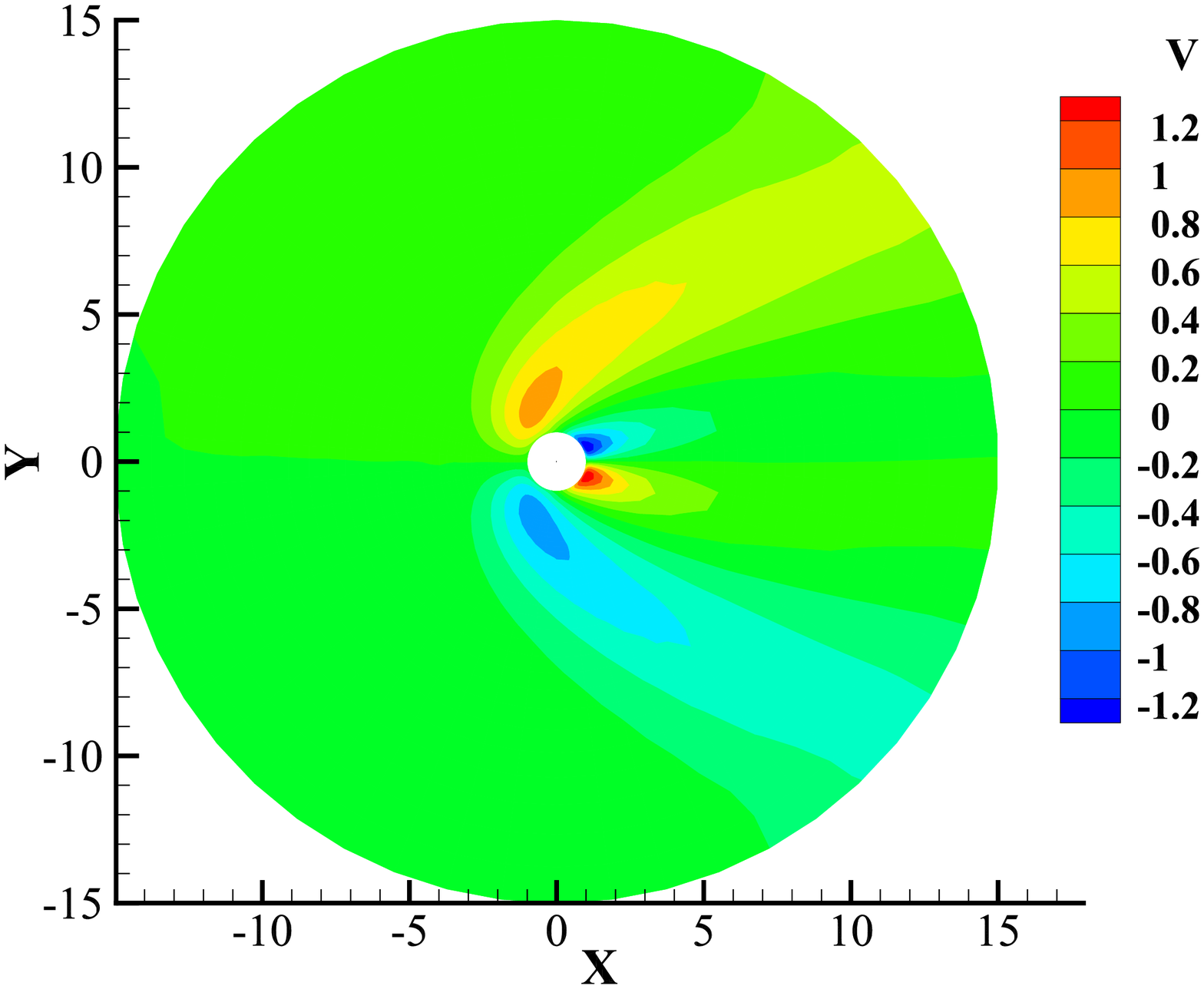}
}\hspace{0.05\textwidth}%
\subfigure[\label{Fig:Ma5Kn1_T}]{
\includegraphics[width=0.45\textwidth]{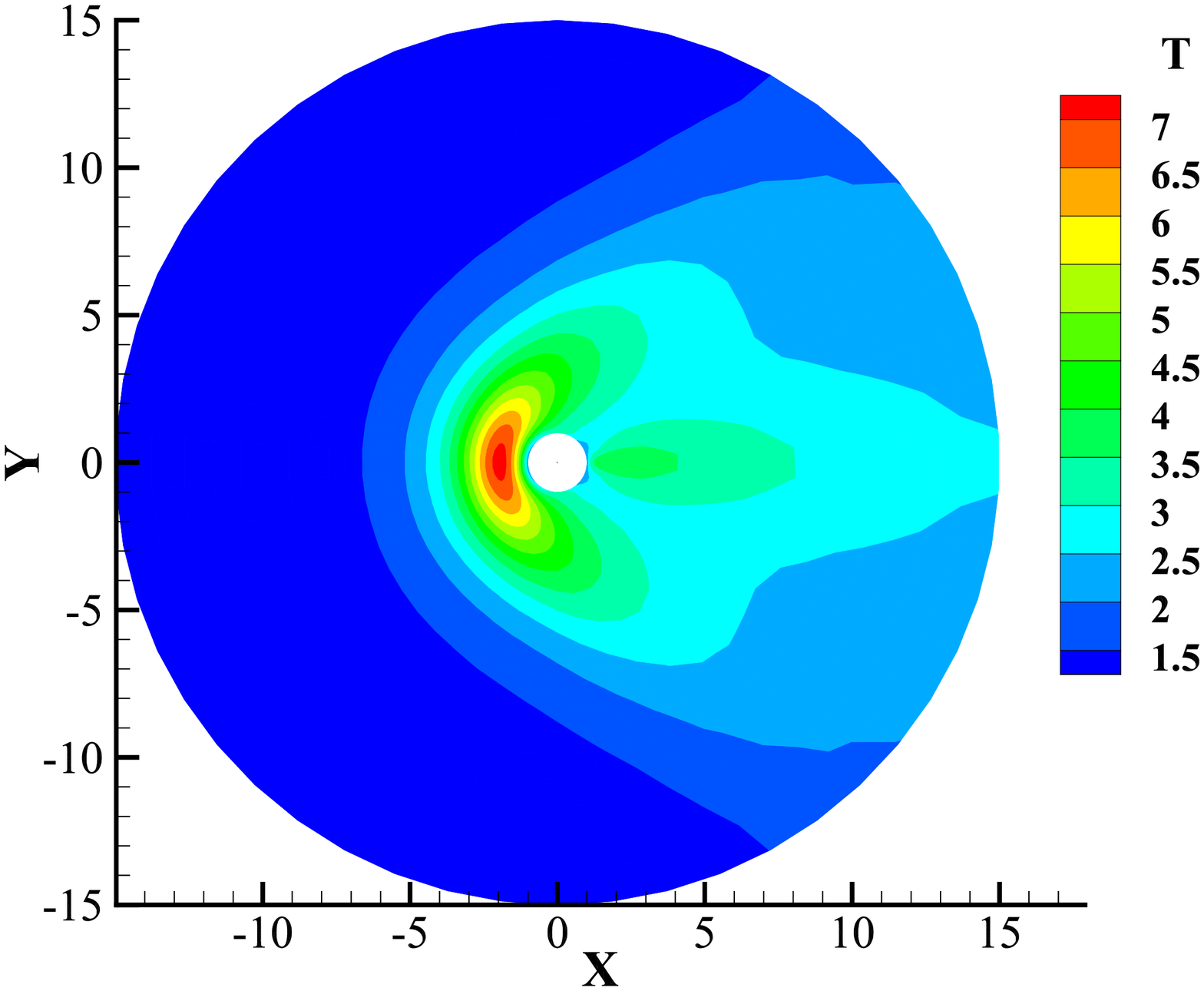}
}
\caption{\label{Fig:Ma5Kn1} The flow field of the hypersonic cylinder flow with Ma$=5$ and Kn=$1$, (a) density contour, (b) U-velocity contour, (c)V-velocity contour, (d) temperature contour}
\end{figure}

\begin{figure}
\centering
\subfigure[\label{Fig:Ma5Kn10_wall_force}]{
\includegraphics[width=0.45\textwidth]{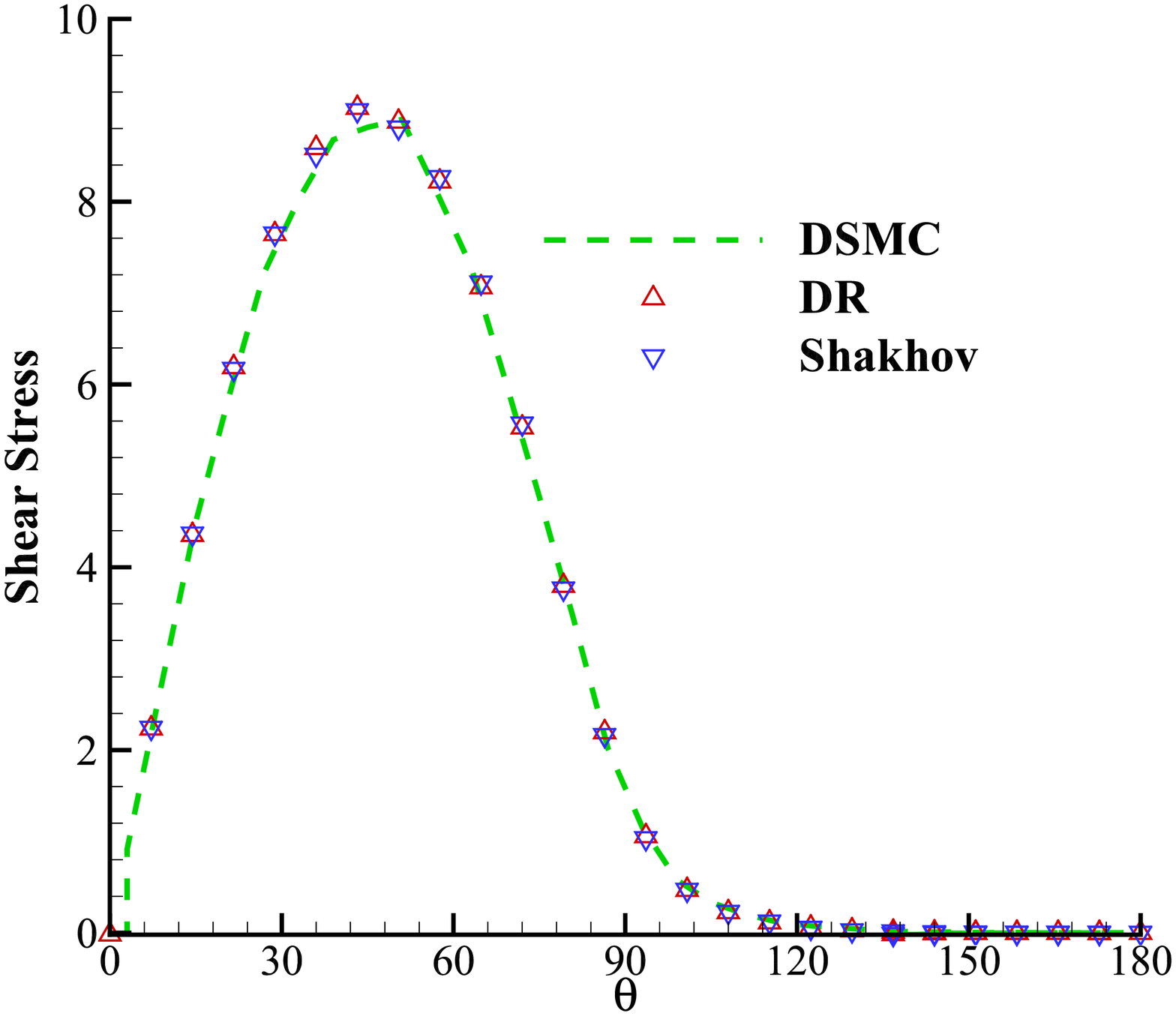}
}\hspace{0.05\textwidth}%
\subfigure[\label{Fig:Ma5Kn10_wall_heat}]{
\includegraphics[width=0.45\textwidth]{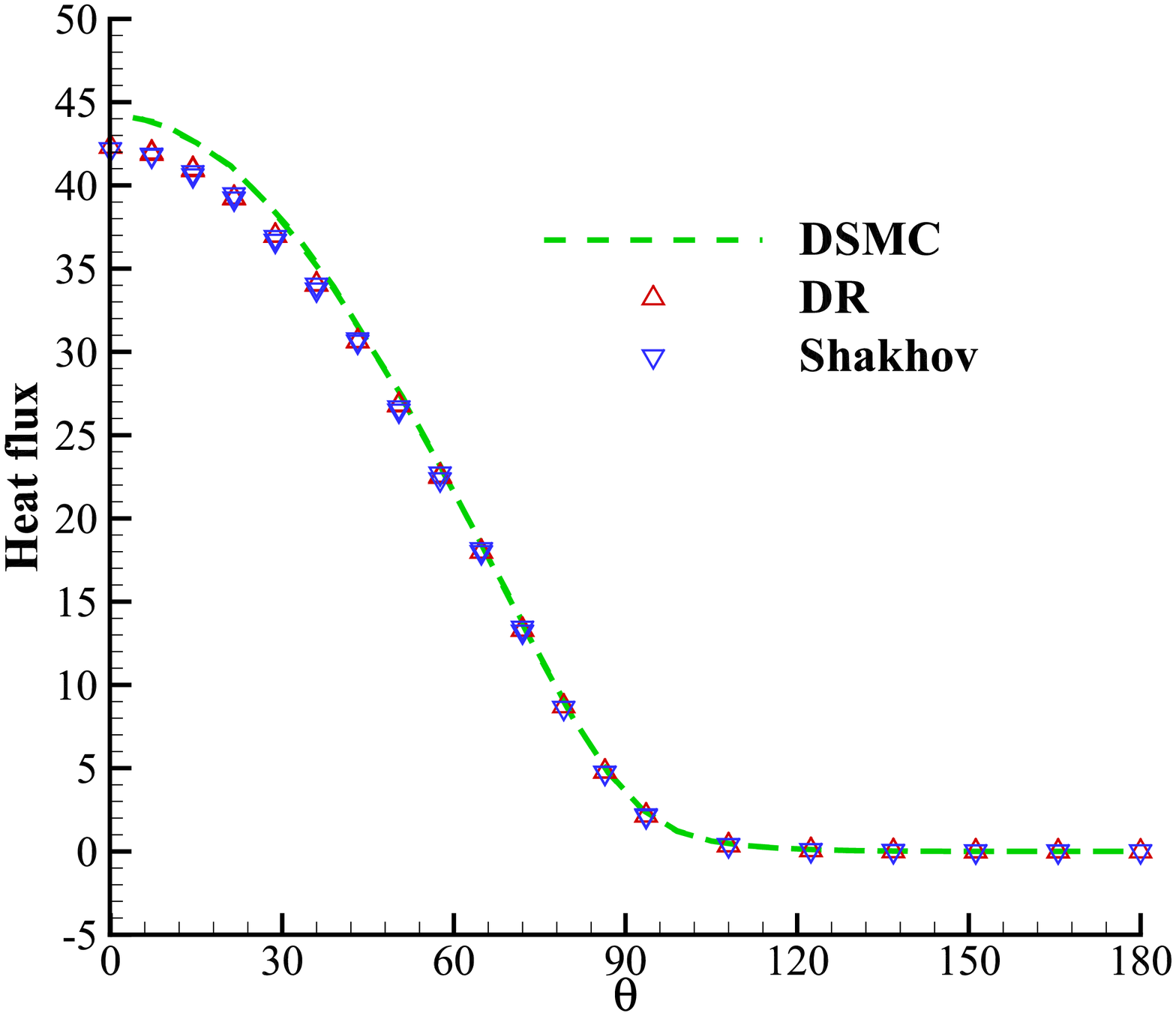}
}
\caption{\label{Fig:Ma5Kn10_wall} The friction force and heat flux on the wall of hypersonic cylinder flow with Ma$=5$ and Kn=$10$, (a) friction force, (b) heat flux}
\end{figure}

\begin{figure}
\centering
\subfigure[\label{Fig:Ma5Kn1_wall_force}]{
\includegraphics[width=0.45\textwidth]{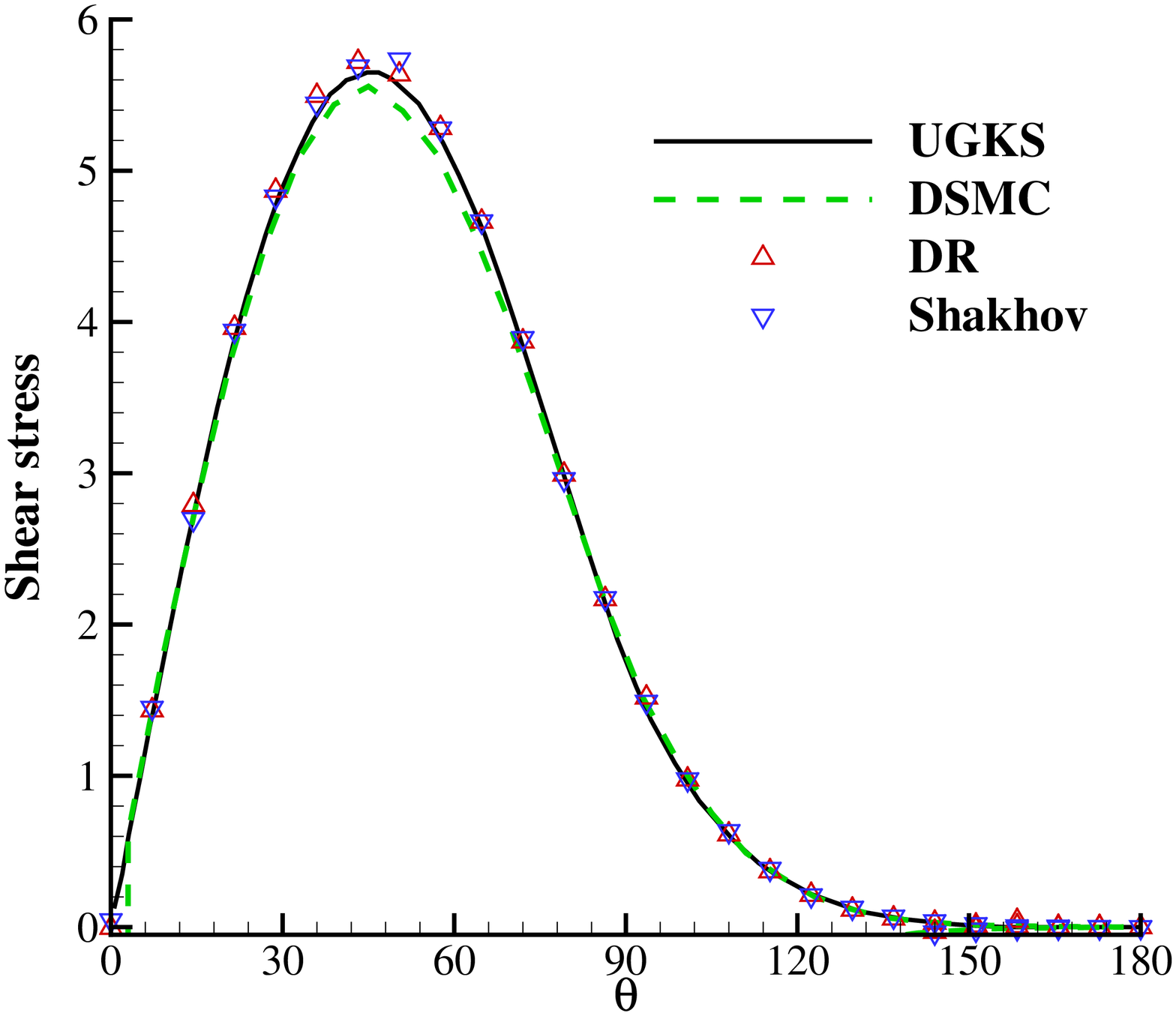}
}\hspace{0.05\textwidth}%
\subfigure[\label{Fig:Ma5Kn1_wall_heat}]{
\includegraphics[width=0.45\textwidth]{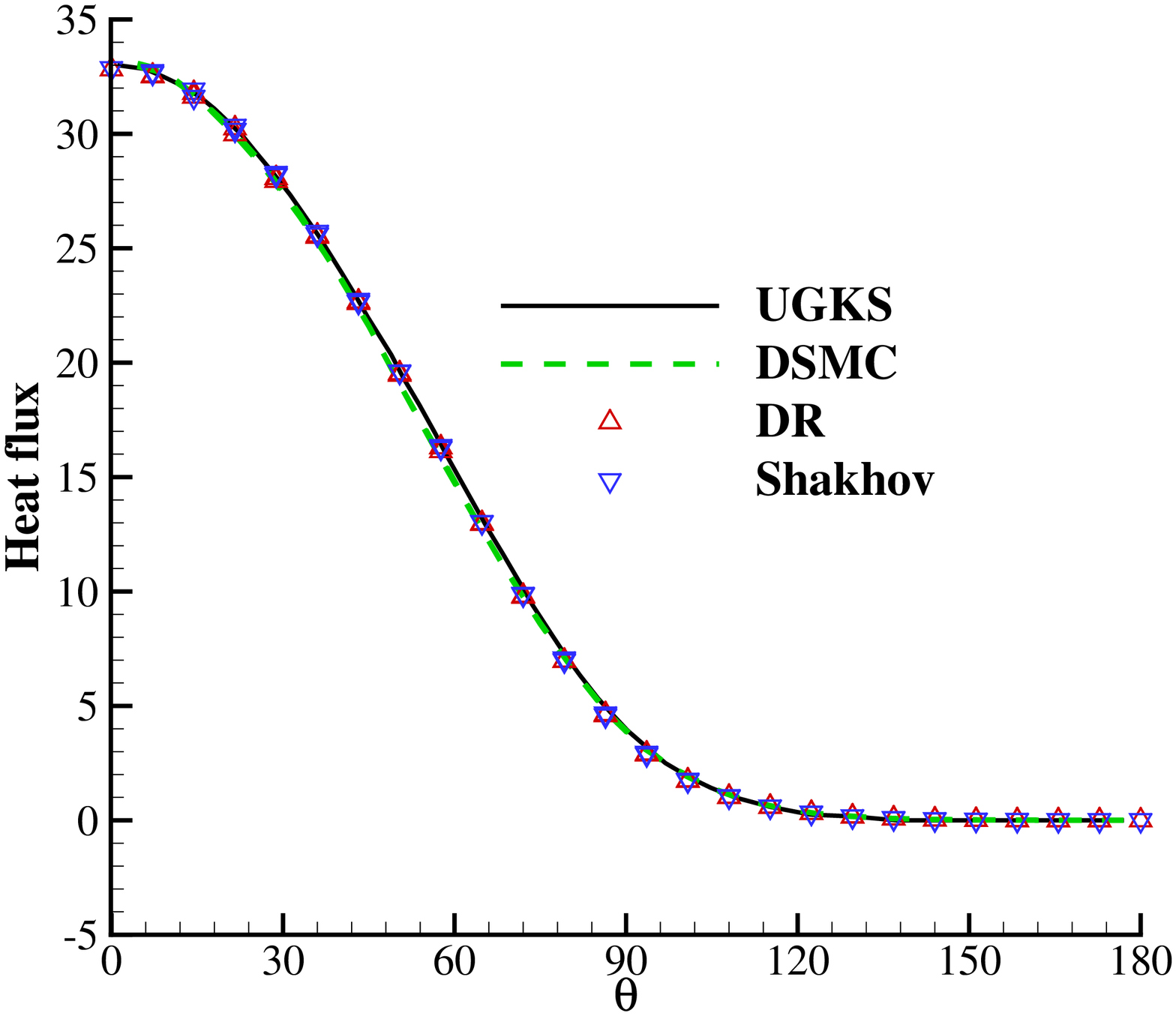}
}
\caption{\label{Fig:Ma5Kn1_wall} The friction force and heat flux on the wall of hypersonic cylinder flow with Ma$=5$ and Kn=$1$, (a) friction force, (b) heat flux}
\end{figure}

Fig.~\ref{Fig:Ma5Kn10} and Fig.~\ref{Fig:Ma5Kn1} illustrate the density, U-velocity and temperature contours predicted by the present DR method for Ma$=5$ cases with Kn=$10$ and $1$. The detailed density, U-velocity and temperature along the front stagnation line (the front central horizontal line) for Ma$=5$ cases with Kn=$10$ and $1$ are shown in Fig.~\ref{Fig:Ma5Kn10_line} and Fig.~\ref{Fig:Ma5Kn1_line}, respectively. The particle method with DR and that with Shakhov are used for prediction, and the benchmark data is from the DSMC and UGKS data in Ref.~\cite{Huang2012A}. The profiles predicted by DR and Shakhov match well with each other, and match well with those from UGKS. There are deviations of these three methods from the DSMC in temperature profiles, because of the deviation of model equation from the Boltzmann one. More specifically, Shakhov model equations are designed that its relaxation rates of stress (second order moment) and heat flux (third order moment) are the same with the Boltzmann equation, and the relaxation rate of entropy or higher order moments are left without consideration, which should be considered in the case of highly non-equilibrium distribution. In the construction work of Shakhov model and the construction work of DR process in this paper, since the considered relaxation rates have the same order, and the same Hermitian distribution is used , the numerical behaviors of theses two methods are very similar. If other distribution forms or high order Hermitian distribution are imposed in DR, the similarity will not exist. This reason is further discussed in Sec.~\ref{sec:disscuss}. The friction force and heat flux on the wall of cylinder are illustrated in Fig.~\ref{Fig:Ma5Kn10_wall} and Fig.~\ref{Fig:Ma5Kn1_wall}, where both the DR and Shakhov method match well with the benchmark data. The predictions on solid wall are better than those in the inner flow filed, because the non-equilibrium effect in the boundary layer is weaker than that around or in the shock waves.

\begin{figure}
\centering
\subfigure[\label{Fig:Ma20Kn01_line_rho}]{
\includegraphics[width=0.45\textwidth]{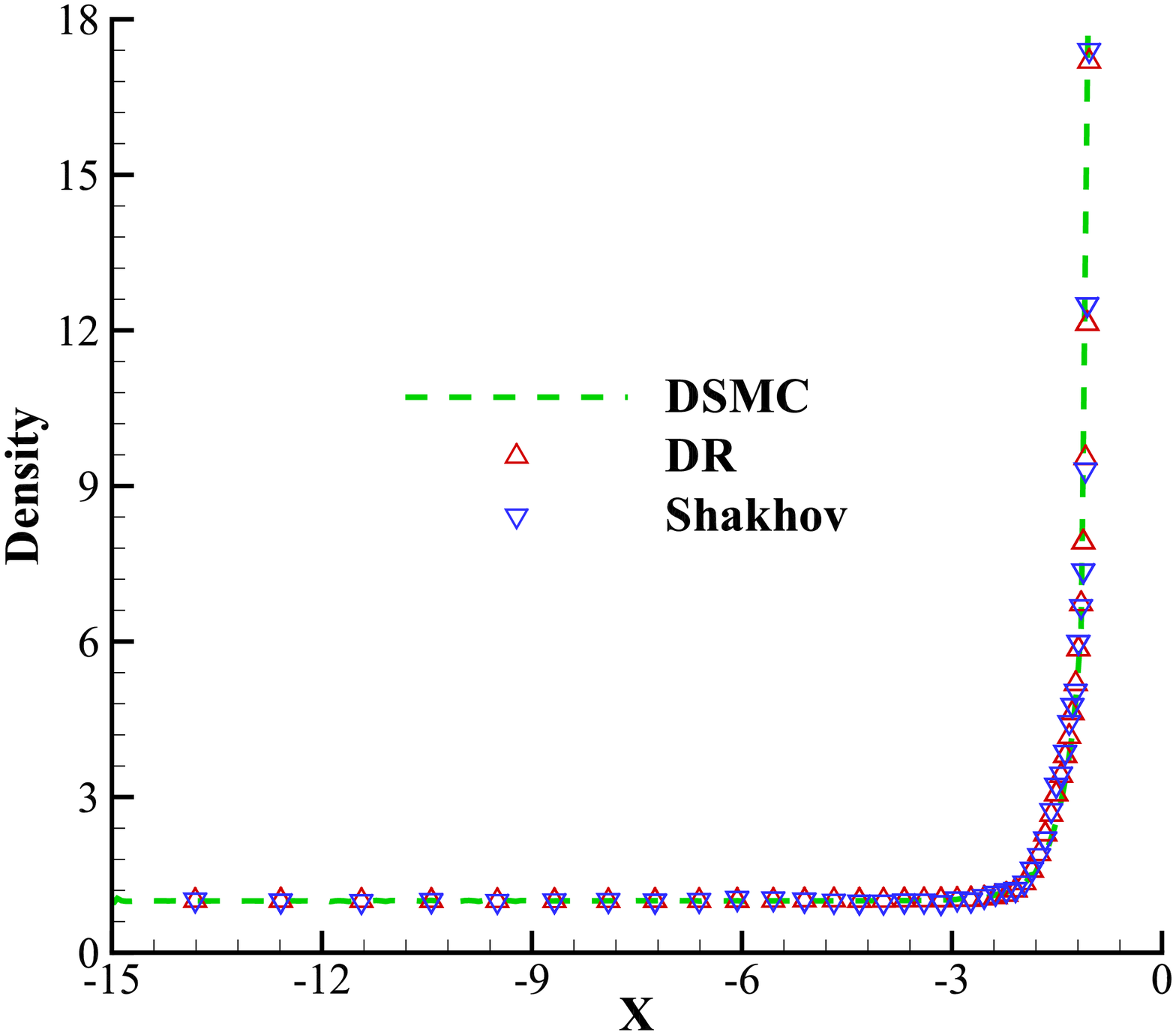}
}\hspace{0.05\textwidth}%
\subfigure[\label{Fig:Ma20Kn01_line_V}]{
\includegraphics[width=0.45\textwidth]{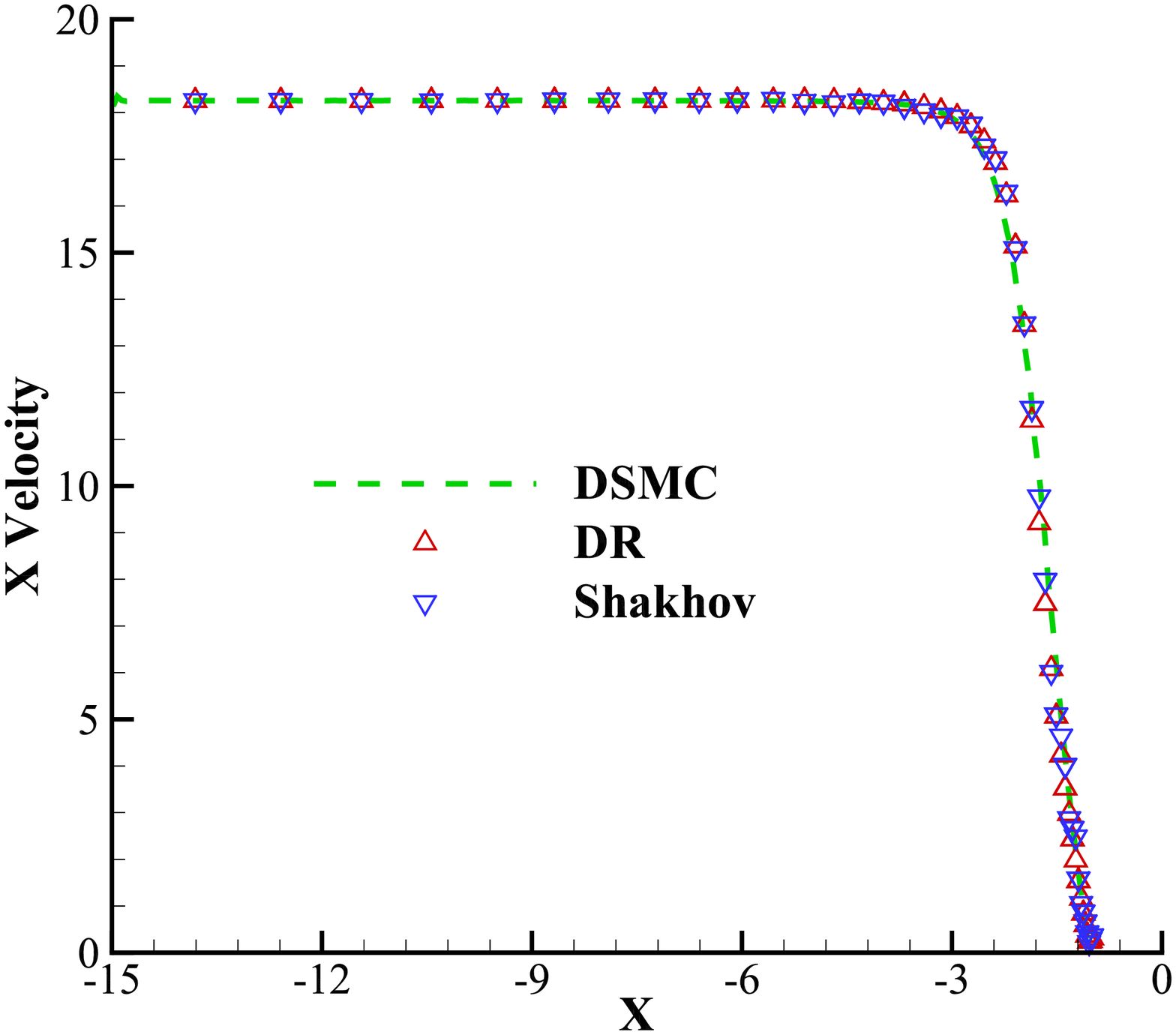}
}\\
\subfigure[\label{Fig:Ma20Kn01_line_T}]{
\includegraphics[width=0.45\textwidth]{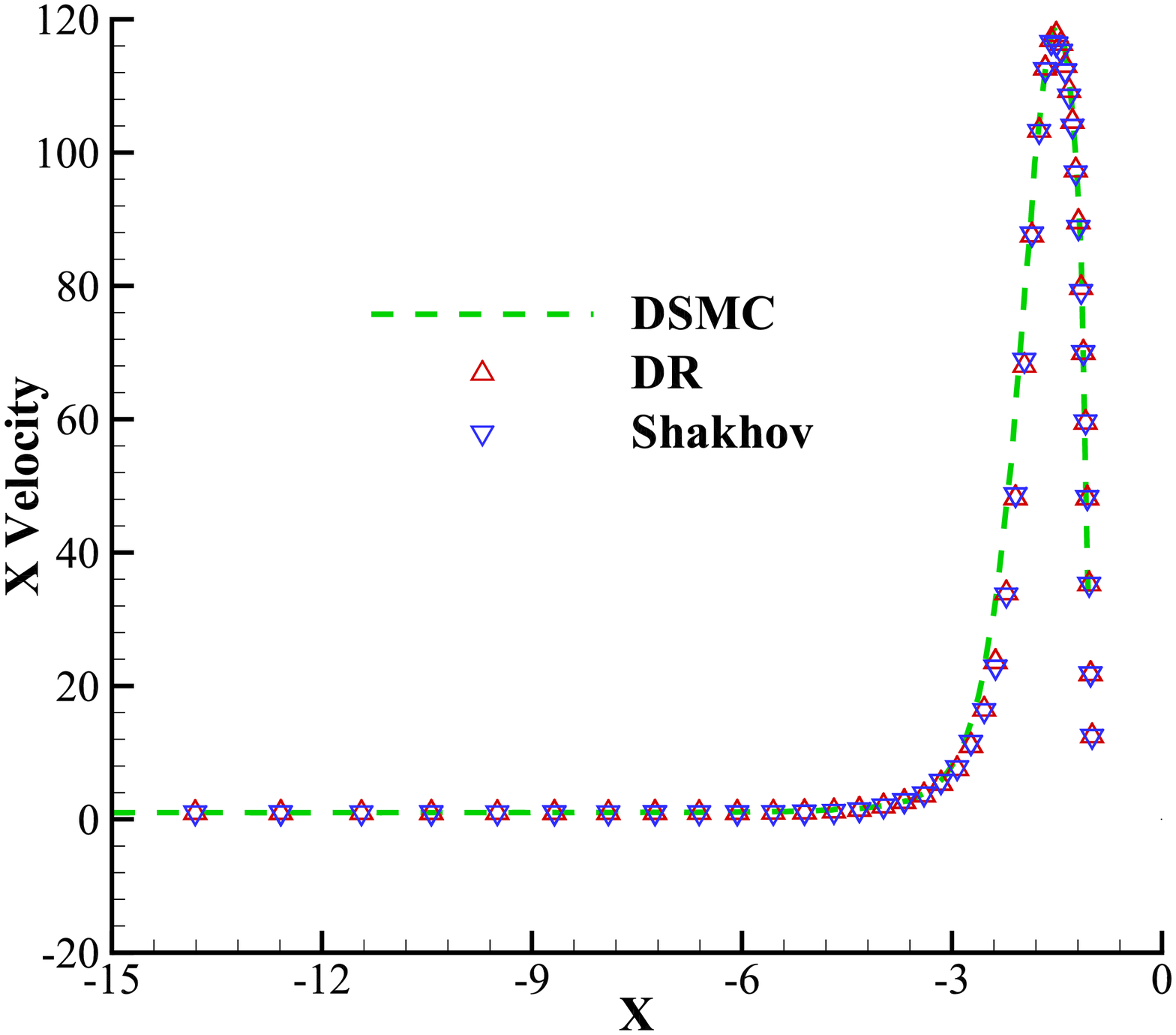}
}\hspace{0.05\textwidth}%
\caption{\label{Fig:Ma20Kn01_line} The macroscopic variables along the stagnation line of the hypersonic cylinder flow with Ma$=20$ and Kn=$0.1$, (a) density, (b) U-velocity, (c) temperature}
\end{figure}

\begin{figure}
\centering
\subfigure[\label{Fig:Ma20Kn01_rho}]{
\includegraphics[width=0.45\textwidth]{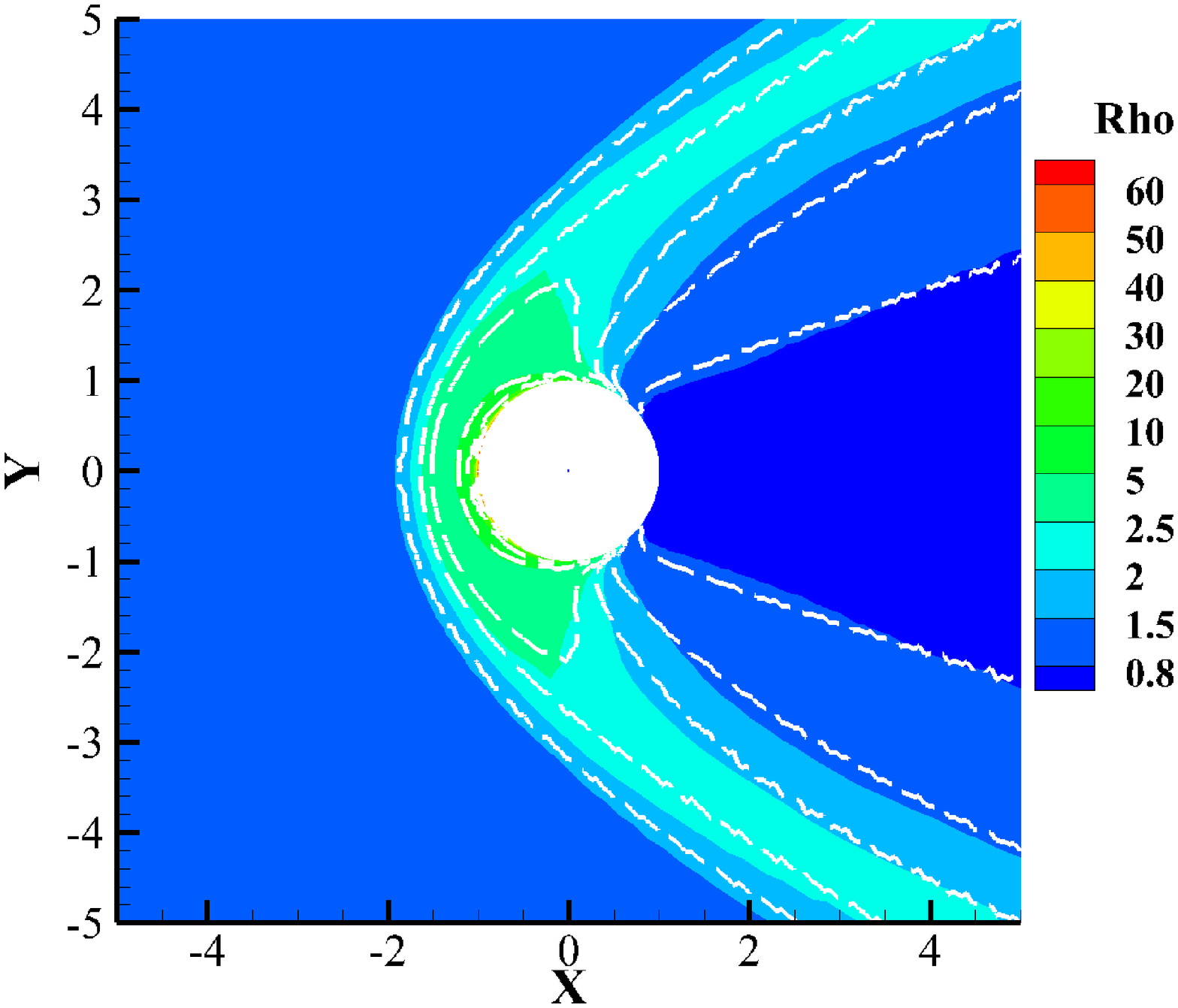}
}\hspace{0.05\textwidth}%
\subfigure[\label{Fig:Ma20Kn01_V}]{
\includegraphics[width=0.45\textwidth]{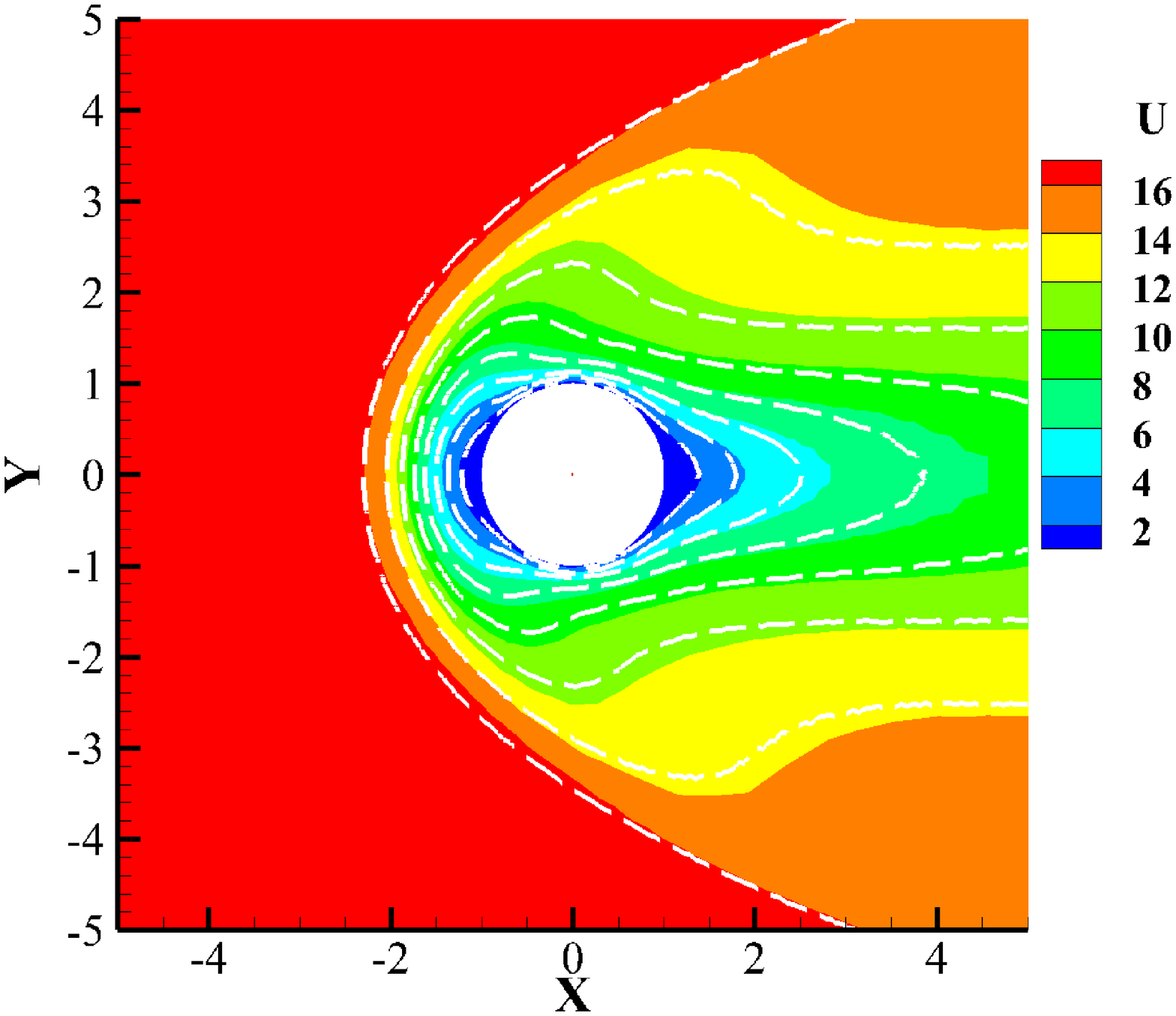}
}\\
\subfigure[\label{Fig:Ma20Kn01_V2}]{
\includegraphics[width=0.45\textwidth]{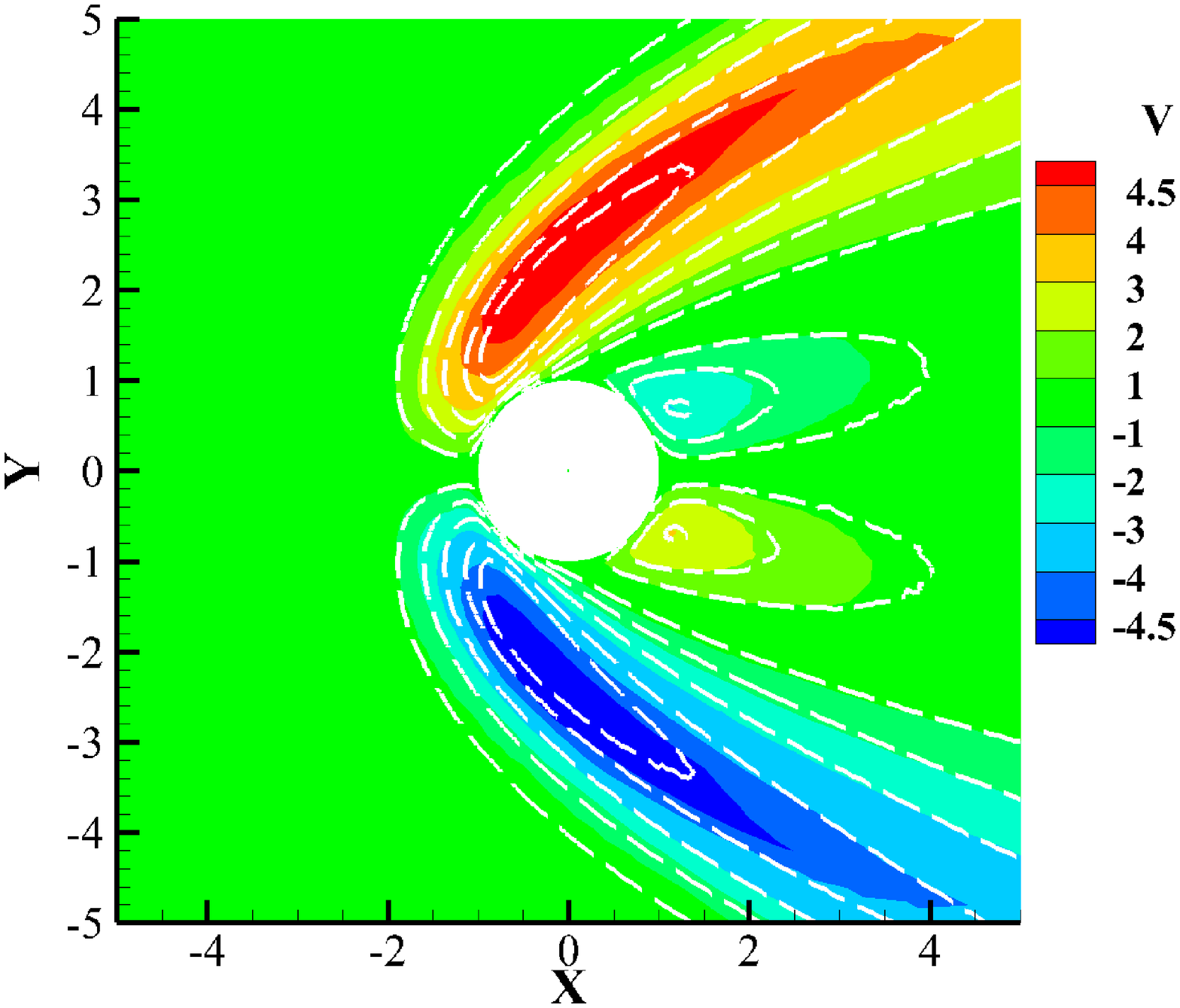}
}\hspace{0.05\textwidth}%
\subfigure[\label{Fig:Ma20Kn01_T}]{
\includegraphics[width=0.45\textwidth]{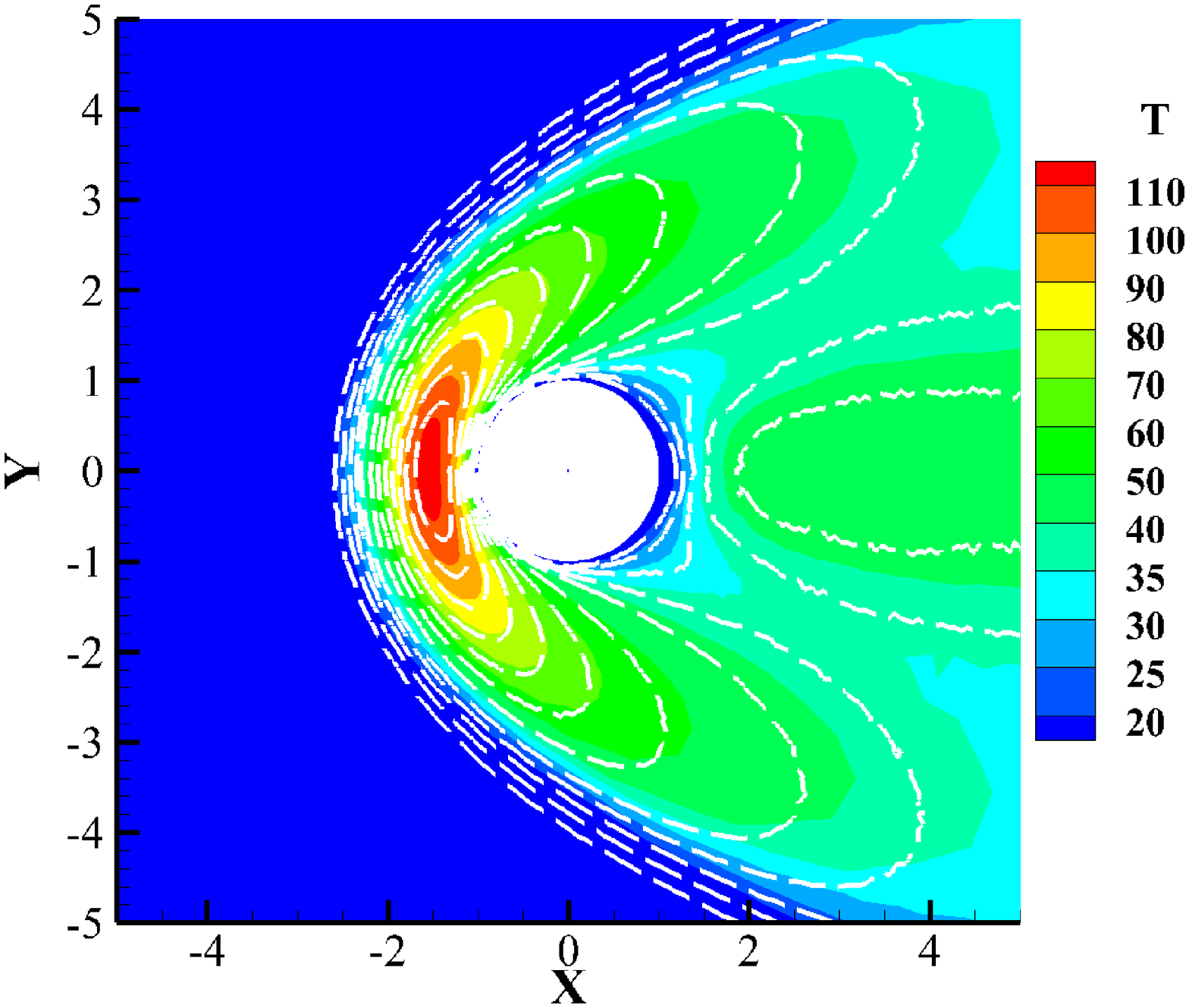}
}
\caption{\label{Fig:Ma20Kn01} The flow field of the hypersonic cylinder flow with Ma$=20$ and Kn=$0.1$, (a) density contour, (b) U-velocity contour, (c) V-velocity contour, (d) temperature contour. The white dashed contours are the DSMC data, and the contours with colored band are the DR predictions}
\end{figure}

Fig.~\ref{Fig:Ma20Kn01} illustrates the density, U-velocity and temperature contours predicted by the present DR method for Ma$=20$ cases with and Kn=$0.1$. These macroscopic variables along the stagnation line for Ma$=20$ case with Kn=$0.1$ are shown in Fig.~\ref{Fig:Ma20Kn01_line}. The particle methods with DR and that with Shakhov are used for prediction, and the benchmark data is from DS2V code~\cite{ds2v}. The results predicted by the present DR method match well with the Shakhov predictions and the benchmark DSMC data, and the corresponding deviations are similar to those of Ma$=5$ case.

\section{\label{sec:conclusion}Discussion and conclusion}
\subsection{Discussion}\label{sec:disscuss}
In the above test cases except the homogenous relaxation, the numerical results given by particle methods with DR process and Shakhov model almost overlap with each other. The reason can be explained as follows:

From the statistical perspective, the values of macroscopic physical variables for colliding molecules can be:
\begin{equation}
\begin{aligned}
&\rho^{*} {\rm{ = }}\frac{\tau }{{\tau {\rm{ + }}\Delta t}}{\rho ^i},\\
&{\bf{U}}^{*} = {{\bf{U}}^i},\\
&T^{*} = {T^i},\\
&{\bf{S}}^{*} = 0,\\
&{\bf{Q}}^{*} = \frac{{\Delta t}}{{\tau  + \Delta t}}\left( {1 - \Pr } \right){{\bf{Q}}^i}.
\end{aligned}
\end{equation}
Inserting them into the Hermitian distribution of colliding molecules (Eq.~\ref{eq:gstar}), this distribution can be rewritten as:
\begin{equation}
\begin{aligned}
{g^*}_{H}\left( {\bf{\xi }} \right) = &\frac{{\Delta t}}{{\tau  + \Delta t}}{g^i}_{M}\left( {\bf{\xi }} \right)\\
&\left\{ {1{\rm{ + }}\left( {\frac{{\tau {\rm{ + }}\Delta t}}{{\Delta t}}} \right)\frac{1}{{R{T^i}{P^i}}}\left[ {\frac{{\Delta t\tau \left( {1 - \Pr } \right)}}{{\left( {\tau  + \Delta t} \right)\left( {\tau  + \Pr \Delta t} \right)}}{\bf{Q}}^i \cdot {\bf{c}}^{\rm{i}}\left( {\frac{{{{\left| {{{\bf{c}}^i}} \right|}^2}}}{{5R{T^i}}} - 1} \right)} \right]} \right\},
\end{aligned}
\end{equation}
where ${{\Delta t}}/{\left({\tau  + \Delta t}\right)}$ is the portion of colliding particles. If we further assume $\Delta t \ll \tau$  (this assumption exactly holds in free molecular limit, and is accurate enough for rarefied flow), the Hermitian distribution becomes
\begin{equation}
\begin{aligned}
{g^*}_{H}\left( {\bf{\xi }} \right)&= \frac{{\Delta t}}{{\tau  + \Delta t}}{g^i}_{M}\left( {\bf{\xi }} \right)\left\{ {1{\rm{ + }}\frac{1}{{R{T^i}{P^i}}}\left[ {\left( {1 - \Pr } \right){\bf{Q}}^i \cdot {\bf{c}}^{\rm{i}}\left( {\frac{{{{\left| {{{\bf{c}}^i}} \right|}^2}}}{{5R{T^i}}} - 1} \right)} \right]} \right\}\\
&{\rm{= }}\frac{{\Delta t}}{{\tau  + \Delta t}}{g^i}_{S}\left( {\bf{\xi }} \right),
\end{aligned}
\end{equation}
which is the Shakhov distribution function multiplied by ${{\Delta t}}/{\left({\tau  + \Delta t}\right)}$ ($g_{S}$ denotes the Shakhov equilibrium distribution). It means that, for this single component case, the effect of DR treatment can be reduced to that of the particle-BGK method with Shakhov model. But if the Hermitian distribution is not chosen as the mathematical form, there is no such reduction.

Although the DR treatment is designed to avoid the construction of Boltzmann model equation, especially for the complex thermal non-equilibrium gas mixtures. Since the DR process directly calculates the macroscopic relaxation, its numerical behavior is better than that sampled from the Shakhov equilibrium state in the case of large iteration time such as the homogenous relaxation case.

\subsection{Conclusion}
In this paper, the Direct Relaxation (DR) approach directly calculates the macroscopic variables after relaxation and those for colliding particles, according to which the after-collision velocity is sampled. This modified process makes the particle-BGK not restricted by the model equations, and the relaxation rate that constrains the model equations can be directly used in numerical schemes. By conducting the 0-D relaxation case, 1-D Sod shock tube and high non-equilibrium shock structure cases, 2-D low speed cavity case and 2-D hypersonic cylinder case, the numerical results of present method match well with those of the UGKS, the DSMC and the particle-BGK method with Shakhov model. Noted that the distribution used in this paper is the Hermitian one where the negative distribution may appear when the heat flux is large, the more precise distribution and sampling technology will be considered in the future works, and DR methods will be extended to gas mixtures.

\begin{acknowledgments}
The authors thank Prof. Kun Xu, Dr. Yajun Zhu and Dr. Yipei Chen at Hong Kong University of Science and Technology (HKUST), and Prof. Chang Liu at Institute of Applied Physics and Computational Mathematics for discussion in constructing multi-scale numerical methods based on wave-particle methods and direct modeling philosophy. Sha Liu thanks Prof. Jun Zhang at Beihang University and Prof. Fei Fei at Huazhong University of Science and Technology for discussion about the multi-scale particle methods. The present work is supported by the National Numerical Wind-Tunnel Project of China and National Natural Science Foundation of China (Grants No. 11902266, No. 12072283 and No. 11902264).
\end{acknowledgments}

\section*{DATA AVAILABILITY}
The data that support the findings of this study are available from the corresponding author upon reasonable request.

\section*{Reference}
\bibliographystyle{elsarticle-num}
\bibliography{SPDR}
\clearpage

\clearpage

\begin{figure}
\centering
\subfigure[\label{Fig:0D_stress}]{
\includegraphics[width=0.45\textwidth]{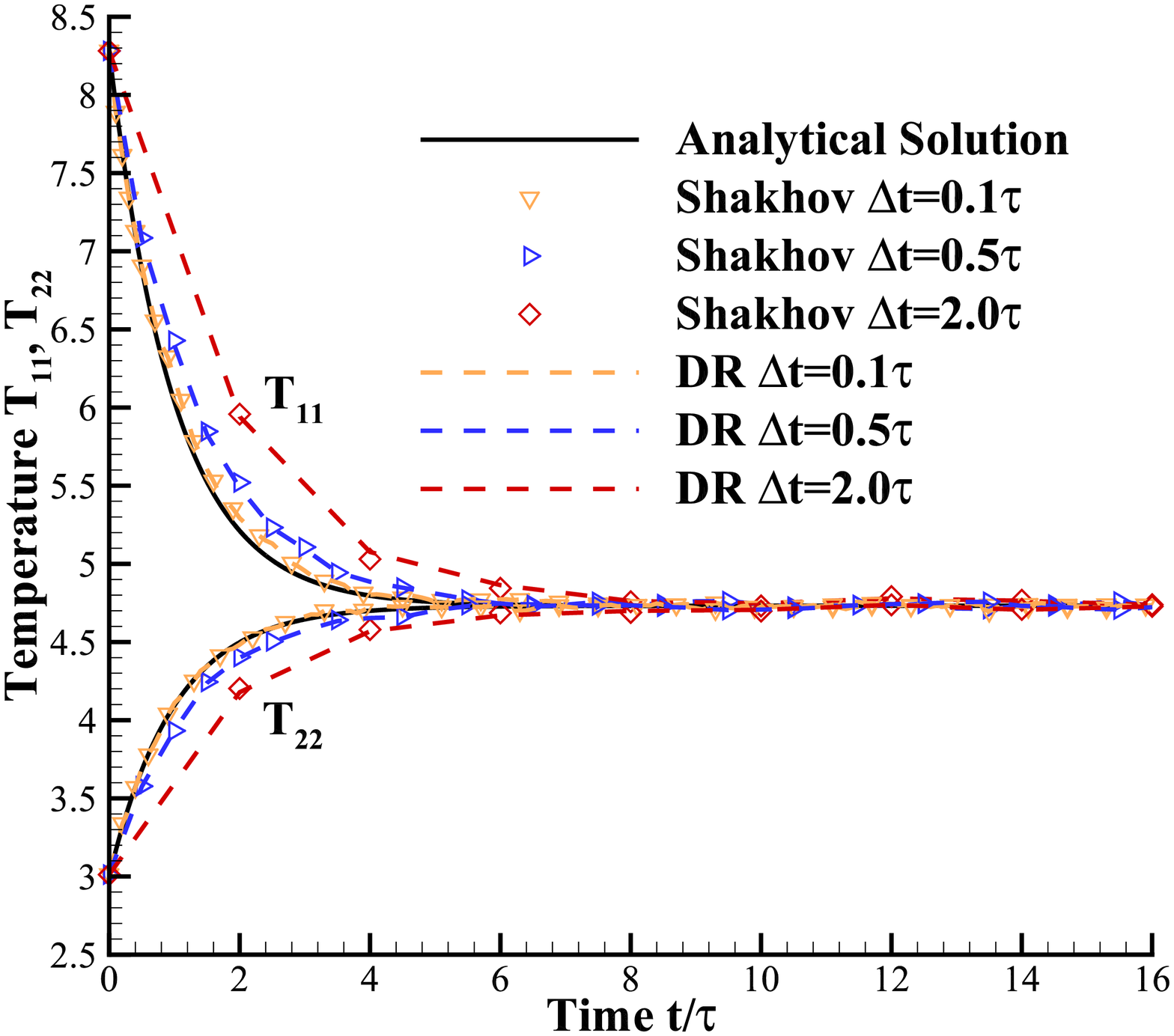}
}\hspace{0.05\textwidth}%
\subfigure[\label{Fig:0D_heat}]{
\includegraphics[width=0.45\textwidth]{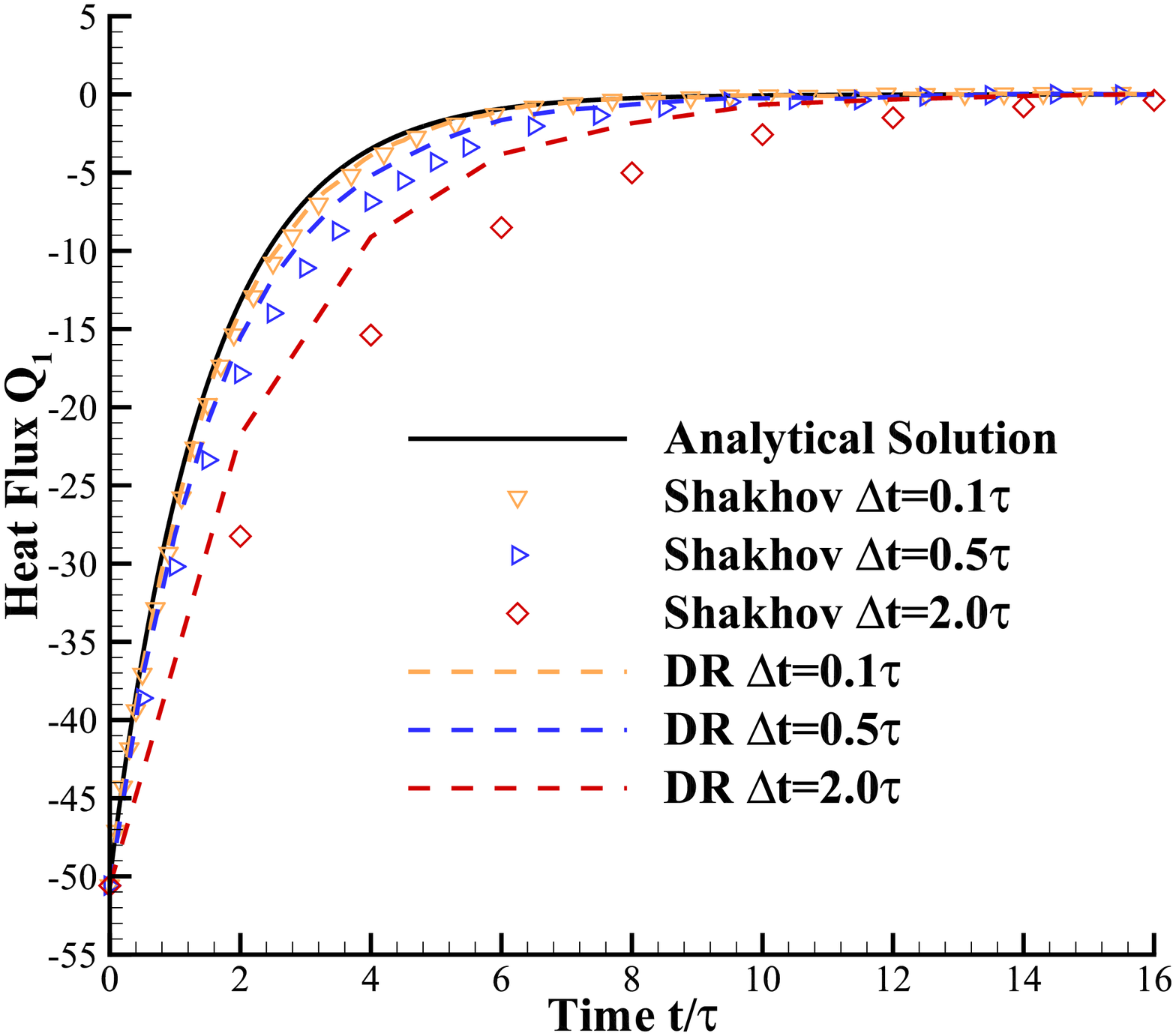}
}
\caption{\label{Fig:0D} The time evolution of anisotropic temperature and heat flux prediction by DR and Shakhov, (a) anisotropic temperature, (b) heat flux}
\end{figure}

\end{document}